\documentclass{aa}  

\usepackage{graphicx}
\usepackage{txfonts}
\usepackage{natbib,twoopt}
\usepackage[breaklinks=true]{hyperref} 
\bibpunct{(}{)}{;}{a}{}{,} 
\makeatletter
\newcommandtwoopt{\citeads}[3][][]{\href{http://adsabs.harvard.edu/abs/#3}%
{\def\hyper@linkstart##1##2{}%
\let\hyper@linkend\@empty\citealp[#1][#2]{#3}}}
\newcommandtwoopt{\citepads}[3][][]{\href{http://adsabs.harvard.edu/abs/#3}%
{\def\hyper@linkstart##1##2{}%
\let\hyper@linkend\@empty\citep[#1][#2]{#3}}}
\newcommandtwoopt{\citetads}[3][][]{\href{http://adsabs.harvard.edu/abs/#3}%
{\def\hyper@linkstart##1##2{}%
\let\hyper@linkend\@empty\citet[#1][#2]{#3}}}
\newcommandtwoopt{\citeyearads}[3][][]%
{\href{http://adsabs.harvard.edu/abs/#3}
{\def\hyper@linkstart##1##2{}%
\let\hyper@linkend\@empty\citeyear[#1][#2]{#3}}}
\makeatother
\usepackage{tabularx}
\usepackage{xcolor}

\def\HI{\ion{H}{I}}

\def\OIII{[\ion{O}{III}]}

\newcommand{\couno}{{CO(1-0)}}

\newcommand{\kms}{$\,$km~s$^{-1}$}

\newcommand{\ergscm}{$\,$erg$\,$s$^{-1}\,$cm$^{-2}$}

\newcommand{\Jyb}{Jy beam$^{-1}$}
\newcommand{\mJyb}{mJy beam$^{-1}$}

\newcommand{\whz}{W~Hz$^{-1}$}
\newcommand{\msun}{{${\rm M}_\odot$}}

\newcommand{\eg}{\mbox{e.g.}}
\newcommand{\ie}{\mbox{i.e.}}

\newcommand{\forn}{\mbox Fornax~A}

\newcommand{\ngcsix}{\mbox NGC~1316}

\newcommand{\pl}{{\em{Planck}}}
\newcommand{\meer}{{MeerKAT}}

\hypersetup{draft}

\begin{document} 

   \title{The flickering nuclear activity of \forn}

   	\author{F. M. Maccagni\inst{1},
   	M. Murgia\inst{1},
	P. Serra\inst{1},
	F. Govoni\inst{1},
	K. Morokuma-Matsui\inst{2},
	D. Kleiner\inst{1},
	S. Buchner\inst{3},
	G. I. G. J\'ozsa\inst{3,4,5},
	P. Kamphuis\inst{6},
	S. Makhathini\inst{4},
	D. Cs. Moln\'ar\inst{1},
	D. A. Prokhorov\inst{7},
	A. Ramaila\inst{3,4},
	M. Ramatsoku\inst{1,4},
	K. Thorat\inst{3,4},
	O. Smirnov\inst{3,4}}
   \institute{INAF -- Osservatorio Astronomico di Cagliari, via della Scienza 5, 09047, Selargius (CA), Italy
   \and
    Institute of Astronomy, Graduate School of Science, The University of Tokyo, 2-21-1 Osawa, Mitaka, Tokyo 181-0015, Japan
	\and
	South African Radio Astronomy Observatory, Black River Park, 2 Fir Street, Observatory, Cape Town, 7925, South Africa
	\and
	Department of Physics and Electronics, Rhodes University, PO Box 94, Makhanda 6140, South Africa
	\and
	Argelander-Institut für Astronomie, Auf dem H\"ugel 71, D-53121 Bonn, Germany
	\and
	Ruhr-University Bochum, Faculty of Physics and Astronomy, Astronomical Institute, 44780 Bochum, Germany
	\and
	School of Physics, University of the Witwatersrand, Private Bag 3, 2050 Johannesburg, South Africa
	\\
	\email{filippo.maccagni@inaf.it}
	}
              
   \date{Submitted 7 October 2019 / Accepted 20 November 2019}

  \abstract 
   {We present new observations of \forn\ taken at $\sim 1$~GHz with the MeerKAT telescope and at $\sim6$~GHz with the Sardinia Radio Telescope (SRT). The sensitive (noise $\sim 16\,\mu$\Jyb), high resolution ($\lesssim 10\arcsec$) MeerKAT images show that the lobes of \forn\ have a double-shell morphology, where dense filaments are embedded in a diffuse and extended cocoon. We study the spectral properties of these components by combining the MeerKAT and SRT observations with archival data between $84$ MHz and $217$ GHz. For the first time, we show that multiple episodes of nuclear activity must have formed the extended radio lobes. The modelling of the radio spectrum suggests that the last episode of injection of relativistic particles into the lobes started $\sim 24$ Myr ago and stopped $12$ Myr ago. More recently ($\sim 3$ Myr ago), a less powerful and short ($\lesssim 1$ Myr) phase of nuclear activity generated the central jets. Currently, the core may be in a new active phase. It appears that \forn\ is rapidly flickering. The dense environment in which Fornax A lives has lead to a complex recent merger history for this galaxy, including mergers spanning a range of gas contents and mass ratios, as shown by the analysis of the galaxy's stellar- and cold-gas phases. This complex recent history may be the cause of the rapid, recurrent nuclear activity of \forn.}

   \keywords{galaxies: individual: Fornax A, NGC 1316 --
                galaxies: jets -- 
                galaxies: active --
                radio continuum: galaxies --
                	radiation mechanisms: non-thermal
               }

   \titlerunning{The flickering activity of \forn}
   \authorrunning{Maccagni et al.}
   
   \maketitle

%

\section{Introduction}
\label{sec:intro}

Active galactic nuclei (AGN) are associated with the accretion of material onto a super-massive black hole (SMBH) at the centre of their host galaxy. The energy released by the AGN into the surrounding interstellar medium through radiation and/or relativistic jets of radio plasma, can drastically change the fate of its host galaxy by removing, or displacing, the gas in the galaxy and preventing it from cooling to form new stars~\citep[see, for example,][]{fabian2012,mcnamara2012,bolatto2013,fluetsch2019}. This mechanism, commonly called `AGN feedback', likely plays a fundamental role in regulating the star formation of the host galaxy as well as the observed relation between the masses of the bulge and of the SMBH~\citep[\eg][]{bower2006,croton2006,booth2009}. Numerical simulations of galaxy evolution indicate that only multiple phases of nuclear activity may prevent the hot circum-galactic gas from cooling back, and therefore explain the rapid quenching of star formation in early-type galaxies~\citep{ciotti2010,ciotti2018}. The timescale of the different phases of activity may depend on the environment~\citep{hogan2015}. 

Different observations of AGN at optical and radio wavelengths have suggested that the phase of accretion onto the SMBH, \ie\ the active phase, is short compared to the lifetime of the galaxy, and that it may be recurrent~\citep[\eg][]{woltjer1959,marconi2004,saikia2009,shabala2016,kuzmicz2017,morganti2017}. In the optical bands, the multiple episodes of nuclear activity are sometimes identified by the presence of `light echoes', \ie\ clouds of gas ionised by an AGN in the outskirts of a galaxy without an active core~\citep[see, for example,][]{lintott2009,jozsa2009,comerford2017}. In the radio band, multiple phases of activity can be identified by the presence of both a flat-spectrum core, indicating a recent activity, and a large-scale diffuse emission around the AGN, with a steep-spectrum brightening at low frequencies ($\lesssim 1$ GHz), tracing the remnant of a past activity~\citep[][]{jamrozy2004,parma2007,shulevski2012,brienza2016}. In some cases, jets related to the nuclear activity of the present epoch may also be found~\citep[\eg][]{jones2001,saikia2009,shulevski2012}.

The radio emission of AGN allows us to measure the duty-cycle of the nuclear activity. In particular, the steepening of the radio spectrum is often interpreted as radiative ageing of the electron population in the relativistic plasma~\citep[see, for example,][]{carilli1991,komissarov1994,parma1999,murgia1999,orru2010,murgia2011,murgia2012,harwood2013,kolokythas2015}.

In nearby AGN different studies traced the history of injection of relativistic particles from the SMBH into the radio jets and lobes, through the pixel-by-pixel study of the spectral index and break-frequency maps of the AGN radio spectrum~\citep[\eg][]{murgia2010a,murgia2010b,orru2010,deGasperin2012}. In particular, these studies related the energetic output of each episode of AGN activity to the fate of the host galaxy and its surrounding environment~\citep[][]{gizani2003,stanghellini2005,deGasperin2014,brienza2018}. 

In this paper, we study the radio spectrum of \forn, the third brightest nearby radio galaxy \citep[$D_L=20.8 \pm 0.5$ Mpc;][]{cantiello2013}\footnote{Throughout this paper we use a $\Lambda$CDM cosmology, with Hubble constant $H_0$ = 70\kms\ Mpc$^{-1}$ and $\Omega_\Lambda = 0.7$ and $\Omega_{\rm M} = 0.3$. At the distance of \forn\ the image scale is 101 parsec/arcsec.} after Centaurus A and \mbox{M 87}, to determine the time-scale and the duty-cycle of its nuclear activity.

\subsection{\forn}
\label{sec:fornaxA}

\forn\ is one of the most fascinating radio sources in the local Universe because of its filamentary extended radio lobes~\citep[$\sim 1.1^\circ$][]{ekers1983,fomalont1989,bernardi2013}. To the south of the galaxy, a `bridge' of synchrotron emission connects the two lobes. In the centre, two radio jets are embedded in the host galaxy ($r\lesssim 6$ kpc) and exhibit an s-shaped morphology \citep[][]{geldzhaler1978,geldzahler1984}. The emission of the jets extends all the way to the centre of the galaxy at angular resolutions $\gtrsim 1$\arcsec~\citep[][]{geldzahler1984}. Most of the radio emission is produced in the extended lobes. At $1.4$~GHz, their total flux density is $121$ Jy while that of the jets (including the galaxy centre) is $\sim300$ mJy~\citep[][]{fomalont1989}. The very central flux density is $\sim100$~mJy and $\sim 30$ mJy at $1.4$~GHz and $4.8$~GHz, respectively (at the resolution of $\sim 1$\arcsec) while it is $\sim 7$ mJy at $15$~GHz~\citep[resolution $\sim0.1$\arcsec;][]{geldzahler1984}. However, no central emission is detected at $2.2$~GHz and $8.4$~GHz down to $3$ and $6$~mJy ($1$~sigma upper limits) at a resolution of $90$ and $27$ mas, respectively ~\citep{jones1994,slee1994}. A summary of the properties of \forn\ is shown in Table~\ref{tab:forna}.

\forn\ is hosted by the giant early-type galaxy \ngcsix, which is the brightest member of a galaxy group at the outskirts of the Fornax cluster, likely falling into it~\citep{drinkwater2001}. The brightest cluster galaxy NGC 1399 is located $~\sim 4^\circ$ to the north-east of Fornax A. \ngcsix\ shows clear indications of a past major merger event, that likely formed the several tails and loops at the outskirts of the stellar body~\citep{schweizer1980,grillmair1999,mackie1998,carlqvist2010,galametz2012,duha2016}. Deep photometric observations indicate that \ngcsix\ may be in a later phase of mass assembly, where smaller satellites recursively accrete into the galaxy~\citep{iodice2017}. The major merger event likely brought large amounts of dust, cold molecular gas~\citep{horellou2001,roussel2007,galametz2014,morokuma2019} and neutral hydrogen~\citep[][]{horellou2001,serra2019} in the centre and around the galaxy.
 
Based on the spread in age of the globular clusters hosted by \ngcsix\, the merger is estimated to have occurred $1$ -- $3$ Gyr ago~\citep{schweizer1980,goudfrooij2001,sesto2016,sesto2018}, and it has been suggested to have possibly triggered the nuclear activity of \forn~\citep[\eg\ ][]{ekers1983,fomalont1989,mckinley2015}. Nevertheless, large uncertainties remain on the timescale of formation of the radio lobes. Moreover, this past merger event does not well explain the properties of the central emission~\citep{geldzhaler1978,geldzahler1984}, nor the soft X-ray cavities between the lobes and the host galaxy~\citep{lanz2010}. 

Crucial information on the pressure of the medium through which the radio lobes are expanding could be provided by high energy observations, but the extent of the radio lobes of \forn\ and their proximity does not allow a complete mapping of the X-ray halo surrounding the radio source. Nevertheless, some regions of the West lobe of \forn\ are bright at high energies showing $\gamma$-ray emission~\citep{ackermann2016}. The soft X-ray spectrum of the West lobe shows hints of diffuse thermal emission that may trace material entrained by the expanding radio lobes~\citep{seta2013}, which may also explain the presence of the low-polarisation patches in the filaments of the lobes~\citep{anderson2018}. ~\citet{iyomoto1998} show that the X-ray spectrum of the nucleus of \forn\ suggests that the AGN is currently inactive, while \citet{lanz2010} identify two X-ray cavities between the host galaxy and the radio lobes. 

\citet{mckinley2015} studied the total flux density of \forn\ in the radio, X-ray and $\gamma$-ray frequencies, providing information on the energy that must have been injected by the AGN into the surrounding inter-galactic medium (IGM). To gain further insights on how the AGN formed the giant radio lobes and the central emission of \forn, it is crucial to spatially resolve these components over a broad radio band, and trace the differences in their flux density distributions.

In this paper, we present a new \meer~\citep[][]{jonas2016} interferometric observation of \forn\ in the L-band  ($900$-$1710$ MHz) and a new Sardinia Radio Telescope (SRT) single-dish observation in the C-band (5800-6700 MHz). We use these observations, along with archival observations from other radio telescopes, to study the main properties of the flux density of \forn\ between $84$~MHz and $217$~GHz and to understand the mechanisms and timescales of formation of the lobes and of the jets of this AGN. The paper is structured as follows: in the next section we present the data used to study the radio emission of \forn. In Sect.~\ref{sec:results} we show the radio spectrum of the lobes and of the central emission. Section~\ref{sec:coreProp} focuses on the properties of the radio jets emission determined from the high resolution \meer\ observation. In Sect.~\ref{sec:spModAll} we infer the main physical parameters that characterise the flux density of the lobes and central emission. In Sect.~\ref{sec:spiBrGen} we discuss the spectral flux index and break-frequency maps obtained from the \meer\ images. In Sect.~\ref{sec:disc} we relate the properties of the radio emission of \forn\ to the nuclear activity that formed it, providing indications on how the extended radio lobes may have expanded in the IGM and suggesting a timeline for the nuclear activity. Section~\ref{sec:conc} presents a summary of the main results of this paper.

\begin{figure*}
	\begin{center}
		\includegraphics[trim = 0 0 0 0, width=\textwidth]{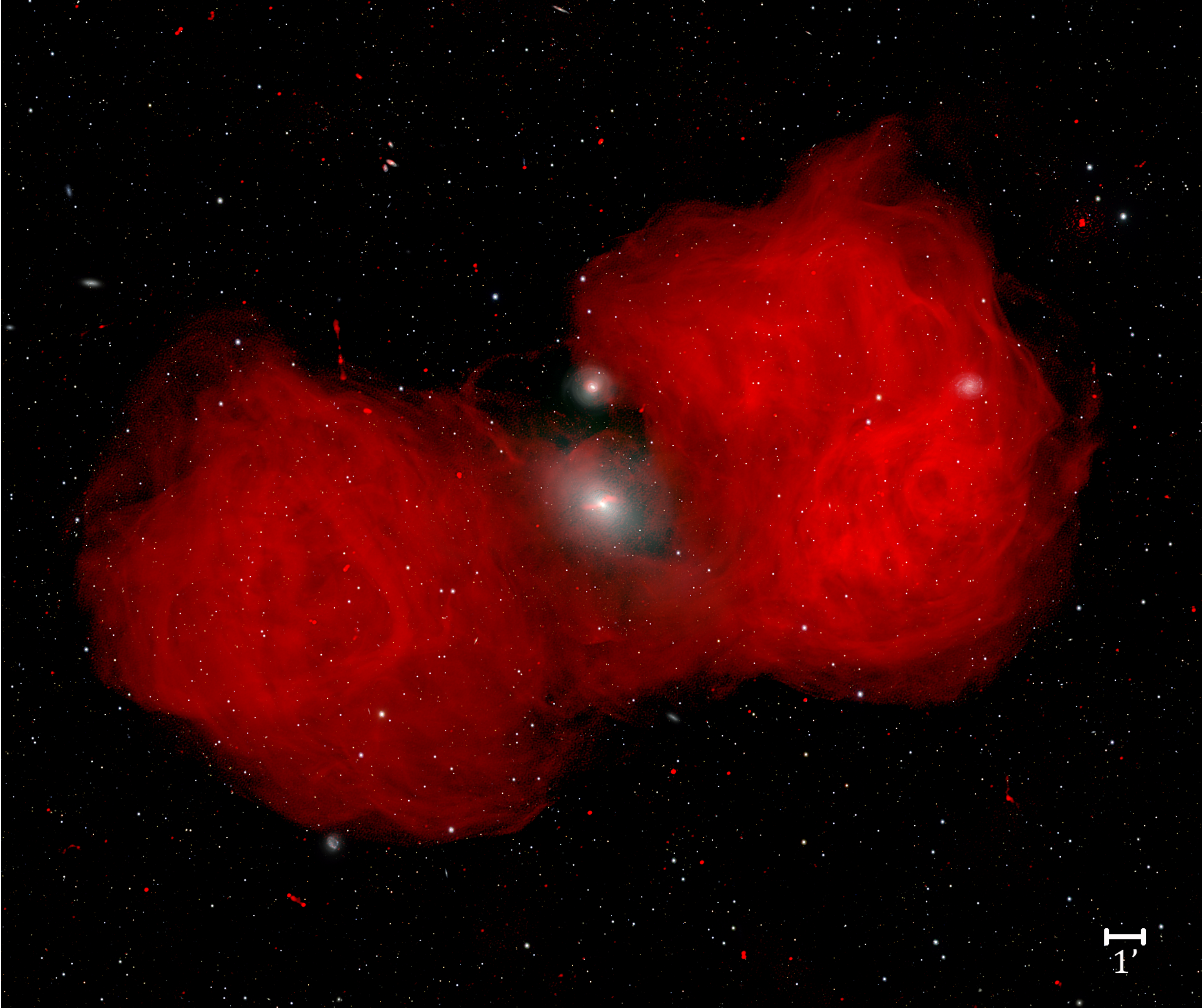}
		\caption{\forn\ seen by MeerKAT at 1.44 GHz. The radio emission is in red, as well as the background and foreground sources. The 3-colour composite image (from the {\em gri} bands) of the same field of view is taken from the Fornax Deep Survey~\citep{iodice2017}.}
		\label{fig:fornaxA} 
	\end{center}
\end{figure*}

\begin{table}[tbh]
	\caption{General properties of \forn}
	\centering
	\label{tab:forna}
	\begin{tabularx}{\columnwidth}{l X c}  
		\hline\hline							     
		Parameter 								&  Value 				&	Ref  \\
		\hline
			Right Ascension $\alpha$ (J2000) 	&  $03^\mathrm{h}22^\mathrm{m}41.7^\mathrm{s}$ & (*) \\ 
			Declination $\delta$ (J2000) 		&$-37^\mathrm{d}12^\mathrm{m}30^\mathrm{s}$ & (*)\\
			$D_{\rm L}$ 					  	&  20.8 Mpc 		  	& (1)\\
			$S_{1.4\,\rm GHz}\, (\rm total)$	& $128$ Jy				& (*)	\\			
			$S_{1.4\,\rm GHz}\, (\rm peak)$		& $125$ mJy				& (*)	\\
			$P_{1.4\,\rm GHz}\, (\rm lobes)$	& $1\times 10^{25}$~\whz	& (*)	\\
			$P_{1.4\,\rm GHz}\, (\rm peak)$		& $5.5\times 10^{22}$~\whz& (*)	\\
			Lobes radius						& $20\arcmin-28\arcmin$ (122-170 kpc) & (*) \\
			$B_{\rm eq}\,\,({\rm Lobes}\, r<170 kpc)$ & $3.0\, \mu$G	& (*)	\\ 
			Magnetic field, $B_{\rm av}$ 		& $2.6\pm 0.3\, \mu$G	& (2,3) \\
			IC magnetic field, $B_{\rm IC}$		& $3.2\, \mu$G 			& (3) \\
			$B/B_{\mathrm IC}$					& $0.8$ 				&	(3)  \\
	\hline                           
	\end{tabularx}
	\tablebib{(*) this work (1) \citet{cantiello2013} (2) \citet{anderson2018} (3) \citet{mckinley2015}. }
\end{table}

\section{Multi-wavelength observations}
\label{sec:dRed}

The goals of our study are to characterise the AGN activity history that created the large radio lobes and the central emission of \forn. The first goal requires, over a wide range of frequencies, images sensitive to low-surface-brightness emission over a wide field of view, since the lobes are faint and extend for $\sim 1^\circ$ in the sky. To achieve the second goal we need images with $\sim 10\arcsec$ angular resolution to resolve the central emission.
 
Over the wide frequency range required to infer the history of the nuclear activity, most available datasets are not suitable for both goals because typically those with a large field of view do not have adequate angular resolution -- the only exception being the new MeerKAT data presented here. 

 For these reasons, we select two different sets of observations to analyse the radio emission in the lobes and in the centre of \forn. To study the lobes we need wide field of view observations sensitive to the diffuse emission of the lobes (\ie\ good $uv$-coverage on the short baselines), while arcsecond resolution is not needed. Hence, between $84$ and $200$~MHz we choose observations from the GaLactic and Extra-galactic All-sky MWA survey \citep[GLEAM;][]{hurleywalker2017} of the Murchinson Wide Field Array. We use the \meer\ observation to generate images of \forn\ at $1.03$ and $1.44$~GHz. At $1.5$~GHz we choose archival Very Large Array (VLA) observations \citep{fomalont1989}. Between $5.7$ and $6.9$~GHz we use the new observation of the SRT. Between $70$~GHz and $217$~GHz, we select images of \forn\ from the final release of the \pl\ foreground maps \citep[][]{planck2018IV}.

The second set is needed to study the central emission, hence we select observations with arcsecond resolution: the MeerKAT images at $1.03$ and $1.44$~GHz, archival VLA observations at $4.8$ and $15$~GHz \citep{geldzahler1984}, and an observation at $108$~GHz taken with the Morita Array of the Atacama Large Millimeter and sub-millimeter Array (ALMA)~\citep[][]{morokuma2019}.

The main properties of all observations considered in this paper are summarised in Table~\ref{tab:forObs}. In the following sections, we provide further details on the new MeerKAT, SRT  and ALMA observations (see Sects.~\ref{sec:dRedMeerKAT}~\ref{sec:dRedSRT} and~\ref{sec:dRedALMA}, respectively). Details on the reduction of archival observations are given in Appendix~\ref{appendix:contIms}.

\begin{table*}
	\caption{Main properties of the observations}
	\centering
	\label{tab:forObs}
	\begin{tabularx}{\textwidth}{X X X c c}  
		\hline\hline							     
		Telescope	& Frequency 	& Bandwidth & Spatial Resolution	& Image r.m.s.	\\
					&  [GHz]		& [MHz]		& 					& [\mJyb]		\\
		\hline
		MWA$^\dag$	& 0.084				& 31&$5.0\arcmin\times4.7\arcmin$		&	256\\
         			& 0.105				& 31&$3.6\arcmin\times3.4\arcmin$		&	102\\
         			& 0.20				& 61&$2.3\arcmin\times2.2\arcmin$		&	28.1\\
  		VLA$^\dag$	& 0.32				& 26&$1.5\arcmin\times0.6\arcmin$		&	3.18\\	
		\meer\		& 1.03				&100&$11.2\arcsec\times9.1\arcsec$&0.03	\\	
					& 1.44				&120&$6.8\arcsec\times5.8\arcsec$		& 	0.02\\
		VLA$^\dag$	& 1.50				& 12&$14\arcsec\times14\arcsec$		&	0.15\\
  		VLA$^\star$	& 4.86 				& 100&$3.9\arcsec\times3.9\arcsec$		&	0.08\\	
		SRT$^\dag$	& 5.80				& 100&$3.3\arcmin\times3.3\arcmin$		&	7.19\\	
					& 5.90				& 200&$3.2\arcmin\times3.2\arcmin$		&	7.71\\	
					& 6.10				& 200&$3.1\arcmin\times3.1\arcmin$		&	6.79\\	
					& 6.30				& 200&$3.0\arcmin\times3.0\arcmin$		&	6.73\\	
					& 6.50				& 200&$2.9\arcmin\times2.9\arcmin$		&	6.84\\	
					& 6.70				& 200&$2.8\arcmin\times2.8\arcmin$		&	6.95\\	
					& 6.87				& 145&$2.7\arcmin\times2.7\arcmin$		&	6.05\\	
  		VLA$^\star$	& 14.9 				&  100&$4.1\arcsec\times4.1\arcsec$	&	0.12\\					
		\pl$^\dag$	& 30.0				& 6000&$32\arcmin\times 32\arcmin$		&	482\\	
					& 40.0				& 4800&$27\arcmin\times27\arcmin$		&	336\\					
					& 70.0				& 14$\times 10^3$&$13\arcmin\times 13\arcmin$&	108\\					
					& 100				& 32$\times 10^3$&$9.7\arcmin\times 9.7\arcmin$&	81.8\\		
		ALMA - Morita Array$^\star$		& 108 & $8\times10^3$ 	&$18.1\arcsec\times 9.0\arcsec$			& $0.21$	\\	
		\pl$^\dag$	& 143				& 46$\times 10^3$&$7.2\arcmin\times 7.2\arcmin$&	36.9\\
					& 217				& 65$\times 10^3$&$5.0\arcmin\times 5.0\arcmin$&	26.5\\			
		\hline                           
	\end{tabularx}
	\tablefoot{$^\star$ observations considered only for the analysis of the radio emission in the central kpc.\newline $^\dag$ observations considered only for the analysis of the radio emission in the East and West lobe.}
\end{table*}

\subsection{MeerKAT: $1.03$ GHz and $1.44$ GHz }
\label{sec:dRedMeerKAT}
\meer\ is an interferometric radio telescope built in South Africa as a precursor for the `small dish plus single-pixel, wide-band feed' component of the Square Kilometer Array~\citep[][]{jonas2016,camilo2018}. In preparation for the MeerKAT Fornax Survey~\citep[][]{serra2016}, whose aim is to study galaxy evolution in the Fornax cluster, we analyse a \meer\ commissioning observation of \forn.

This observation was taken on June 2, 2018 with a reduced array of 40 antennas and the SKARAB-4K correlator (4096 channels with resolution of approximately 209 kHz) over the full MeerKAT bandwidth, $0.86$ -- $1.71$ GHz. The total observing time on target was 7.8 hours.

A complete description of the reduction of this observation can be found in~\cite{serra2019}. Here, we focus on how we generated the continuum images of \forn\ between $0.98$ and $1.08$ GHz and between $1.38$ GHz and $1.50$ GHz, needed for the purposes of this paper. 

\begin{figure*}
	\begin{center}
		\includegraphics[trim = 0 0 0 0, width=\textwidth]{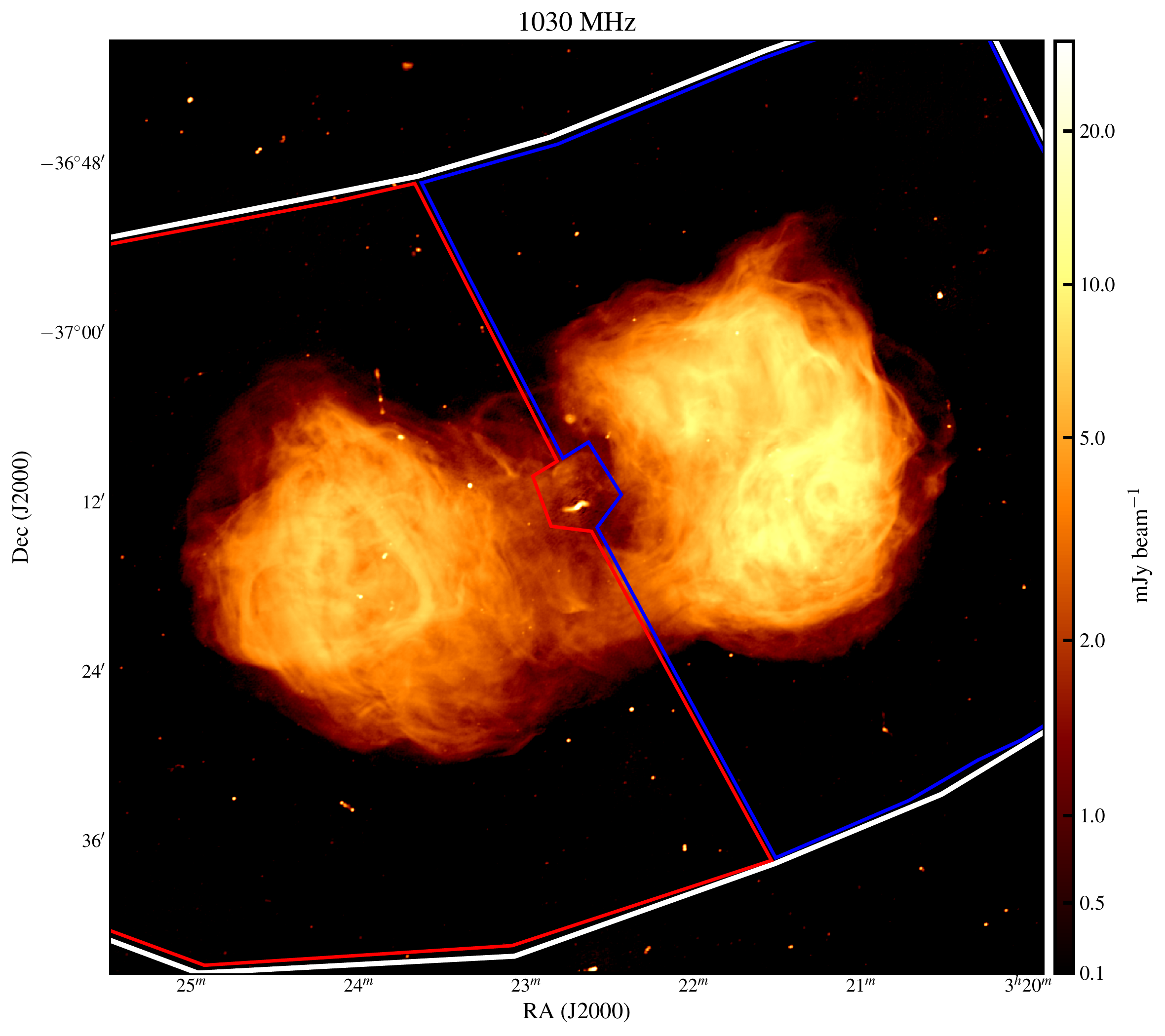}
		\caption{\forn\ seen by MeerKAT at 1.03 GHz. The white contours mark the region where we measure the total flux density of the source, the red and blue contours the regions of the East and West lobe. The synthesized beam of the image is $11.2\arcsec\times9.1\arcsec$.}
		\label{fig:fornaxAFlRegions} 
	\end{center}
\end{figure*}

The reduction of this observation was done with a new pipeline that is being developed for the reduction of continuum and spectral interferometric observations (Makhathini et al. in prep)\footnote{\url{https://github.com/ska-sa/meerkathi}}. The pipeline is set up in a modular fashion using the platform-independent radio interferometry scripting framework {\ttfamily stimela}\footnote{\url{https://github.com/SpheMakh/Stimela}}. This means that for each step of the data reduction, such as calibration, flagging, self-calibration, we use tasks from different radio astronomical packages. Imaging was performed in Stokes I using {\ttfamily WSclean} \citep{offringa2014}. Multi-scale cleaning~\citep{offringa2017} was performed within a mask marking the regions of \forn\ and of the other sources in the field. In the first imaging step, we generated a clean mask from the VLA observations at $1.5$ GHz~\citep[][~see Appendix~\ref{appendix:contIms}]{fomalont1989}. In the subsequent steps, the clean mask was recursively improved by running the source finder {\ttfamily SoFiA}~\citep{serra2015} on the cleaned image. We iteratively imaged with Briggs weighting {\tt robust}$=-0.5$, and we calibrated, using {\ttfamily MeqTrees}~\citep{noordam2010}, solving for frequency-independent gain phase  with a solution time interval of $2$ minutes. Shorter time intervals were found to produce noisier gain solutions with no improvement in the images, while longer time intervals failed to capture the gain variations in the data and caused low-level artefacts.

Given the large extent of the lobes of \forn\ ($1.1^\circ$), in order to accurately study the properties of their radio spectrum it is crucial to correct the images with an accurate primary beam model. We created the primary beam Jones matrix of MeerKAT for any frequency of L-band observations, using {\ttfamily eidos}~\citep[][]{asad2019}~\footnote{\url{https://github.com/ratt-ru/eidos}}. From the Jones matrix we produced the images of the real and imaginary parts of the primary beam self-correlations (xx and yy). Their combination results in the image of the primary beam in the same field of view and with the same pixel size as the continuum images. We obtained the primary beam corrected images by dividing the continuum images by the primary beam images at the corresponding frequency.

Figure~\ref{fig:fornaxA} shows the radio emission of \forn\ seen by MeerKAT at 1.44 GHz overlaid on the deep photometric image from the Fornax Deep Survey (FDS) 3-colour composite (from the {\em gri} bands) \citep{iodice2017}. The average noise in the $2^\circ$ field of view is $16$ $\mu$\Jyb. The dynamic range of the image (defined as the ratio between the peak emission of \forn\ and the noise in the image) is $\sim 7000$. This, along with the complete {\em uv}-coverage of MeerKAT over short ($\lesssim 50$ m), as well as long ($\gtrsim 2$ km), baselines allows us, for the very first time, to detect with high resolution ($6.8\arcsec\times5.8\arcsec$, PA$=117^\circ$) and signal to noise (S/N$>20$) the diffuse emission at the edges of the radio lobes. In the same short ($\sim8$ hours) observation, we also spatially resolve the central emission and the filaments of the lobes. Figure~\ref{fig:fornaxAFlRegions} shows the radio emission of \forn\ at $1.03$ GHz. The image has similar resolution ($11.2\arcsec\times9.1\arcsec$, PA$=119^\circ$) and sensitivity (noise $\sim28$ $\mu$\Jyb) as the image at $1.44$ GHz. Only in the proximity of bright ($\gtrsim 15$~m\Jyb) sources at about 1 degree from the phase centre (where the response of the MeerKAT antennas drops) the noise increases because of direction-dependent calibration effects. We did not attempt to correct for such effects.

The images obtained from the MeerKAT observation allow us to reveal the double-shell structure of the lobes of \forn, where the bright filaments appear embedded in a diffuse cocoon. For the first time, we are able to study the spectral properties of this double-shell structure and understand how the lobes of \forn\ may have formed.

\subsection{SRT: 5.7 GHz -- 6.9 GHz}
\label{sec:dRedSRT}
 
\forn\ was observed with the SRT on January 26 and February 7, 2017. These observations were conducted in the context of the development of the spectral-polarimetric wide-field imaging of the SARDARA backend \citep[SArdinia Roach2-based Digital Architecture for Radio Astronomy;][]{melis2018}. The total observing time on target was $4.5$ hours. We performed several on-the-fly (OTF) mappings in the equatorial frame in both right ascension and declination. We imaged a field of view of $1.2^\circ \times 1.2^\circ$ using four RA and three Dec scans. The average angular resolution of the SRT at this frequency is $FWHM = 3.0\arcmin$ so we set the telescope scanning speed to $6\arcmin$ s$^{-1}$ and the scan separation to $42\arcsec$ to properly sample the beam. The correlator configuration was set to $1024$ frequency channels (2.2-MHz- wide) for a total bandwidth of 2300 MHz in full-Stokes mode. We set the Local Oscillator to 5600 MHz and used a filter to select the frequency range $5700$ -- $6945$ MHz, which is relatively free from strong radio frequency interference.

Data reduction was performed with the proprietary Single-dish Spectral-polarimetry Software \citep[{\ttfamily SCUBE};][]{murgia2016}. Bandpass and flux density calibration were performed by observing respectively \mbox{3C 138} and \mbox{3C 295}, assuming the flux density scale of \cite{perley2013a}. Persistent radio frequency interference (RFI) were flagged and we applied the gain-elevation curve correction to account for the gain variation with elevation due to the telescope structure gravitational stress change. In band C, SRT observations  are only moderately affected by atmospheric absorption which depends on the weather conditions at the telescope site and on the elevation of the source. At the time of the observations, the opacity ($\tau$) was $0.008$ and $0.009$, respectively. Since, \forn\ had elevation ($el$) $\lesssim 15^\circ$, the opacity correction to the flux density is:  

\begin{equation}
\label{eq:op}
	k = \bigg({e^{\frac{-\tau}{\sin(el)}}}\bigg)^{-1} = 1.05
\end{equation}

We performed the polarisation calibration by correcting the instrumental polarisation and the absolute polarisation angle.The on-axis instrumental polarisation was determined through observations of the bright unpolarised source \mbox{3C 84}. The leakage of Stokes I into Q and U is in general less than $2\%$ across the band, with a r.m.s. scatter of $0.7-0.8\%$. We fixed the absolute position of the polarisation angle using as reference the primary polarisation calibrator \mbox{3C 138}. The difference between the observed and predicted position angle according to \cite{perley2013b} was determined, and corrected channel-by-channel. 

\begin{figure*}
	\begin{center}
		\includegraphics[trim = 0 0 0 0, width=\textwidth]{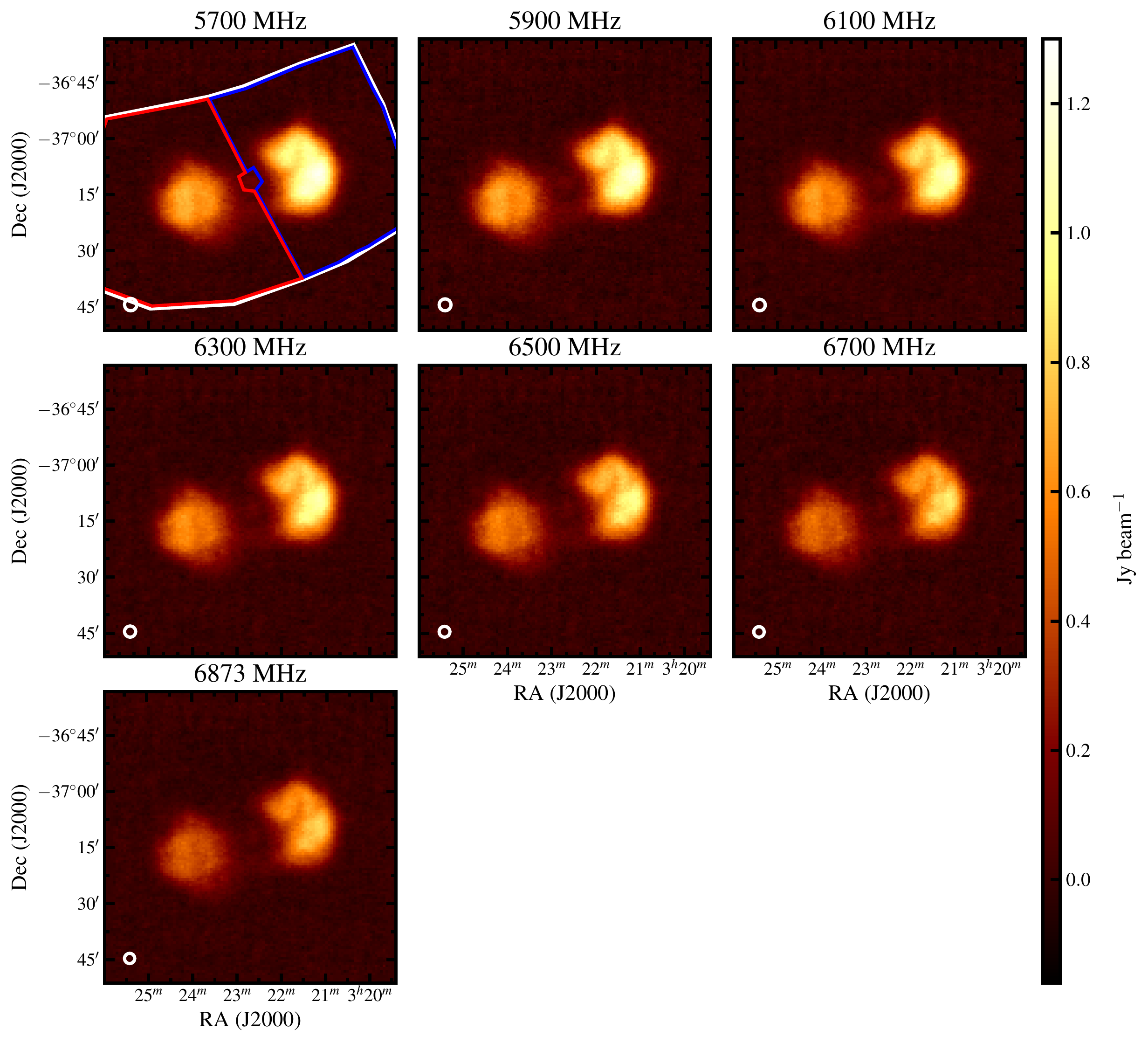}
		\caption{\forn\ seen between $5.7$ and $6.87$ GHz by the SRT. In the top left panel, the contours mark the regions where we measure the total flux density and the flux density of the lobes (in white, red and blue, respectively). The synthesized beam of the images is shown in white in the bottom left corner of each panel.}
		\label{fig:fornSRT} 
	\end{center}
\end{figure*}

All frequency cubes obtained by gridding the scans along the two orthogonal axes (RA and Dec) were then stacked together to produce full-Stokes I, Q, U images of an area of $1.2$ square degree centred on \forn. In the combination, the individual image cubes were averaged and de-stripped by mixing their Stationary Wavelet Transform (SWT) coefficients \citep[see][, for details]{murgia2016}. For the purposes of this work we further averaged the total intensity spectral cube into seven sub-bands of about $200$ MHz in width. The resulting images are shown in Fig.~\ref{fig:fornSRT}, their noise varies between $7.2$ and $6.0$~\mJyb (see Table~\ref{tab:forObs}). We use these observations only to analyse the emission of the radio lobes of \forn, since the SRT beam does not resolve the central radio emission of \forn.

\subsection{ALMA: 108 GHz}
\label{sec:dRedALMA}

Observations of \forn\ centred at 108 GHz were carried out during cycle 5 as part of an ALMA survey of $65$ galaxies in the Fornax cluster (PI Kana Morokuma-Matsui, project code 2017.1.00129.S)

\begin{figure*}
	\begin{center}
		\includegraphics[width=\textwidth]{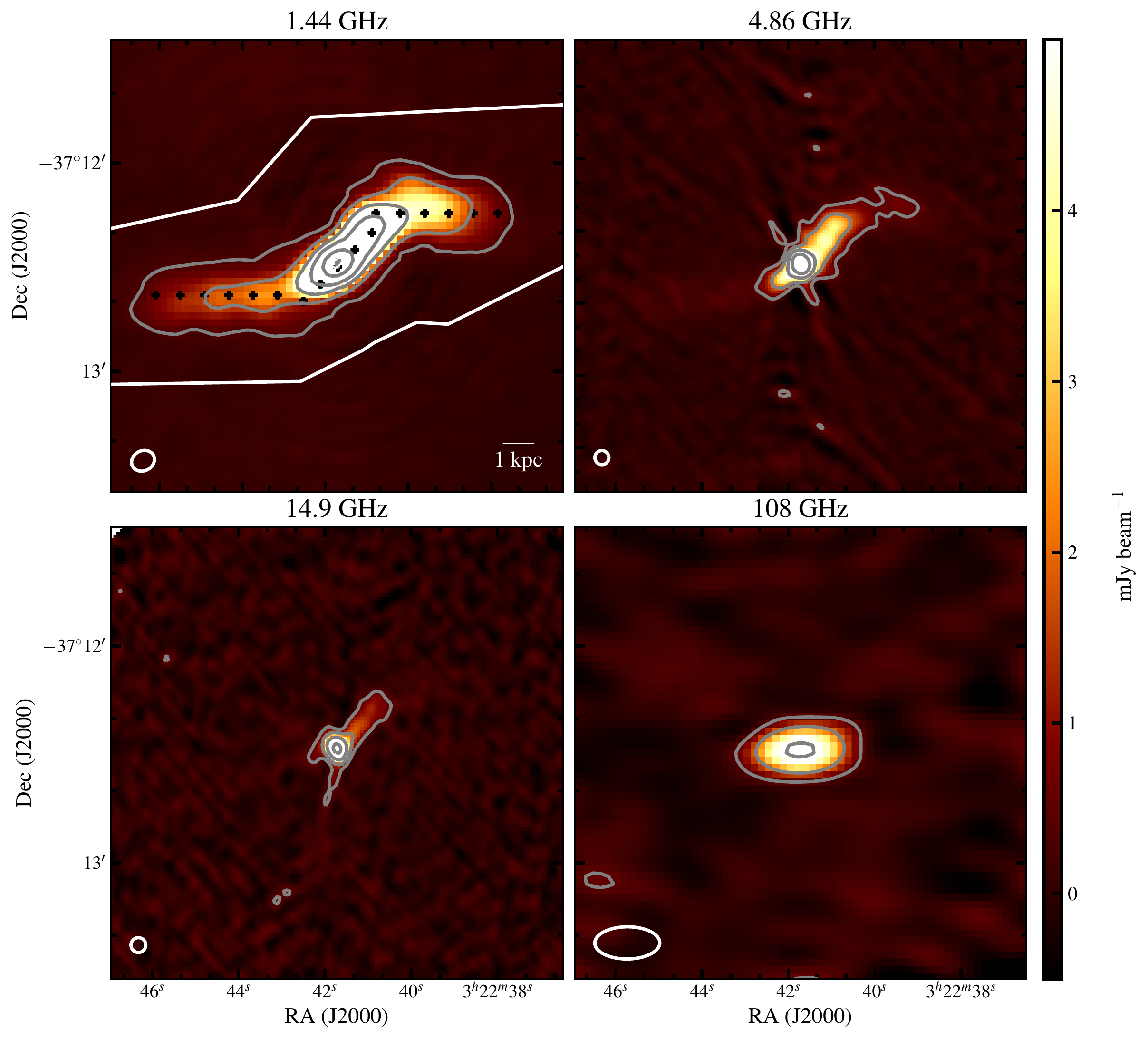}
		\caption{The central emission of \forn\ seen at 1.44 GHz by \meer\ ({\em top left panel}), at 4.86 GHz ({\em top right panel}), 14.9 GHz ({\em bottom left panel}) by the VLA and at 108 GHz ({\em bottom right panel}) by ALMA. The PSF of the images is shown in white in the bottom left corner. Contour levels start at $0.6$~\mJyb, increasing by factors of 3. In the top left panel, the black dots show where we measured the variations in flux density and width along the jets (see Sect.~\ref{sec:coreProp} for further details).}
		\label{fig:fornCore} 
	\end{center}
\end{figure*}

These observations were taken during an observing campaign that lasted from October 16 to December 21, 2017, and consist of 10 different pointings of the ALMA-Morita array within the innermost $2\arcmin\times3\arcmin$ of \forn\ ($\sim12\times18$ kpc). The chosen array configuration has baselines ranging from 8.9 m to 48.9 m. The main goal of these observations was to observe the distribution and kinematics of the molecular gas in \ngcsix, at traced by the \couno\ line at $\nu_{\rm rest} = 115.27$ GHz~\citep{morokuma2019}. Three 2-GHz wide spectral windows were dedicated to continuum studies (centred at 113.39 GHz , 103.27 GHz and 101.39 GHz, respectively). The continuum image shown in Fig.~\ref{fig:fornCore} (bottom right panel) was generated considering all four spectral windows of the observation, where the frequency range of the detected $^{12}$CO (1-0) line emission and of the tentative detection of the CN hyper-fine transitions at $112.85$ GHz have been flagged. This image allows us to estimate the flux density of the continuum emission in the innermost kpc of \forn\ in the millimetre band.

The calibration of the Morita-Array observations was conducted with the ALMA pipeline version 40896~\citep[Pipeline CASA51-P2-B;][]{petry2012} at the Joint ALMA Observatory (JAO). The continuum image was generated using {\ttfamily CASA}. In particular, we used the task {\tt tclean} with {\tt Briggs} weighting and {\tt robust} parameter 0.5. We performed cleaning using the {\tt hogbom} algorithm within a mask selected by standard {\tt auto-threshold} and mosaicking options ({\tt sidelobethreshold} = 1.25, {\tt noisethreshold} = 5, {\tt meanbeamfrac} = 0.1, {\tt lownoisethreshold} = 2.0, {\tt negativethreshold} = 0.0, {\tt gridder} = `mosaic'). The synthesized beam is $18.1\arcsec\times9.0\arcsec$ ($\sim 1.8\times 0.99$~kpc) with $PA=-89^\circ$, and the noise is $0.21$~\mJyb.

\section{Radio flux density of \forn}
\label{sec:results}
In this section we measure the flux density distribution of the radio lobes of \forn\ between $84$ MHz and $217$ GHz, and of the central emission between $1.03$ and $108$ GHz. 

\subsection{The lobes}
\label{sec:sedLobe}

We analyse the properties of the radio lobes of \forn\ separately. The regions enclosing the East and West lobe are shown in Fig.~\ref{fig:fornaxAFlRegions}. We define the regions large enough to cover the radio lobes at all frequencies. The figure also shows the region where we measure the total flux of \forn\ (that is dominated by the emission of the two lobes). We use this measurement to compare our results to those in the literature~\citep[][]{mckinley2015,perley2017}, where the analysis of the flux density distribution was performed on the sum of the emission of the two lobes (see Appendix~\ref{appendix:contIms} for further details). 

Before measuring the flux, all images are corrected for the primary beam response and re-projected to the same reference frame. In each image, the noise per beam is equal to the dispersion of the signal in the field of view, once all sources have been excluded. This is a reliable measurement of the noise only if it has constant power spectrum density (\ie\ white noise). This is the case for all considered observations except for the images taken from the full-sky \pl\ foreground maps. The background of the \pl\ images is not flat, but has large non-zero patches. We include those large-scale fluctuations of the background in our estimate of the \forn\ flux uncertainties. The origin of those patches could be related to the background subtraction process. To obtain a reliable estimate of the error on the \pl\ flux densities of \forn, we extract the images on a large field of view ($\sim 4$ degrees), and we measure the error on the flux density as the dispersion of the flux density of $100$ independent regions in the field of view of shape and size equal to the regions where we measure the flux density of the lobes (Fig.~\ref{fig:fornaxAFlRegions}), Further details on the reduction and analysis of the \pl\ observations are given in Appendix~\ref{sec:dRedPlanck}.

The total flux density of \forn\ and the flux densities of the East and West lobes with errors are shown in Table~\ref{tab:sedLobes}. At each frequency, the error on the flux density is the combination between the noise in the image (shown in Table~\ref{tab:forObs} for each observation) and the error due to the uncertainties in the calibration of the observations, which is $20\%$ for MWA~\citep[][]{mckinley2015}, $15\%$ for \pl~\citep{planck2018IV}, $5\%$ for SRT~\citep[][]{egron2017,battistelli2019} and $3\%$ for VLA~\citep[][]{perley2017}, respectively. For the \meer\ observation we estimate a flux-calibration error of $5\%$, to allow for uncertainties in the flux of the gain calibrator. Archival observations from the VLA and MWA are on the \citet{baars1977} flux density scale, \pl\ observations are on an absolute flux density scale~\citep{planckScale}, while \meer, SRT and ALMA observations have been calibrated according to~\citet{perley2017}. Discrepancies caused by these different scales are typically on the order of $\sim 1$--$3\%$, and are within the errors we assume on the flux density measurements.

\begin{figure*}
	\begin{center}
		\includegraphics[width=0.44\textwidth]{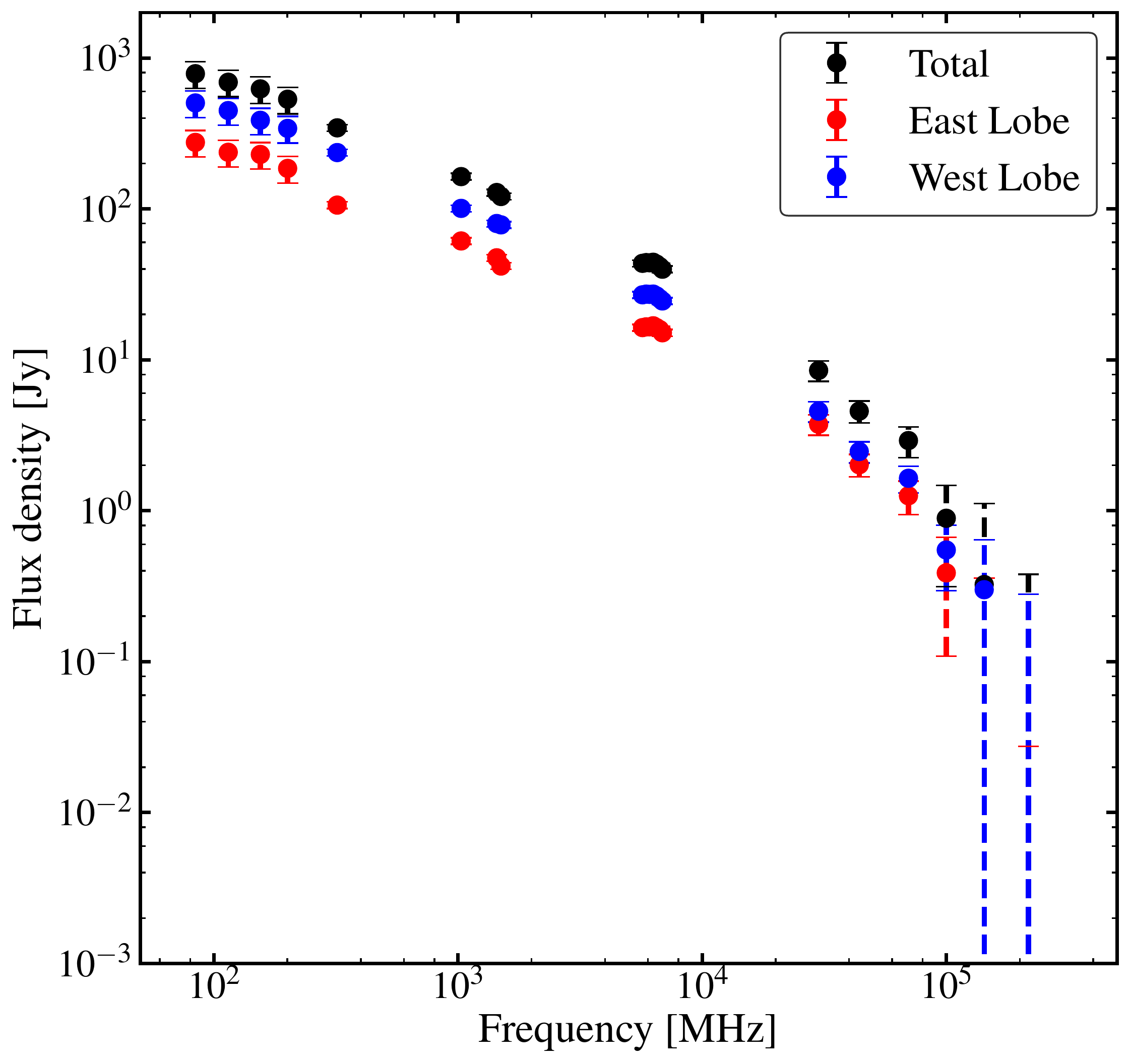}
		\includegraphics[width=0.44\textwidth]{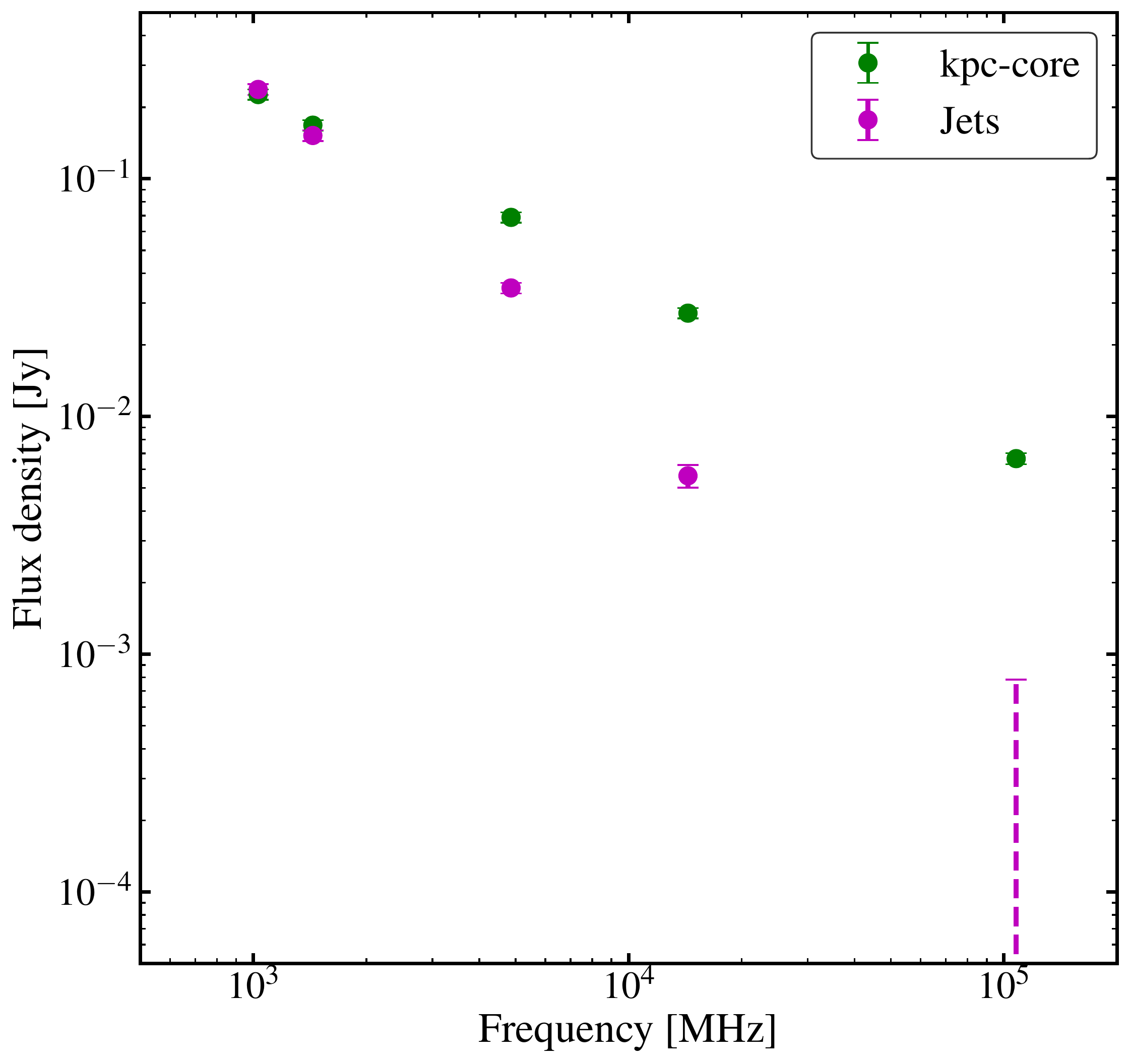}
		\caption{{\em Left panel}: Total flux density distribution of \forn\ between $84$ MHz and $217$ GHz (in black) and of the East and West lobes (in red and blue, respectively). The dashed error bars represent $1\sigma$ upper limits at the frequencies where the lobes are not detected. {\em Right panel}: flux density distribution of the kpc-core (green) and of the jets (magenta) of the central emission of \forn\ between $1.03$ and $108$ GHz.}
		\label{fig:sedLobes} 
	\end{center}
\end{figure*}

Figure~\ref{fig:sedLobes} shows the flux density of the radio lobes and the total radio emission of \forn. As shown in Fig.~\ref{fig:sedLit} and in Table~\ref{tab:sedTot}, over the frequency range in common, the measurements of this work are consistent with the measurements of \citet{mckinley2015} and \citet{perley2017}. Further details are given in Appendix~\ref{appendix:sed}.

\begin{table}[tbh]
	\caption{Total flux density of \forn, and of the East and West lobe.}
	\centering
	\label{tab:sedLobes}
	\begin{tabularx}{\columnwidth}{X c c c }  
		\hline\hline		
		Frequency	& East Lobe			& West Lobe		& Total		\\
		$[$GHz$]$	&  [Jy]				&	[Jy]		& [Jy]		\\
		\hline
		0.84			&	$276\pm55$	& $504\pm 101$ 	&	$788\pm158$		\\
		0.118			&	$237\pm48$	& $448\pm 89$	&	$692\pm138$		\\
        0.154			&	$229\pm46$	& $387\pm 77$	&	$624\pm125$		\\
        0.200			&	$185\pm37$	& $341\pm 68$	&	$533\pm106$		\\
		0.320			&	$106\pm5$	& $236\pm 12$	&	$345\pm17$		\\
		1.03			&	$61\pm3$		& $101\pm 5$	& 	$163\pm8$		\\
		1.44			&	$47\pm2$		& $80\pm 4$		&	$128\pm6$		\\
		1.50			&	$41\pm2$		& $78\pm 4$		&	$121\pm6$		\\
		5.80			&	$16\pm1$		& $26\pm 1$		&	$44\pm2$		\\
		5.90			&	$16\pm1$		& $27\pm 1$		&	$44\pm2$		\\
		6.10			&	$16\pm1$		& $27\pm 1$		&	$44\pm2$		\\
		6.30			&	$16\pm1$		& $27\pm 1$		&	$44\pm2$		\\
		6.50			&	$16\pm1$		& $26\pm 1$		&	$43\pm2$		\\
		6.70			&	$15\pm1$		& $25\pm 1$		&	$41\pm2$		\\
		6.87			&	$15\pm1$		& $24\pm 1$		&	$39\pm2$		\\
		30				&	$3.7\pm0.6$	& $4.6\pm 0.7$	&	$8.5\pm1.3$		\\
		44				&	$2.0\pm0.3$	& $2.5\pm 0.4$	&	$4.6\pm0.7$		\\
		70				&	$1.2\pm0.3$	& $1.7\pm 0.4$	&	$2.9\pm0.5$		\\
		100				&	$0.4\pm0.3$	& $0.6\pm 0.2$	&	$0.9\pm0.4$		\\
		143				&	$(0.4)$		& $0.3\pm 0.3$	&	$0.3\pm0.5$		\\
		217				&	$(0.3)$		& $(0.6)$		& 	$(0.7)$			\\
		\hline                           
	\end{tabularx}
	\tablefoot{$1\sigma$ upper limits are shown within parenthesis, at the frequencies where the lobes are not detected.}
\end{table}

\subsection{The central emission}
\label{sec:sedCore}

The central radio emission of \forn\ is clearly visible in the \meer\ images (see Fig.~\ref{fig:fornaxA},~\ref{fig:fornaxAFlRegions} and Fig.~\ref{fig:fornCore}). At $4.8$ GHz and $14$ GHz, this emission is still visible in VLA images (see the top right and bottom left panel of Fig.~\ref{fig:fornCore}). The ALMA image does not show significant extended emission beyond the central synthesized beam ($18.1\arcsec\times9.0\arcsec$, see the bottom right panel of Fig.~\ref{fig:fornCore}). Given the morphology of the central emission, we measure its flux density by dividing it into two parts, the central unresolved component (hereafter, the {\em kpc-core}) and the extended component forming the emission (the {\em jets}).

To measure these flux densities, we regrid all images to a common frame of reference and we convolve them to the lowest resolution of the sample ($18.1\arcsec\times18.1\arcsec$). We define a region including the emission in all images (see the top left panel of Fig.~\ref{fig:fornCore}). Since the kpc-core is unresolved, its flux density is equal to the peak flux within the central synthesized beam. The flux density of the jets is the sum of the emission in the selected region minus the flux density of the kpc-core. Table~\ref{tab:sedCore} shows the flux densities of both components at all selected frequencies. The spectrum of the two components is shown in the right panel of Fig.~\ref{fig:sedLobes}. The spectral shape of the extended component is much steeper than the one of the kpc-core.

\begin{table}[tbh]
	\caption{Flux density of the central emission of \forn. We refer to the unresolved component as kpc-core and to the extended component as jets.}
	\centering
	\label{tab:sedCore}
	\begin{tabularx}{0.7\columnwidth}{X c c}  
		\hline\hline		
		Frequency	& kpc-core			& Jets		\\
		$[$GHz$]$	&  [mJy]				&	[mJy] \\
		\hline
		1.03		&	$156\pm8$ & $238\pm 12$ \\
		1.44		&	$105\pm5$ & $152\pm 8$	\\
		4.86		&	$43\pm2$	  & $35\pm 2$	 			 \\
		14.4		&	$20\pm1$  & $6\pm 1$			 	 			 	\\
		108			&	$6.7\pm0.4$   & $(0.0007)$ \\
		\hline                           
	\end{tabularx}
	\tablefoot{$1\sigma$ upper limit is shown within parenthesis at $108$ GHz, where the jets are not detected.}
	\end{table}
	
\section{Properties of the jets of \forn}
\label{sec:coreProp}

We analyse the morphology and brightness distribution of the central emission of \forn\ that we can infer from the high resolution observation at 1.44 GHz (see the top left panel of Fig.~\ref{fig:fornCore}). In the previous Section, we separated the central emission into two components, the unresolved kpc-core and the extended jets. Another way to decompose the central emission consists of considering as {\em jet} the emission in the north-west of the core and as {\em counter-jet} the emission in the south-east. Both jet and counter-jet expand symmetrically away from the nucleus with a position angle of  $135^\circ$ (north through east). At the distance of $\sim 2.5$ kpc, both jets bend to the east-west direction, then fade below the $5\sigma$ detection limit ($\sim 80~\mu$\Jyb) at distances above $6$ kpc from the nucleus.

To characterise the jets we measure their variations of transverse size and surface brightness with increasing distance from the AGN. We fit a single Gaussian to the one-dimensional profiles extracted at intervals of $8\arcsec$ (approximately half the beam of the image) in the direction perpendicular to the jet expansion. This gives us the measured profile width ($w_{\rm 0}$) and peak ($I_{\rm 0}$) as a function of radius. We correct these values for the effect of the beam width ($w_b$), as in~\cite[][]{laing1999}: 

\begin{equation}
w = \sqrt{w_{\rm 0}^2-w_b^2}\\
I = I_f\sqrt{w^2/w_b^2+1}
\end{equation}

\noindent obtaining the corrected values of transverse size ($w$) and surface brightness $I$ as a function of radius.

\begin{figure*}
	\begin{center}
		\includegraphics[width=\textwidth]{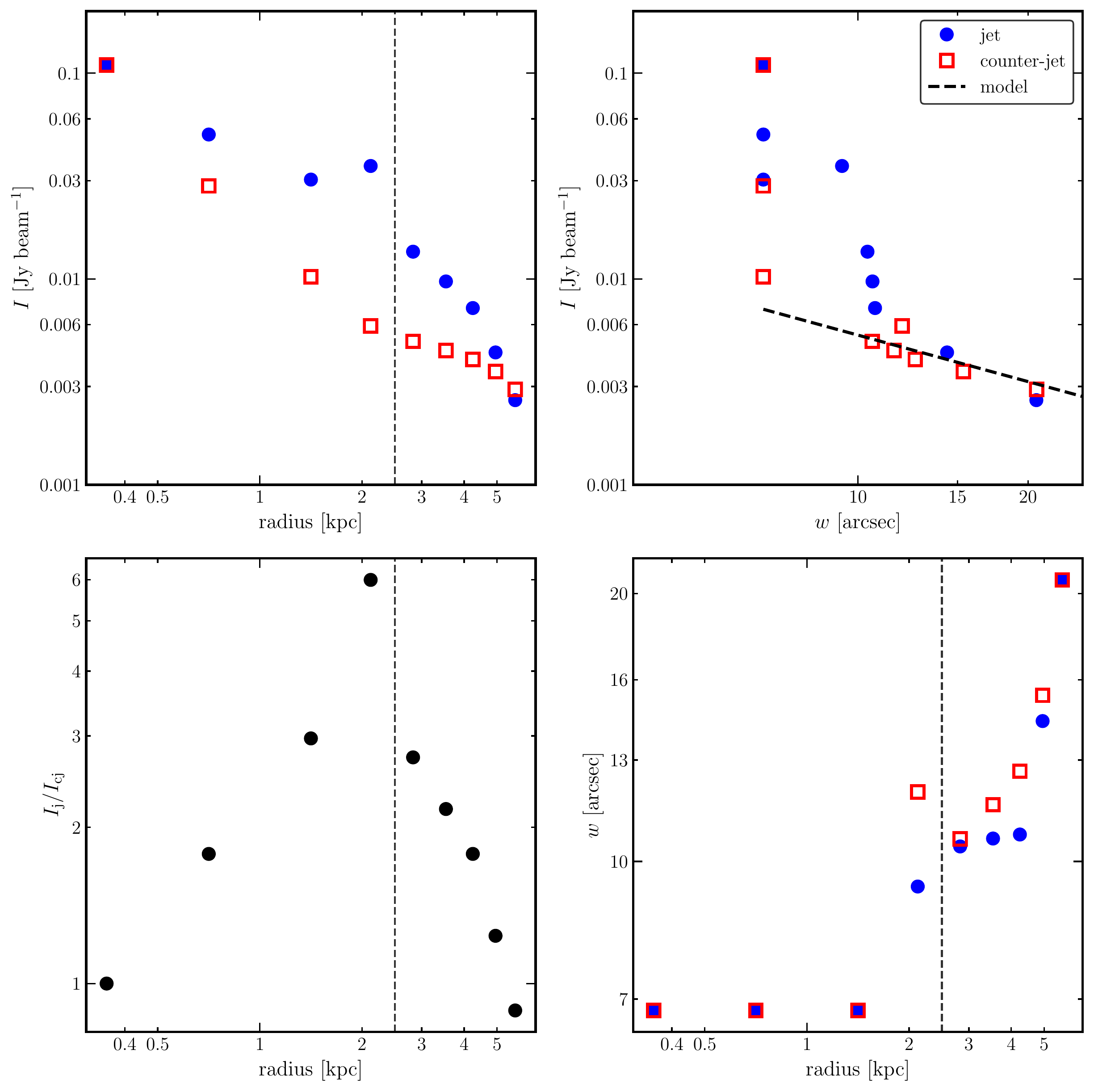}
		\caption{{\em Top left panel}: Surface brightness versus radial distance from the core for the jet (blue circles) and counter-jet of \forn\ (red squares). The dashed vertical line marks the radius where the jets bend in the east-west direction. {\em Top right panel}: Surface brightness of the jet and counter-jet versus transverse size. The dashed line shows the predicted distribution if the jets were adiabatically expanding. {\em Bottom left panel}: Surface brightness ratio between jet and counter-jet versus radial distance from the core. {\em Bottom left panel}: Transverse size of the jet and counter-jet versus radial distance from the core. Colours are as in the top panels. 
		}
		\label{fig:jetCjet} 
	\end{center}
\end{figure*}

In the top left panel of Fig.~\ref{fig:jetCjet}, we show the peak surface brightness of the jet and counter-jet against the radial distance from the core, which shows that the jet is typically brighter than the counter jet. The bottom left panel of the figure, shows that in the innermost $\sim 2.5$~kpc, the jet has approximately between two and six times the surface brightness of the counter-jet. This ratio reaches its maximum value in proximity of where the jets bend ($I_j/I_{\rm cj}\sim 6$), then it decreases with distance from the nucleus. Overall the brightness ratio is always lower than $4$ suggesting that the jets are `two-sided' \citep{bridle1984,bridle1991}. 

The bottom right panel of the figure shows the transverse size of the jet and counter jet with distance from the core in logarithmic scale. The ratio between the size of the jet and counter-jet is maximal where the jets bend, and then decreases with distance from the centre.

In the top right panel of Fig.~\ref{fig:jetCjet}, we show the brightness versus the transverse size of the jet and counter-jet, and the model (dashed black line) that these quantities should follow if the jets were undergoing adiabatic expansion at constant velocity and spectral index \citep[a power law with spectral index $\beta\sim-3.4$;][]{laing1999,parma1999}. The main outliers from this curve are points tracing the innermost part of the jets ($r\lesssim 5$), before they bend in the east-west direction. After this bend, it is possible that the jets are freely expanding. The central emission may have been generated by precessing jets. S-symmetric bent jets are a typical signature for this mechanism~\citep{monceau-baroux2014,donohoe2016,krause2019}.

Knowing the spectral index of the jets (see Sect.~\ref{sec:spiCore}), and making an assumption on their velocity ($\beta$), it is possible to estimate their orientation ($\theta$) with respect to the plane of the sky ($\theta=0$). The ratio of the flux density of the jets depends on $\theta$, $\beta$ and their spectral index as~\citep[][]{bridle1984,parma1987}: 

\begin{equation}
\frac{I_{\rm j}}{I_{\rm cj}} = \bigg(\frac{1+\beta\cos\theta_0}{1-\beta\cos\theta_0}\bigg)^{3+\alpha}
\end{equation}

\noindent Assuming $\alpha_{\rm inj}$ (jets) $= 0.6$ (as we infer from our model) the orientation angle is $<\theta>\leq30^\circ$, for values of $\beta$ between $0.4$ and $0.9c$~\citep[which are realistic velocities for jets expanding in the ISM;][]{parma1987,bridle1991,churazov2001}. Likely, the orientation of the jets is approximately in the plane of the sky, and the jet in the north-west is coming towards the observer.\\

Assuming that the radio emission contains relativistic particles confined in a uniformly distributed magnetic field in energy equipartition conditions~\citep[][]{kardashev1962,bridle1984}, it is possible to estimate the magnetic field strength (in Gauss) as: 

\begin{equation}
B_{\rm eq} = 9.6\cdot10^{-12}\bigg(\frac{P_{1.4 \,\rm GHz}(1+k)}{V^3}\bigg)^{2/7}
\end{equation}

\noindent where the radio power at 1.4 GHz ($P_{1.4 \,\rm GHz}$) is expressed is in $W/Hz$, $k$ is the ratio between the energy of protons and electrons in the radio emitting region (assumed equal to 1), and $V$ is the volume of the emitting region in kpc$^3$.

We determine the equipartition magnetic field of the kpc-core and of the jets (see Table~\ref{tab:jets}). We assume that an upper limit to the volume of the kpc-core region is given by the volume of an ellipsoid with axes equal to the beam of the image at $108$ GHz, which is the lowest resolution with which we image the kpc-core ($x=18\arcsec,\, y = 9\arcsec,\, z =18\arcsec$). The magnetic field of equipartition of the kpc-core is $B_{\rm eq,\, kpc-core}\gtrsim 50\, \mu$G. The jets extend in the plane of the sky for approximately 6 kpc in each direction (see the upper left panel of Fig.~\ref{fig:jetCjet}). We assume that the jets occupy a cylindrical volume of length 12 kpc and width given by the average size of the jet and counter-jet ($20\arcsec,\,2.02$~kpc). This gives us an upper limit to the magnetic field of $B_{\rm eq,\, jet} \gtrsim 23\, \mu$G.

The total energy of a synchrotron source is given by the energy of the relativistic particles plus the energy of the magnetic field in which they are embedded. The energy is minimum when the contributions of magnetic fields and relativistic particles are approximately equal, \ie\ equipartition condition~\citep[][]{pacholczyk1970}. Hence, from the equipartition magnetic field we can estimate the minimum energy of the jets: $B_{\rm eq} =\sqrt{24\pi/7\, u_{min}}$. We estimate $u_{\rm min,\, kpc-core}\gtrsim 3.2\times10^{-11}$ erg cm$^{-3}$ and $u_{\rm min,\, jets}=4.9 \cdot 10^{-12}$ erg cm$^{-3}$. If the jets were confined by the external medium, these minimum energies would require the ISM to have $n_e T_e\sim4\times10^4$~K cm$^{-3}$. In the innermost $200\arcsec$ (21.7 kpc) of \forn, X-ray observations have observed the presence of two cavities, broadly aligned with the jets \citep{isobe2006,lanz2010}. The {\em Chandra} spectrum of the X-ray emission in the innermost $15$ kpc provides an estimate of the temperature and density of the ISM  of $0.77$ keV ($8.9\times10^6$ K) and 0.4 cm$^{-3}$~\citep{nagino2009}, exceeding the minimum $n_e\,T_e$ value required for jets confinement by two orders of magnitude. Therefore, the pressure in the central region of \forn\ is sufficient to confine the jets. The properties of the jets that we derived in this Section are summarised in Table~\ref{tab:jets}.

\begin{table}[tbh]
	\caption{Mean properties of the jets and kpc-core of \forn\ inferred from the \meer\ observation}
	\centering
	\label{tab:jets}
	\begin{tabularx}{0.9\columnwidth}{l c}  
		\hline\hline							     
		Parameter 										&  Value 			\\
		\hline
		$P_{1.4\,\rm GHz}$ (kpc-core)					& $5.5\times10^{22}$~\whz	\\
		$P_{1.4\,\rm GHz}$ (jets)					& $1.1\times10^{23}$~\whz	\\
		$<FWHM>\,\, ({\rm jet})$						& $14 \arcsec$	\\
		$<FWHM>\,\, ({\rm counter-jet})$					& $20 \arcsec$	\\
		$<B_{\rm eq}>$ (kpc-core)	& $\gtrsim 50\, \mu$G \\
		$<B_{\rm eq}>$ (jets)	& $\gtrsim 23\, \mu$G\\
		$<u_{\rm min}>$ (kpc-core)	& $\gtrsim 2.3\times10^{-10}$ erg cm$^{-3}$ \\
		$<u_{\rm min}>$ (jets)	& $\gtrsim 4.9\times10^{-11}$ erg cm$^{-3}$\\
		$<\Theta>$ (opening angle) 			& $\leq30^\circ$  	\\
		\hline                           
	\end{tabularx}
\end{table}

\section{Spectral analysis of the main components of \forn}
\label{sec:spModAll}

To estimate the time-scale of formation of the radio lobes of \forn\ and disentangle the different phases of the nuclear activity, we study the spectrum of the integrated radio emission of the lobes and of the centre. Assuming that radiative energy losses from synchrotron ($s$) and inverse Compton ($ic$) radiation dominate over expansion losses, and by excluding in situ injection or re-acceleration of the relativistic electrons in the lobes (besides those provided by the central engine through the radio jets), the radio spectrum shows a sharp cut-off whose frequency depends on the age of the radiation and its history of injection~\citep[\eg][]{kardashev1962,pacholczyk1970,slee2001,murgia1999,murgia2011,harwood2013}. A general equation describing the variation of the energy distribution of particles in a confined volume is given by the continuity equation:

\begin{equation}
\label{eq:cont}
\frac{\partial N(\epsilon, t) }{\partial t}+\frac{\partial}{\partial \epsilon}\bigg(\frac{\partial \epsilon}{dt}N(\epsilon, t)\bigg)+\frac{N(\epsilon, t)}{T_{\rm conf}} = Q(\epsilon, t)
\end{equation}

\noindent where $N(\epsilon, t)$ is the number of particles of energy $\epsilon$ at the time $t$. $N(\epsilon, t)/T_{\rm conf}$ indicates the frequency with which particles can escape the volume and $Q(\epsilon, t)$ represents the continuous injection of particles into the volume till the time $t$. Here, we assume that the relativistic particles injected by the AGN do not escape the radio lobes or jets, $T_{\rm conf}=\infty$. Following~\citet{kardashev1962}, the lobes (or jets) are continuously injected with particles ({continuous injection model}, CI) with energies distributed in a power law:

\begin{equation}
Q(\epsilon,t) = A\epsilon^{-\delta}
\end{equation}

\noindent while the radiative losses can be expressed as:

\begin{equation}
\bigg(\frac{d\epsilon}{dt}\bigg)_{s,ic} = b (B^2 + B^2_{CMB})\epsilon^2 = b_{s,ic}\epsilon^2 \propto B^2\epsilon^2
\end{equation} 

\noindent where $b$ is a generic constant~\citep[][]{pacholczyk1970}. If we assume that inverse Compton losses are due to the cosmic microwave background and act as a magnetic field, $B_{CMB}^2 = 8\pi u_{CMB}$ (where $u_{CMB}$ is the energy density of the radiation), then $b (B^2 + B^2_{CMB}) = b_{s,ic}$. We assume that the distribution of electrons stays isotropic during its losses, which is known as the Jaffe \&Perola approximation~\citep[JP;][]{jaffe1973}. Under this approximation the timescale for continuous isotropisation of the electrons is much shorter than their radiative lifetime, and their losses do not depend on the actual pitch angle but rather on an average performed over all possible angles they travelled since the injection. The losses depend on the magnetic field on the perpendicular direction to the motion of the electrons:

\begin{equation}
b_{s,ic}=2.37\cdot10^{-3}\cdot(2/3)\cdot B_\perp^2
\end{equation}

\noindent Under these assumptions, the number of particles of energy $\epsilon$ at a given time $t$ is given by the solution of Eq.~\ref{eq:cont}:

\begin{equation}
\begin{split}
\label{eq:Ne}
N(\epsilon, t ) &\approx \,\,A \cdot \epsilon^{-\delta} \cdot t &, \qquad\qquad\qquad \epsilon < \epsilon^\star = \frac{1}{b_{s,ic}t}\\
&\approx\frac{A}{\delta-1}\cdot\frac{\epsilon^{-(\delta+1)}}{b_{s,ic}}&, \qquad\qquad\qquad \epsilon \gtrsim \epsilon^\star = \frac{1}{b_{s,ic}t}
\end{split}
\end{equation}

\noindent In the spectrum of the CI model, $J_s(\nu)$, two power law regimes can be identified:

\begin{equation}
\label{eq:Jnu}
\begin{split}
J_s(\nu) &\propto \nu ^{-(\delta-1)/2}  = \nu^{-\alpha_{\rm inj}}&,\qquad\qquad\qquad \nu<\nu_{\rm break} \\
 &\propto \nu ^{-\delta/2}  = \nu^{-(\alpha_{\rm inj}+1/2)} &,\qquad\qquad\qquad\nu\gtrsim\nu_{\rm break} \\
\end{split}
\end{equation}

\noindent At low energies (below the break-frequency, $\nu_{\rm break}$), the spectral index is $\alpha_{\rm inj} = \frac{\delta-1}{2}$, while at high energies ($\nu\gtrsim\nu_{\rm break}$) the index is $\alpha_{\rm high}=\alpha_{\rm inj}+1/2$. The cutoff frequency depends on the time since the injection began,~\ie\ the radiative age of the source $t_s$~\citep[\eg][]{murgia2010b,orru2010,harwood2013}:

\begin{equation}
\label{eq:t_s}
	t_{\rm s}	 = 1610 \frac{B^{0.5}}{B^2+B^2_{\rm CMB}}\frac{1}{[\nu_{\rm break}(1+z)]^{1/2}]}
\end{equation}

A more complicated scenario describing the radiation of the lobes and jets of \forn\ can be the {continuous injection plus turn off model} (CI$_{\rm OFF}$). The injection of high energy particles from the nucleus starts at $t = 0$ and, at the time $t_{\rm CI}$, it is switched off ($Q(\epsilon,t_{\rm CI})= 0$). After that, a new phase of duration $t_{\rm OFF}$ begins (\ie\ the {\em off phase} of the AGN), and $t_{\rm s}=t_{\rm CI} + t_{\rm OFF}$ \citep[\eg][]{slee2001,parma2007,murgia2011}. Compared to the CI model, the spectral shape is characterised by a second break-frequency ($\nu_{\rm break,\, high}$), beyond which the radiation spectrum drops exponentially. This frequency depends on the ratio between the dying phase and the total age of the source ($t_s/t_{\rm OFF}$): 

\begin{equation}
\label{eq:nu_break}
\nu_{\rm break,\, high} = \nu_{\rm break}\bigg(1+\frac{t_{\rm CI}}{t_{\rm OFF}}\bigg)^2=\nu_{\rm break}\bigg(\frac{t_s}{t_{\rm OFF}}\bigg)^2
\end{equation}

\noindent The longer the source has been off, the shorter is the distance between the two break-frequencies.

In the following sections, we determine if the radio lobes and the kpc-core and jets of \forn\ are best described by a continuous injection model (CI) or by a continuous injection plus turn-off model (CI$_{\rm OFF}$). We measure the break-frequency ($\nu_{\rm break}$) injection index ($\alpha_{\rm inj}$), and, eventually, $t_{\rm OFF}/t_{\rm start}$, that define the spectral shape of the different components and we estimate the age of their synchrotron radiation. We study the variations of the spectral shape through the lobes and jets and build the map of the break-frequency to constrain their injection history.

To determine the best-fit models, we use the software package {\tt SYNAGE++}~\citep{murgia1999}, that has been used to measure the age of the relativistic electrons for several radio AGN \citep[\eg][]{parma1999,murgia2003,murgia2010b,orru2010,murgia2011,murgia2012,kolokythas2015}. 

\begin{figure*}
	\begin{center}
		\includegraphics[width=0.44\textwidth]{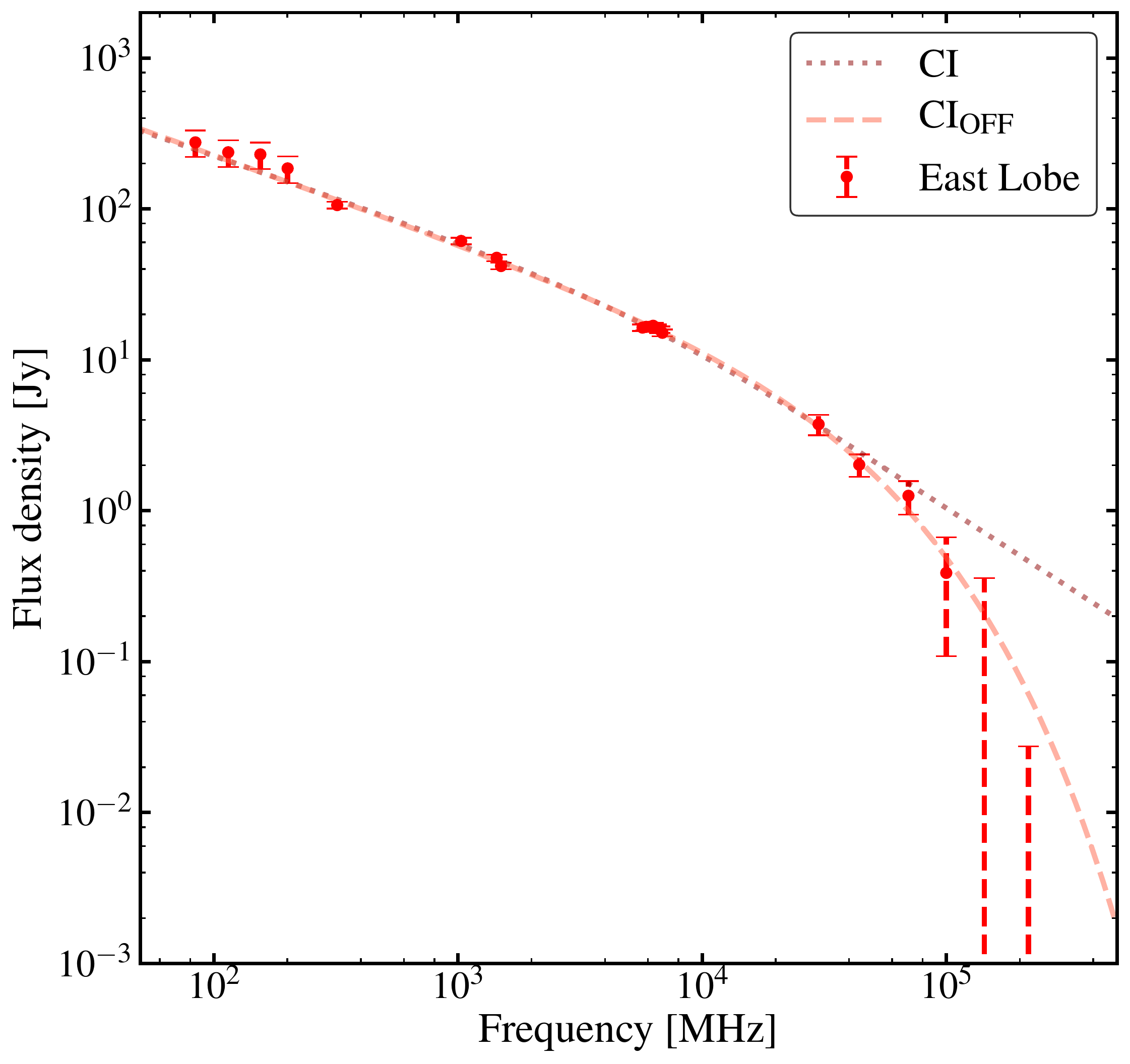}
		\includegraphics[width=0.44\textwidth]{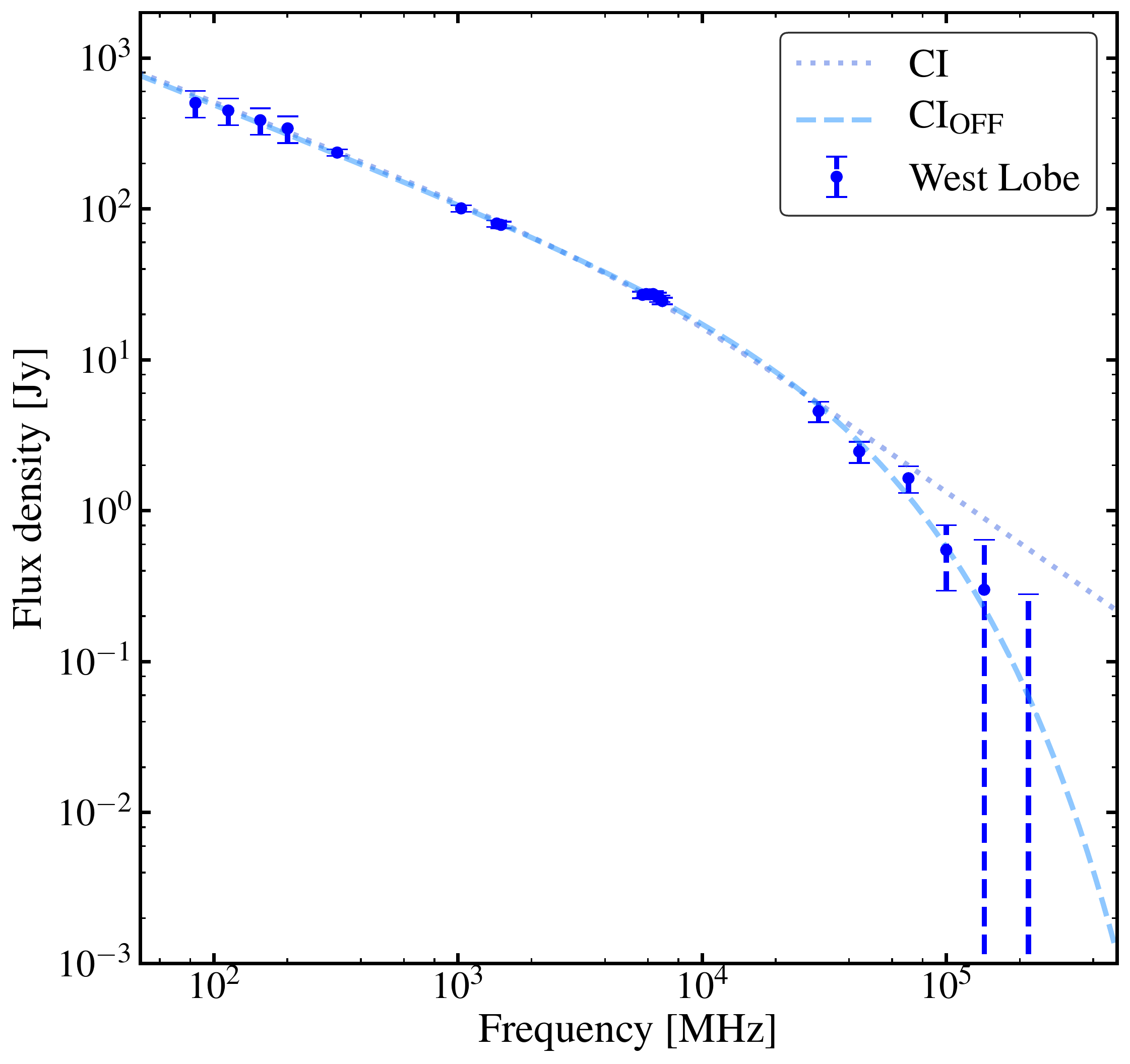}
		\caption{{\em Left panel}: spectrum of the East lobe fitted with a continuous injection model (CI) and a continuous injection with a turn-off (CI$_{\rm OFF}$). {\em Right panel}: spectrum of the West lobe fitted with the same models as in the left panel.}
		\label{fig:modEW} 
	\end{center}
\end{figure*}

\begin{table*}
	\caption{Physical parameters of the radio emission of the different components of \forn\ according to the results of the spectral modelling}
	\centering
	\label{tab:CICIOFFall}
	\begin{tabular}{l | c  c | c  c  | c c  | c}  
		\hline\hline							     
{Component} &\multicolumn{2}{c|}{$\tilde{\chi}^2$} & \multicolumn{2}{c|}{$\alpha_{\rm inj}$} & \multicolumn{2}{c|}{$\nu_{\rm break}$ [GHz]} & $t_{\rm OFF}/t_s$\\
		    & CI & CI$_{\rm OFF}$& CI & CI$_{\rm OFF}$ & CI & CI$_{\rm OFF}$ & CI$_{\rm OFF}$\\
			\hline
			Total 		& 1.29 & 0.38 &$0.59^{+0.01}_{-0.05}$ & $0.62^{+0.03}_{-0.07}$ & $9^{+2}_{-4}$	& $39^{+4}_{-31}$ & $0.59^{+0.23}_{-0.49}$ \\
			East Lobe 	& 1.65 & 1.03 &$0.54^{+0.02}_{-0.06}$ & $0.57^{+0.02}_{-0.10}$ & $8^{+2}_{-4}$ & $32^{+8}_{-27}$ & $0.49^{+0.08}_{-0.42}$\\
			West Lobe 	& 1.80 & 0.45 &$0.63^{+0.02}_{-0.04}$ & $0.63^{+0.02}_{-0.04}$ & $9^{+2}_{-3}$ & $33^{+9}_{-26}$ & $0.53^{+0.18}_{-0.41}$\\	
			kpc-core			& 10.1 & 20.3 &$0.67^{+0.02}_{-0.03}$ & $0.68^{+0.02}_{-0.03}$ & $\gtrsim250$& $\gtrsim250$& - \\
			Jets			& 2.69 & 1.01 &$1.05^{+0.06}_{-0.16}$ & $0.92^{+0.09}_{-0.20}$ & $7^{+2}_{-5}$& $19^{+2}_{-16}$& $0.89^{+0.01}_{-0.71}$  \\
		\hline                           
	\end{tabular}
	\tablefoot{$\tilde{\chi}^2$: reliability of the best-fit model. $\alpha_{\rm inj}$: injection index of the flux density distribution. $\nu_{\rm break}$: break-frequency of the spectrum. $t_{\rm OFF}/t_s$: ratio between the total lifetime of the AGN and the off-phase. Further details about these parameters are are given in Sect.~\ref{sec:spModAll}.}
\end{table*}

\subsection{Spectral modelling of the lobes}
\label{sec:spModLobes}

Figure~\ref{fig:modEW} shows the CI and CI$_{\rm OFF}$ models that best fit the radio spectrum of the East and West lobes. At $\gtrsim 143$ GHz both lobes are undetected, marking a sharp cut-off in the spectrum. The best-fit parameters for both models are listed in Table~\ref{tab:CICIOFFall}. We use the reduced-chi-squared to determine which model is best fits the spectrum. For both lobes, the reduced-chi-squared ($\tilde\chi^2$) of the CI$_{\rm OFF}$ model has values closer to $1$ than the $\tilde\chi^2$ of the CI model. This, along with the sharp cut-off at high frequencies, suggests that the radio spectrum of both radio lobes is best described by the CI$_{\rm OFF}$ model and that, currently, the radio lobes are not being injected with relativistic particles. 

According to the CI$_{\rm OFF}$ model, the injection spectral index of the particles in the East lobe is $\alpha_{\rm inj}= 0.57^{+0.02}_{-0.10}$, the break-frequency is $\nu_{\rm break,\, E}=32^{+8}_{-27}$~GHz and the dying to total age ratio of is $t_{\rm OFF}/t_s=0.49^{+0.08}_{-0.42}$. Assuming that the magnetic field of the lobes is $2.6\pm 0.3\mu$G~\citep{isobe2006,tashiro2009} the radiative age of the lobe is $t_{s,\,\rm E}=25^{+23}_{-19}$ Myr. Likely, in the last $t_{\rm OFF,\, E} = 12^{+2}_{-9}$~Myr ($\sim 49\%$ of the total lifetime of the AGN) the East lobe has not been replenished with relativistic particles.

The results obtained for the West lobe are compatible with the ones of the East lobe, $\alpha_{\rm inj}= 0.63^{+0.02}_{-0.04}$, $\nu_{\rm break,\, W}=33^{+9}_{-26}$~GHz and $t_{\rm OFF}/t_s= 0.53^{+0.18}_{-0.41}$. This gives $t_{s,\,\rm W}=23^{+20}_{-17}$ Myr and  $t_{\rm OFF,\,W} =12^{+4}_{-8}$~Myr. 

The best-fit CI$_{\rm OFF}$ models of the total spectrum and of the West lobe have $\tilde\chi^2\lesssim1$. This likely occurs because the model is not only always compatible with the measured fluxes, within the errors, but it also falls very close to the measured fluxes all through the spectrum. The errors on the flux density at two frequencies observed with the same telescopes are not independent, which could bias the $\tilde\chi^2$ statistics. Nevertheless, for the purposes of this study the $\tilde\chi^2$ is a good indicator to determine which one between the CI or the CI$_{\rm OFF}$ model better fits the observed radio spectra.

The spectral index of the lobes of \forn\ measured between $115$ MHz and $1140$ MHz ~\citep[$\alpha = 0.77\pm 0.05$;][]{mckinley2015} and from the X-ray Inverse Compton emission~\citep[$\alpha = 0.68$;][]{isobe2006,tashiro2009} is overall slightly steeper than the injection index we measure. This occurs because the spectral index of the flux density of the lobes is the exponent for which a single power law  well approximates the flux density distribution, in a limited interval of frequencies below $\nu_{\rm break}$. On the contrary, the injection index describes a double power law with a high frequency cut-off.

A fundamental assumption in estimating the radiative age of the radio source is that the magnetic field within the lobes is constant. In the lobes of \forn\ we assumed $B=2.6\pm 0.3\mu$G, as it has been measured from the simultaneous modelling of synchrotron emission and inverse Compton-scattered X-ray emission~\citep[][]{isobe2006,tashiro2009}. This value agrees with the magnetic field of equipartition ($B_{\rm eq} \sim3.0\,\mu$G), derived assuming a volume of the lobes of $\sim 150$ kpc$^3$), suggesting that the radio lobes are in a state of minimum energy. The value of the inverse Compton magnetic field estimated from the distance of the source is $B_{\rm IC} = 3.25(1+z)^2=3.2\,\mu$G, implying a ratio $B/B_{\rm IC} = 0.8$. Given these conditions, any deviation from the minimum energy assumption has only a moderate impact on the age estimate~\citep[][]{parma1999}.

\subsection{Spectral modelling of the central emission}
\label{sec:spModCore}

\begin{figure*}
	\begin{center}
		\includegraphics[width=0.44\textwidth]{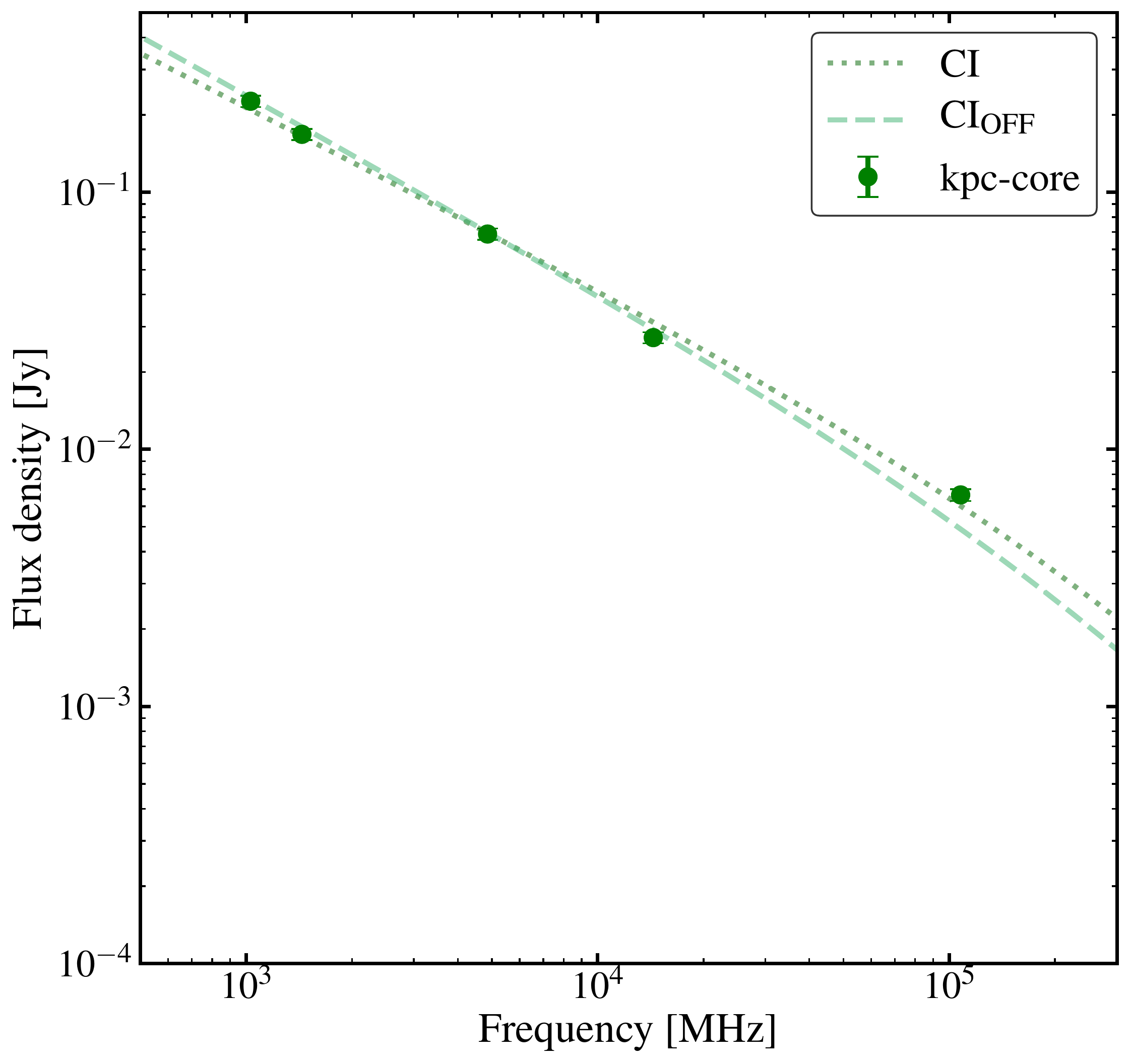}
		\includegraphics[width=0.44\textwidth]{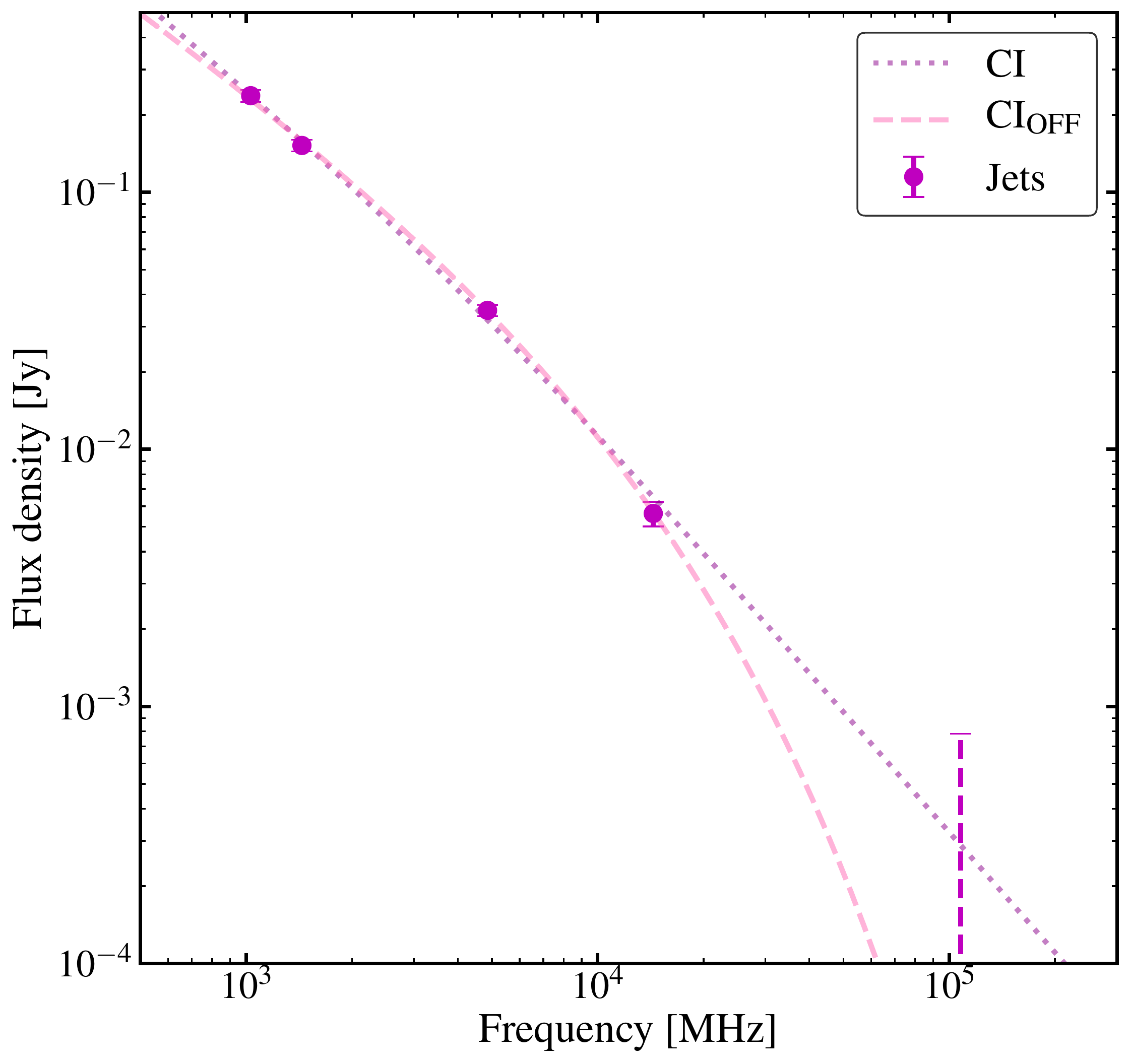}
		\caption{{\em Left panel}: spectrum of the kpc-core of \forn\ (green) fitted with a continuous injection model (CI) and a continuous injection with a turn-off (CI$_{\rm OFF}$). {\em Right panel}: spectrum of the extended central radio jets fitted with a continuous injection (CI) and a continuous injection with a cut-off (CI$_{\rm OFF}$).}
		\label{fig:modCore} 
	\end{center}
\end{figure*}

In the left panel Fig.~\ref{fig:modCore}, we show the results of the best-fit for both the CI and CI$_{\rm OFF}$  model for the flux density of the kpc-core. The right panel shows the results for the jets. The best-fit parameters for both the CI and the CI$_{\rm OFF}$ model are listed in Table~\ref{tab:CICIOFFall}. The reduced chi-squared of the fits suggests that the spectral distribution of the kpc-core is better described by the CI model rather than by the CI$_{\rm OFF}$. The jets, instead, are better fitted by the CI$_{\rm OFF}$. It is possible that the kpc-core is still being injected with high-energy particles, while this is not currently happening in the jets. This result holds as long as the thermal contamination to the 108 GHz flux of the kpc-core, poorly constrained by currently available data, is $\lesssim 80\%$. Previous measurements of the spectral index of the synchrotron emission of the jet of \forn\ between 4.9 and 14.9 GHz~\citep{geldzhaler1978,geldzahler1984}, also showed a steepening of the spectral index moving from the centre to the outer regions of the jets.

Knowing the magnetic field, and the break-frequency of the spectral energy distribution, from Eq.~\ref{eq:t_s}, we estimate the radiative ages of the kpc-core and jets. The kpc-core seems to be currently active and the age of the synchrotron emission is $\sim 1^{+0.3}_{-0.5}$~Myr. By contrast, the jets do not seem to be currently replenished with energetic particles. Their last active phase seems to have occurred $3^{+7}_{-2}$ Myr ago and to have lasted $\lesssim 1^{+6}_{-0.5}$~Myr.

\section{Spectral index and break-frequency maps}
\label{sec:spiBrGen}
In this Section, we use the high resolution observation from MeerKAT to generate the spectral index map of the radio emission of \forn\ between 1.03 GHz and 1.44 GHz. The continuum images at the two frequencies are convolved to a common gaussian beam of $10\arcsec\times10\arcsec$. The spectral index map is computed pixel-by-pixel measuring the intensity ratio between the two images.

\begin{figure*}
	\begin{center}
		\includegraphics[trim = 0 0 0 0, width=\textwidth]{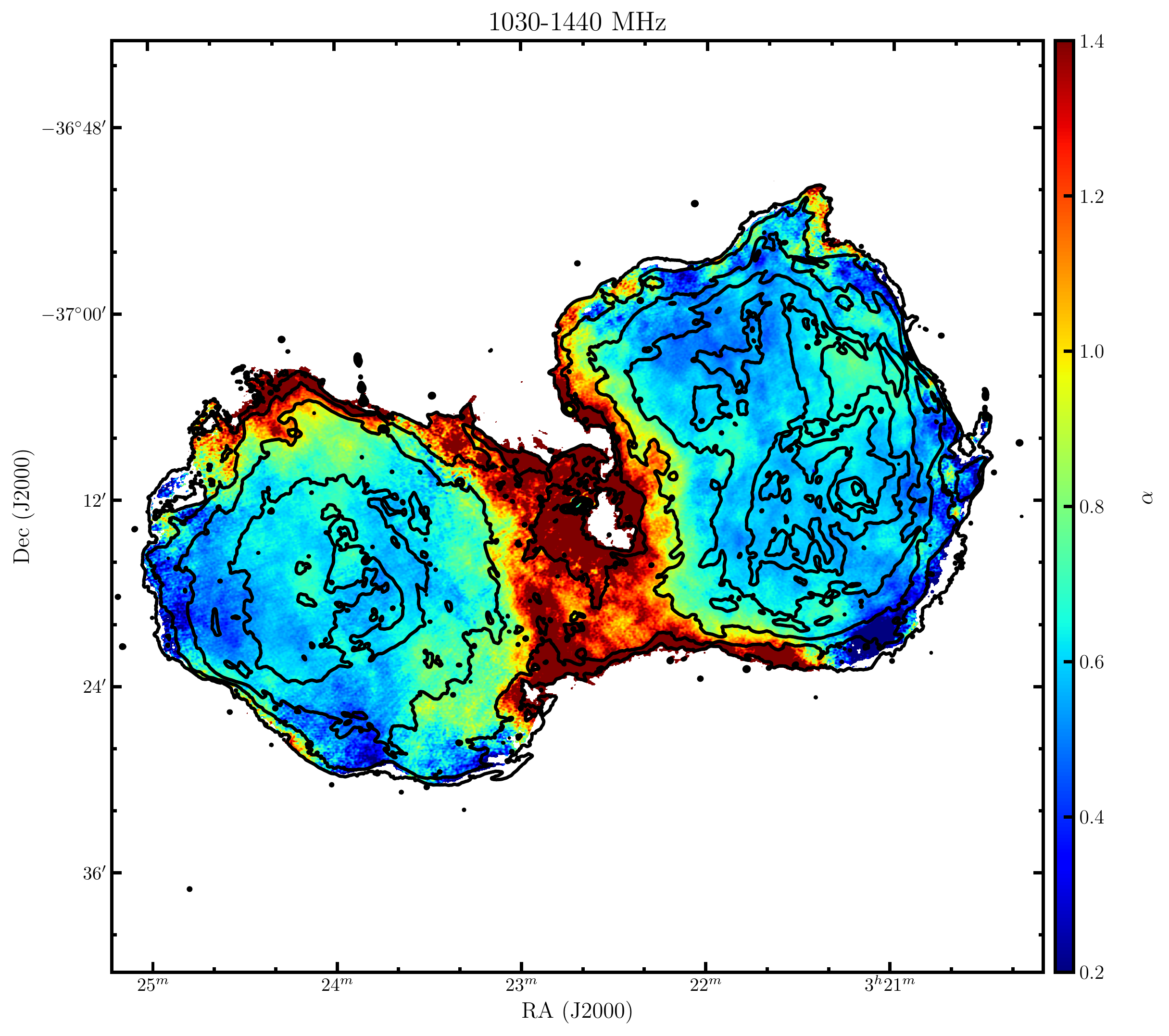}
		\caption{Spectral index map of the radio emission of \forn\ between $1.03$~GHz and $1.44$~GHz. The resolution of the map is $10\arcsec\times10\arcsec$. The contour levels show the radio emission at $1.44$~GHz. They start at $0.8$~\mJyb, increasing by factors of $2$.}
		\label{fig:spIxMeer} 
	\end{center}
\end{figure*}

\subsection{The lobes}
\label{sec:spiBr}

The total intensity map does not show any clear evidence of a hot-spot or a jet expanding through the lobes. Hence, in the spectral index map (Fig.~\ref{fig:spIxMeer}) we identify three main components with different spectral indexes. The centre of the lobes, which has overall spectral index $\sim0.7$. The `bridge' connecting the lobes and the inner edges of the lobes (\ie\ the edges closer to the the host galaxy and in the north of the east lobe and south of the west lobe, respectively) have steep spectrum ($\alpha\sim1.3$) and the outer edges which have flat spectrum ($\alpha\lesssim0.5$). 

\begin{figure*}
	\begin{center}
		\includegraphics[width=0.3\textwidth]{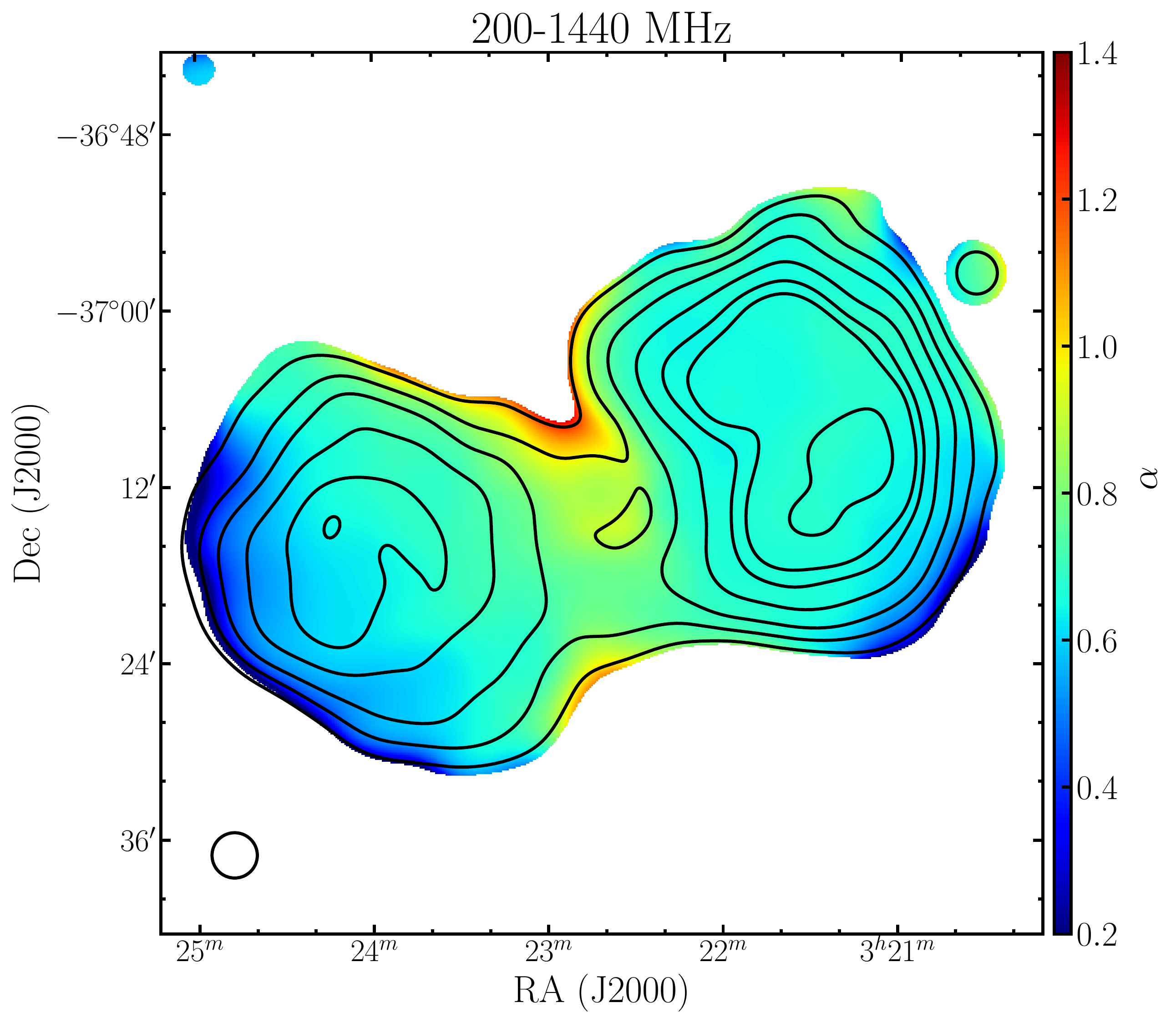}
		\includegraphics[width=0.3\textwidth]{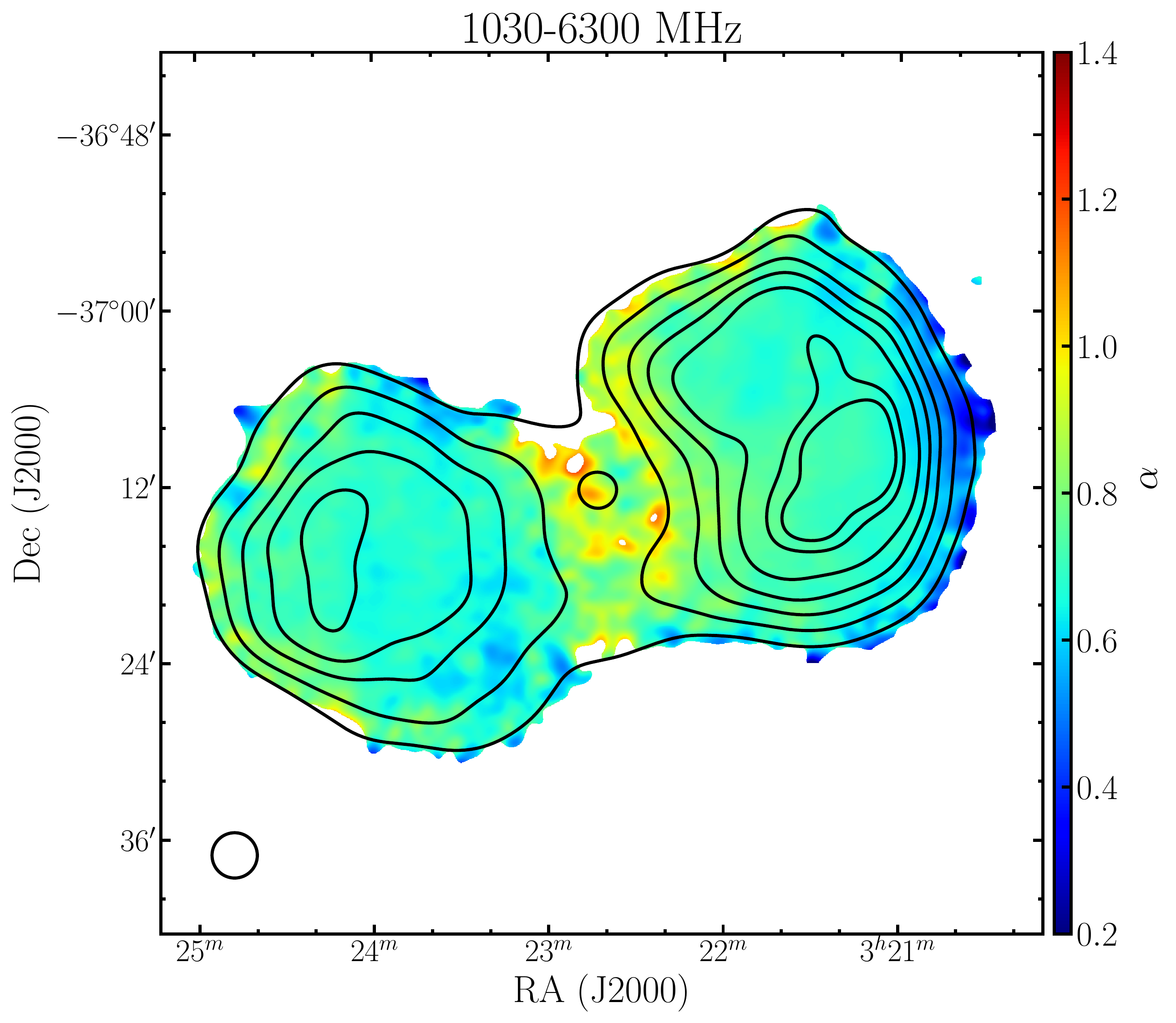}
		\includegraphics[width=0.3\textwidth]{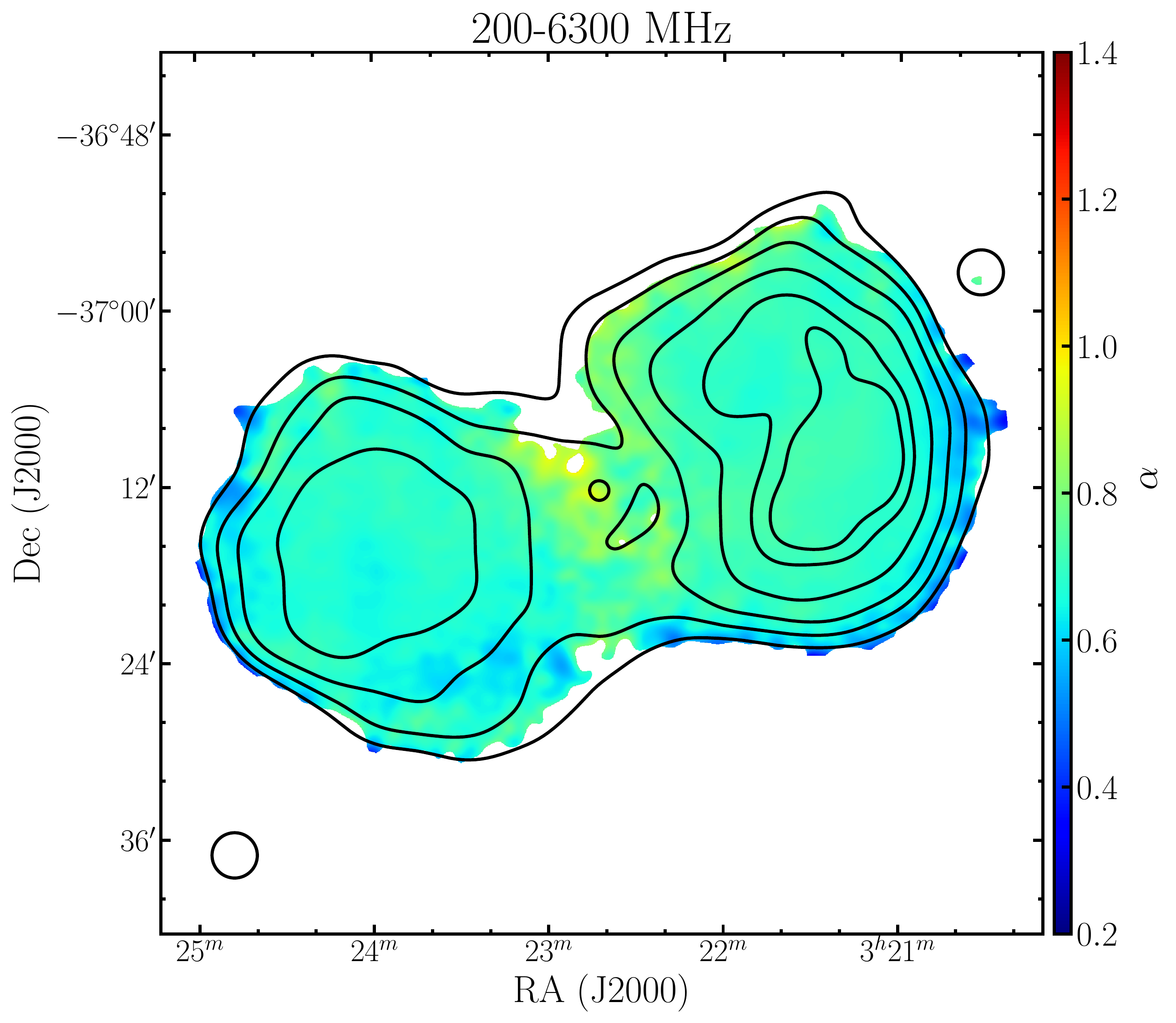}
		\caption{{\em Left panel}: Spectral index map of \forn\ between $200$ MHz and $1.44$ GHz (extracted from MWA and MeerKAT observations). The resolution is the same in all images ($3\arcmin\times3\arcmin$) and is shown in black in the bottom left corner. The colour scale is the same as in Fig.~\ref{fig:spIxMeer}. The contour levels show the radio emission at $1.44$ GHz. Contours start at $0.1$~\mJyb, increasing by factors of 2. {\em Central panel}: spectral index map between $1.03$ GHz and $6.30$ GHz (taken from MeerKAT and SRT observations), the colour-scale of the map is the same as in the left panel. The contour levels show the radio emission at $1.03$ GHz, starting at $0.2$~\mJyb, increasing by factors of 2. {\em Right panel}: spectral index map between $200$ MHz and $6.3$ GHz (from MWA and SRT observations), the PSF and colour-scale of the map is the same as in the top left panel. The contour levels show the radio emission at $200$ MHz, starting at $0.5$~\mJyb, increasing by factors of 2.} 
		\label{fig:spIxCheckMeer} 
	\end{center}
\end{figure*}

The spectral index map between two close frequencies from the same observation can give a limited view of the spectral shape of the synchrotron emission. The calibration of the observation, radio frequency interference and the uv-coverage of the observation are all effects that may contribute to slightly change the flux distribution in the images, and therefore may bias the spectral index map. For these reasons we also create spectral index maps between 200 MHz and 1.44 GHz, between 1.03 GHz and 6.3 GHz, and between 200 MHz and 6.30 GHz. For this analysis we convolve all images to the resolution of $3\arcmin\times3\arcmin$, which matches the resolution of the MWA and SRT images. We do not consider the \pl\ observations because they resolve the lobes with only at most two resolving beams, and this does not allow us to trace the differences in spectral index within the lobes. The results are shown in Fig.~\ref{fig:spIxCheckMeer}. The main spectral features shown in the high resolution map are also recovered in these maps, namely the steep ($\sim 1$) spectral index in the centre and in the bridge, and an overall uniform spectral index in the lobes ($\sim 0.7$). Some regions at the edges of the lobes show a flatter spectrum. 

The CI$_{\rm OFF}$ models provide a good description of the injection history of the lobes of \forn\ considered as a whole. Nevertheless, these components are resolved by our observations, and it is possible to trace within the lobes the differences in radiative age. The continuous injection of particles into the lobes can be thought of as the sum of subsequent injections between the time $0$ and the time $t_{\rm CI}$, whose individual radiative losses are well described by the JP model (see Sec.~\ref{sec:spModAll}). Since the CI$_{\rm OFF}$ model provides a good description of the overall injection history of the lobes, by tracing the spectral differences through the lobes, we expect to identify some regions where the break-frequency of the best-fit JP model has values close to $\nu_{\rm break, \,high}$ and others where the break-frequency is as low as $\nu_{\rm break}$. 

\begin{figure*}
	\begin{center}
		\includegraphics[trim = 0 0 0 0, width=\textwidth]{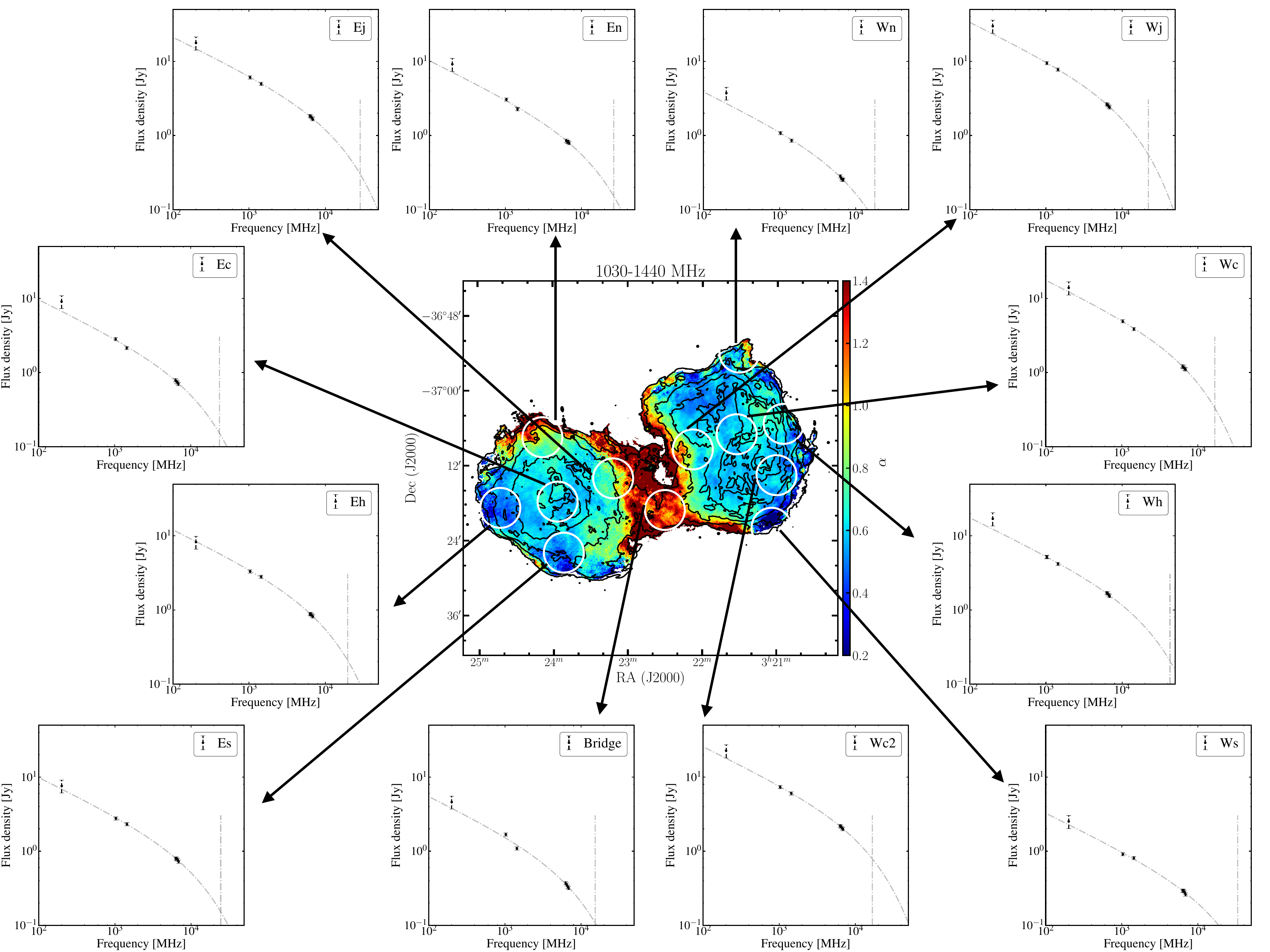}
		\caption{Spectral index map of the radio emission of \forn. The white circles mark the regions where we analyse the spectral differences within the lobes. The panels show the best fit JP model for the spectra of each region (further details are given in Sect.~\ref{sec:spiBrGen}). Regions at the edge of the lobes have higher break frequency than the bridge and the inner regions.}
		\label{fig:pixJP} 
	\end{center}
\end{figure*}

We measure the flux density in different regions of the lobes of \forn\ between 200 MHz and 6.30 GHz on the images convolved at $3\arcmin\times3\arcmin$. The regions have the same size as the PSF and are selected to be located more or less symmetrically in both lobes. We select regions close to the centre ($E_c$, $W_c$, $W_{c2}$), along the apparent direction of expansion of the radio jets ($E_j$, $W_j$), at the edges ($E_h$, $W_h$, $E_s$, $W_s$, $E_n$, $W_n$) and one region in the bridge. These regions are shown in the spectral index map of Fig.~\ref{fig:pixJP}. The panels of this figure show the best-fit JP model to the spectrum of each region. The pattern visible in the $1.03$ - $1.44$ GHz spectral-index image (Fig.~\ref{fig:spIxMeer}) is visible also now that we model the flux density over a broader band even if at lower angular resolution. The regions with higher break are in the north and south edge of the West lobe ($W_n$ and $W_s$) and in the east and south edge of the East lobe (the $E_h$ and $E_s$), while the centre and the bridge show low breaks. The panels in Fig.~\ref{fig:pixJP} confirm that the break-frequency is high throughout the lobes, and that the \pl\ non-detection of the lobes at frequencies $\gtrsim 143$ GHz does not bias this result. Nowhere in the lobes we can observe a pure power law spectrum. This confirms that, likely, the lobes of \forn\ are not currently active.

The regions with flat spectrum (such as $E_h$, $W_s$, $W_n$, $W_{c2}$) correspond to some of the low-polarisation patches identified by \citet{anderson2018}. In these regions, the associated Faraday depth enhancement may be due to magnetised plasma residing in the lobes, and advected from the host galaxy ISM during their expansion. Regions $E_h$ and $W_n$ also coincide with a peak in the hard X-ray distribution~\citep[see Fig.~1b in ][]{kaneda1995}.

\begin{figure*}
	\begin{center}
		\includegraphics[trim = 0 0 0 0, width=\textwidth]{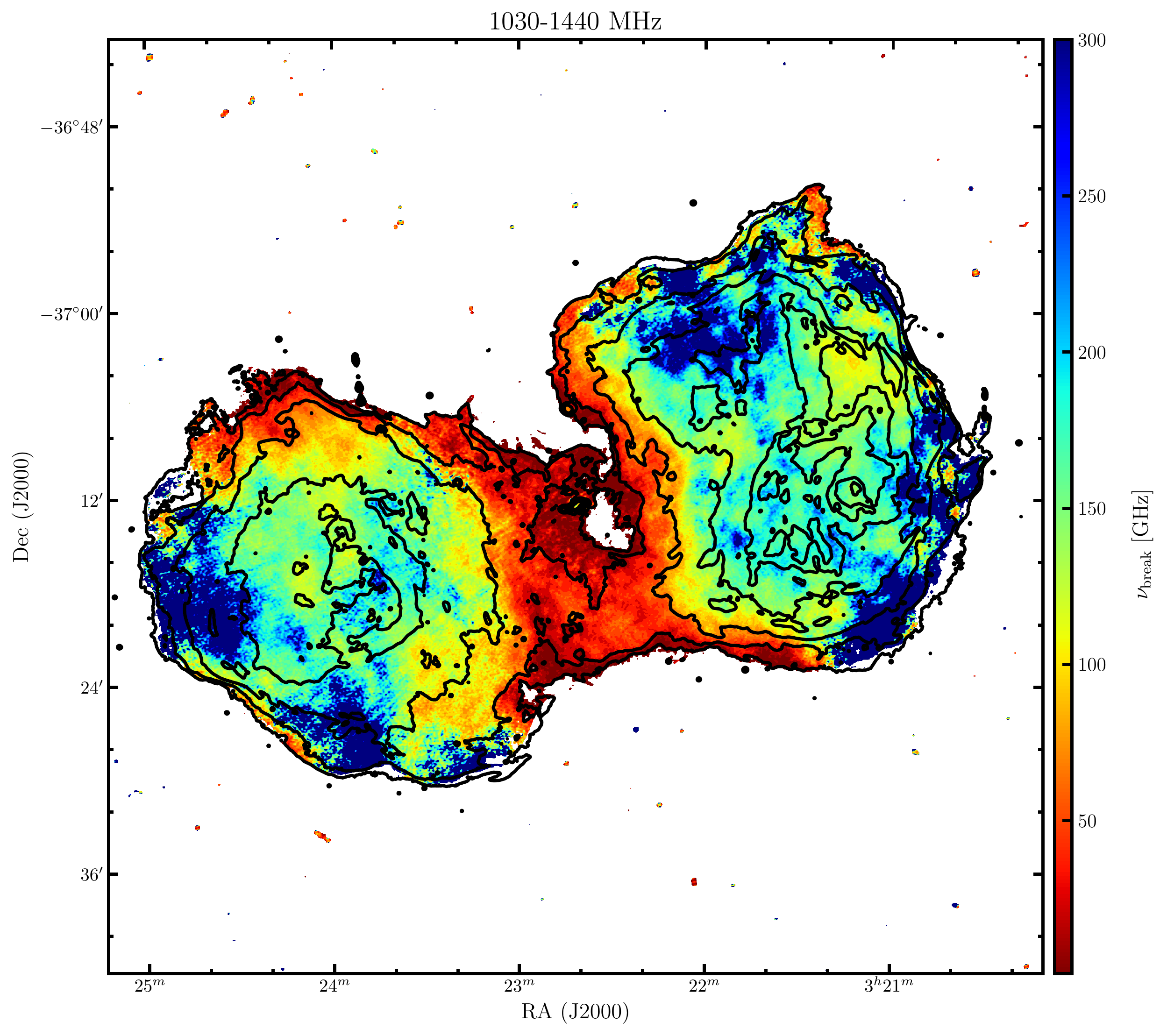}
		\caption{Break-frequency map inferred from the spectral index map shown in Fig.~\ref{fig:spIxMeer}. The PSF of the map is $10\arcsec\times10\arcsec$. Contour levels show the 1.44~GHz radio emission.}
		\label{fig:breakMap} 
	\end{center}
\end{figure*}

It is possible to generate the break-frequency map of the radio emission of the lobes of \forn\ from the spectral index map. \citet{myers1985} show that from Equations~\ref{eq:Ne} and~\ref{eq:Jnu} the radio spectral index between two frequencies is a function of the ratio between the frequency where the spectral index is measured and the break-frequency. The relation between the two-frequency spectral index and the break-frequency diverges for large values of the spectral index or when it is close to zero. For pixels with $\alpha\gtrsim 1.4$ or pixels with $\alpha\lesssim 0.2$ the estimated break-frequencies should be considered as lower and upper limits, respectively. The break-frequency map generated following this method is shown in Fig.~\ref{fig:breakMap}. The outer edges of the lobes have a high break-frequency ($\gtrsim 220 \pm 140$ GHz) while the inner edges and the bridge have the lowest breaks ($\lesssim 50 \pm 5$ GHz). In the centre of the lobes, overall, the break-frequency is $\sim 130\pm 40$ GHz. This value is very close to the $\nu_{\rm break,\, high}$ of the best fit CI$_{\rm OFF}$ model of the total flux density of the lobes.

\subsection{The central emission}
\label{sec:spiCore}

\begin{figure*}
\begin{center}
		\includegraphics[width=0.44\textwidth]{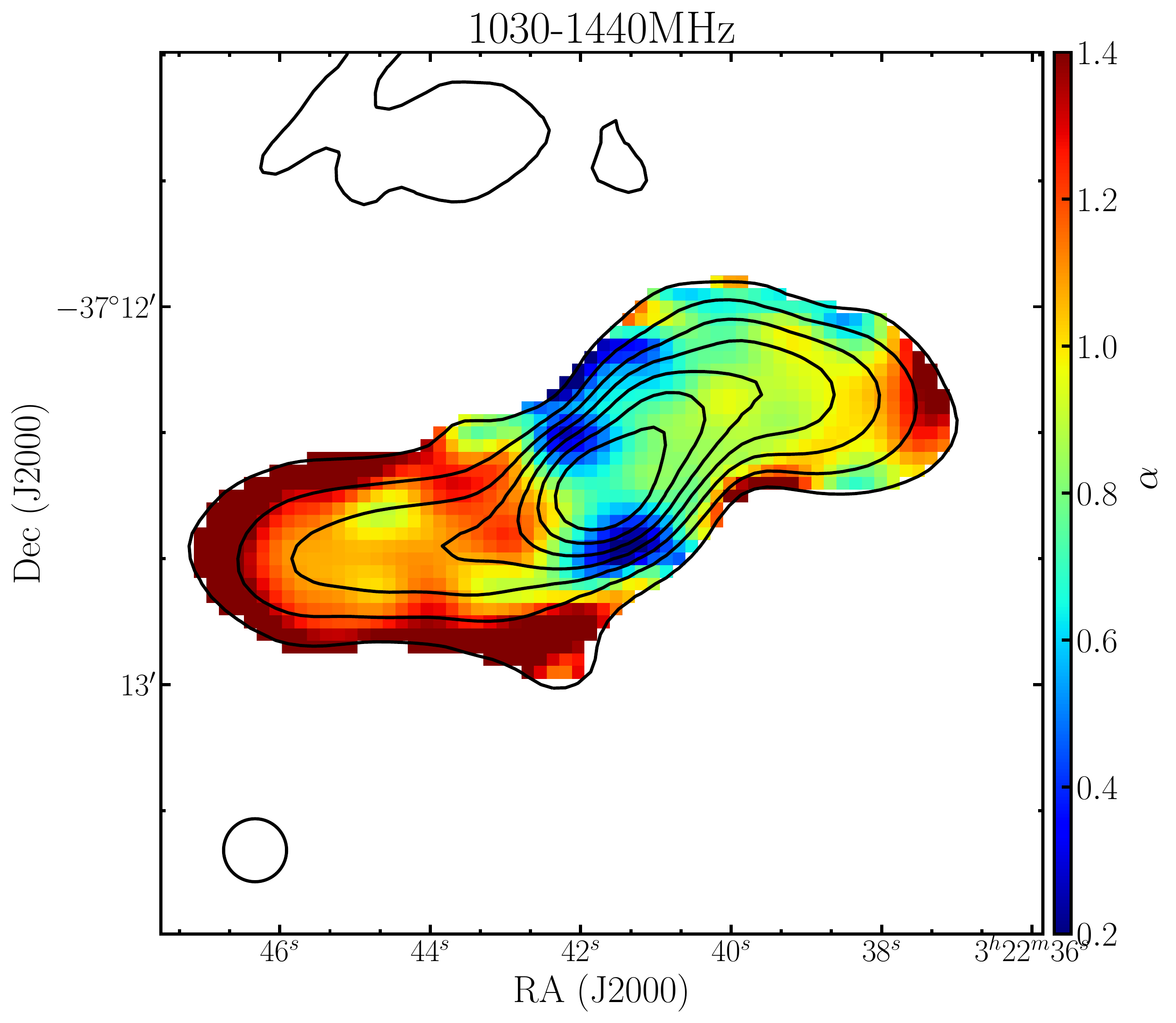}
		\includegraphics[width=0.44\textwidth]{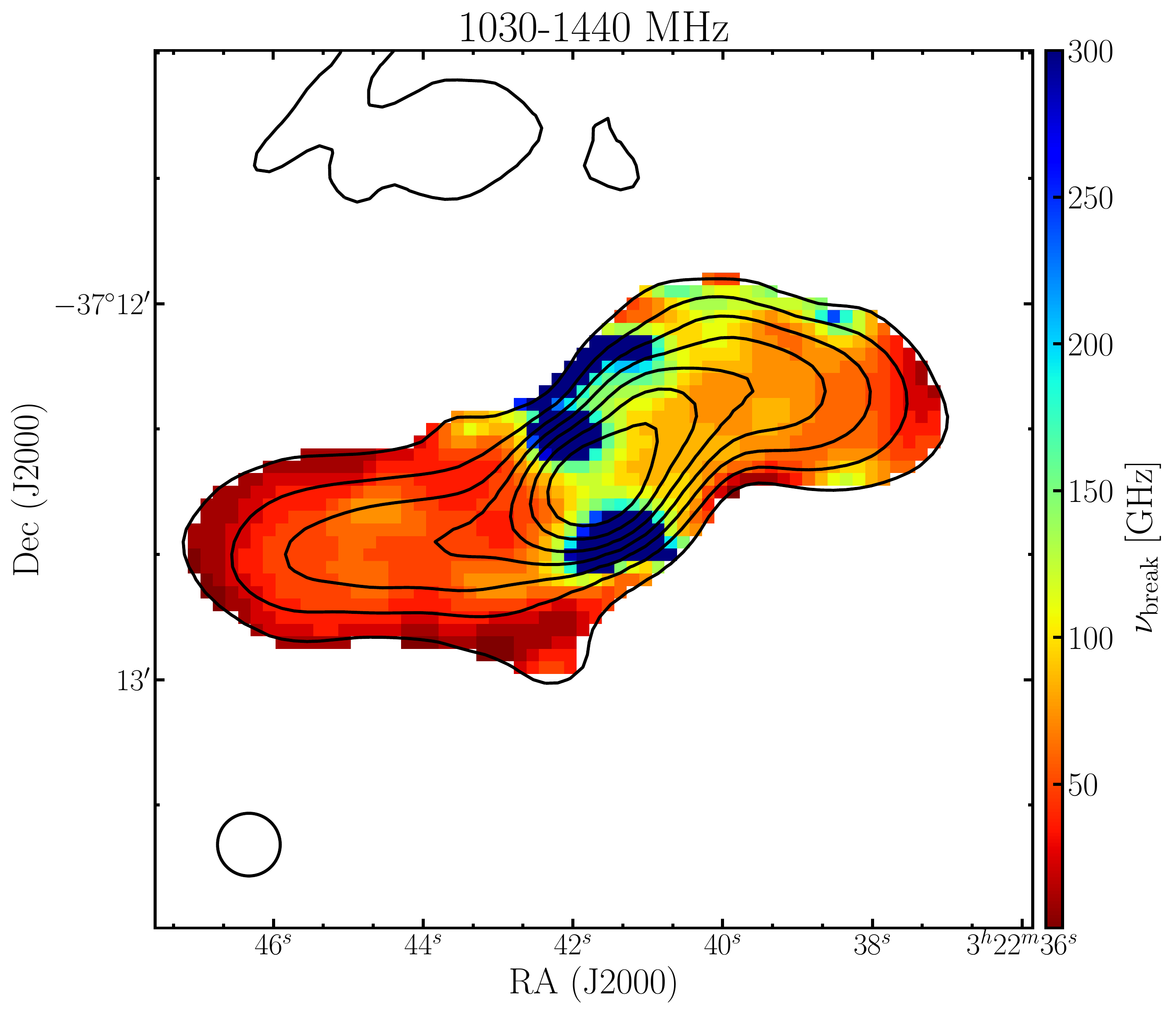}
		\caption{{\em Left Panel}: Spectral index map between $1.03$ and $1.44$~GHz of the central emission of \forn. The resolution ($10\arcsec\times10\arcsec$) is shown in the bottom left corner. The black contours show the $1.03$~GHz flux density, starting at $0.6$~\mJyb\ and increasing by factors of 2. {\em Right Panel}: break-frequency map of the central emission of \forn. The PSF of the map ($10\arcsec\times10\arcsec$) is shown in the bottom left corner. Contour levels are the same as in the left panel.}
		\label{fig:breakMapZoom} 
\end{center}
\end{figure*}

The left panel of Fig.~\ref{fig:breakMapZoom} shows the central region of the spectral index map between 1.03 GHz and 1.44 GHz. Regions with flat spectrum are located close to the kpc-core, while the spectrum steepens moving along the jets. In the right panel, we show the corresponding break-frequency map. The edges of the jets have the lowest break-frequency ($\nu_{\rm break}\lesssim 50$ GHz), while the kpc-core has a high break-frequency. This is in good agreement with the spectra of the jets and kpc-core (Fig.~\ref{fig:modCore}), which suggest that the jets have not been replenished of energetic particles for most of their lifetime. 

The indications given by the map of break-frequency confirm what was previously observed in the spectral index map between the VLA observations at $4.9$ and $14.9$ GHz~\citep{ekers1983,geldzahler1984} that attributed the steepening of the spectral index in the jets to their radiative ageing.
 
\section{Discussion}
\label{sec:disc}

\subsection{The formation of the radio lobes}
\label{sec:lobesForm}

The properties of the flux density spectrum of the lobes of \forn\ are puzzling when compared to their projected size. Typically, lobes extending in the IGM for hundreds of kilo-parsecs are either the remnant of an old nuclear activity, and show a steep spectrum with low break-frequency, or they are being currently injected with relativistic particles, and show a jet or stream of particles connecting the AGN with the lobes. In the previous section, we showed that the most remarkable properties of the lobes of \forn\ are the flat spectral shape and high break-frequency ($\gtrsim20$ GHz) of their radio emission, and that the nuclear activity that was replenishing the lobes with high energy particles was short ($\sim 24$~Myr) and stopped recently ($\sim 12$~Myr ago). Hence, the main open question is how can these large lobes have formed in such a short time?

\begin{figure}
	\begin{center}
		\includegraphics[width=\columnwidth]{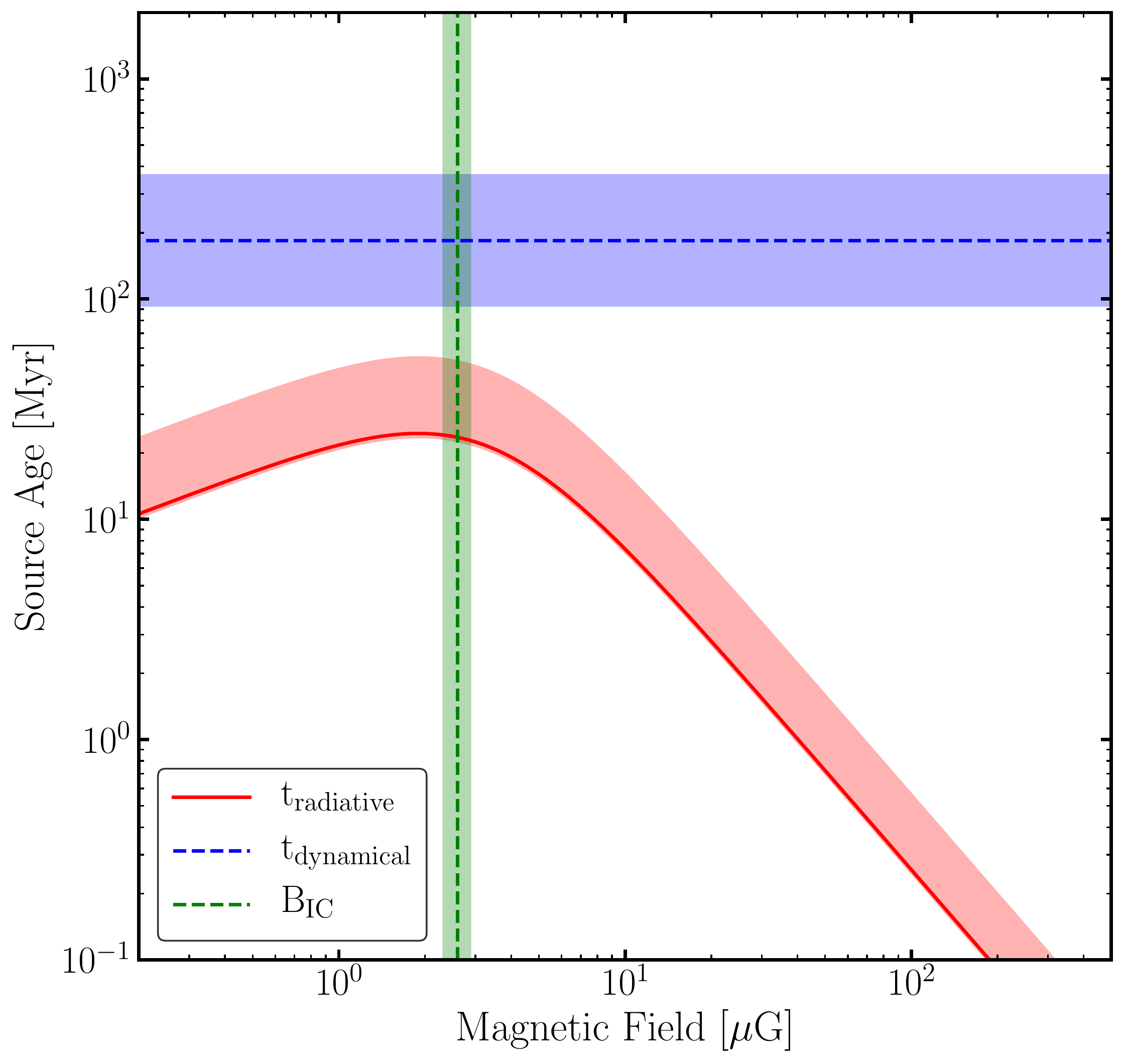}
		\caption{Comparison of radiative and dynamical age of \forn, as a function of the magnetic field. The radiative age (red shaded region) is derived from the spectral break measured in the two lobes. The dynamical age is shown in blue. The green region shows the magnetic field measured in the lobes from X-ray inverse Compton emission ($B_{IC}\sim2.6\pm0.3\mu$G).}
		\label{fig:tsyncOvertad} 
	\end{center}
\end{figure}

We estimate a dynamical age for the lobes of \forn\ by assuming that their expansion was transonic. If the energy density of the relativistic fluid is comparable to the one of the IGM, the lobes may have expanded at the speed of sound:

\begin{equation}
c_s = \sqrt{\Gamma_{\rm gas}\frac{k_BT}{\mu m_h}}
\end{equation}

\noindent where $\Gamma_{\rm gas}=5/3$, $\mu=1$ is the mean molecular weight of the gas, and $k_B$ and $m_h$ are the Boltzmann's constant and the mass of the hydrogen atom, respectively. Given that the temperature ($T$) of the IGM surrounding \forn\ ($60 < r < 350$~kpc) is $\sim2.7\times 10^7$ K \citep[$\sim 0.65-0.77$ keV, as measured from the X-ray emission of the hot halo of \mbox{NGC 1316};][]{isobe2006,nagino2009,gaspari2019}, the speed of sound is $c_s=529$~\kms. If the lobes expanded through the IGM at this speed, it would have taken $\sim 200$~Myr to reach their current size. As shown in Fig.~\ref{fig:tsyncOvertad}, this age is one order of magnitude larger than the radiative age inferred from the break-frequency of the spectrum of the lobes ($\sim 24$~Myr). 

To estimate the error on the dynamical expansion, we assumed that its speed could range between double the speed of sound ($1058$~\kms) and half of it ($\sim264$~\kms); this corresponds to a factor-of-4 change in the temperature of the IGM. The lower limit is close to the buoyancy speed used to estimate the age of the X-ray cavities detected between the lobes and the central emission of \forn~\citep[][]{lanz2010}. The upper limit of the dynamical age of the radio lobes is $\sim 400$~Myr. The lower limit is $\sim 90$ Myr. In order for the dynamical age to be compatible with the radiative age, the temperature of the IGM must be at least one order of magnitude larger ($\sim 10^8$~K), which is not compatible with the X-ray measurements of the IGM of \forn. These high temperatures are typical of cluster environments, such as for example Virgo A, where such a rapid super-sonic dynamical inflation has been observed~\citep[M87;][]{deGasperin2012}. Conversely, \forn\ is located at the edges of the Fornax cluster, where the IGM is much cooler~\citep[][]{gaspari2019}. 

We point out that the radiative age we infer from synchrotron radiation is maximal when the magnetic field is close to the field of equipartition~\citep[$B_{\rm eq}/\sqrt{3}$;][]{murgia2012}. As shown in Fig.~\ref{fig:tsyncOvertad}, the magnetic field in the lobes~\citep[$B_{\rm eq}=2.6\pm 0.3\,\mu$G;][]{isobe2006,tashiro2009} is very close to the value for which the radiative age is maximal.

Figure~\ref{fig:tsyncOvertad} indicates that the radiative age of the lobes of \forn\ and the dynamical age are incompatible. It is possible that in a single powerful episode of nuclear activity, super-sonic jets carved their way through the IGM and shaped the lobes of \forn\ as we see them now. This phase recently stopped, which would explain the absence of jets and or hot-spots and the spectral shape of the lobes. The main argument disfavouring this solution is the axial ratio of the lobes. This is defined as the length of both lobes divided by the total extent of the source~\citep{leahy1984}. Typically, in AGN where the radiative age of the particles of the lobes is similar to their dynamical time of expansion the axial ratio is low~\citep[][]{hardcastle2000,mullin2008}, while it is close to unity in \forn. 

Another possibility is that the lobes of \forn\ formed through multiple episodes of activity and that, currently, we observe the remnant of the last phase, which started approximately $24$~Myr ago and was interrupted $12$~Myr ago. If the lobes, filled with low-energy particles and under-pressured with respect to the surrounding IGM, were already present, because of previous activities, a new nuclear phase may rapidly fill them with new high energy particles which now dominate the  radio emission of the source. This scenario would explain the overall flat radio spectrum of \forn\ and its high break-frequency. The low-energy particles, remnants of the previous phases, even though dominating the number density of the lobes may be negligible in the flux density for two reasons. First, while the source is not active, the particles expand in the IGM losing energy and shifting to lower frequencies. Second, any new injection collects the most energetic particles at the edge of the lobes sweeping away the low energy ones. This may also explain the double-shell structure of the lobes of \forn. The high energy particles collect at the outer edge of the lobes (which shows flat spectrum), while the less energetic (and older particles) are swept away in a back-flow. This is formed by the inner edges of the lobes and the bridge, which both show a steep spectrum (see Fig.~\ref{fig:spIxMeer}). Two separate AGN outbursts have also been proposed by \citet{lanz2010} to explain the location of the X-ray cavities relative to the radio lobes of \forn. 
The formation of the lobes of \forn\ may have been similar to the one of \mbox{Hercules A}. This galaxy shows no clear indication of hot-spots in the lobes, and their double-shell lobe structure has been interpreted as a new activity replenishing high energy particles in the cocoon, remnant of the previous activity~\citep{gizani2003}.

Different episodes of activity would also explain why it is difficult to classify \forn\ as a FRI or a FR II galaxy~\citep{fanaroff1974}. Its low $R$ parameter ($R=S_{\rm kpc-core}/(S_{\rm tot} - S_{\rm kpc-core}) = 4\times10^{-4}$), the non-beamed jets and lack of hot-spots in the lobes would classify \forn\ as a FRI source~\citep{morganti1993}. However, the \OIII\ luminosity is too low for FRI sources~\citep[$7.96\times 10^{-15}$\ergscm;][]{tadhunter1993} and most of the flux is included in the extended lobes, as is common in FRII galaxies. Likely, the first phase of activity that formed the lobes was much more powerful than the activity of the central emission. Another explanation of the difficulty in classifying \forn\ as an FRI/FRII galaxy may be that, because of the peculiar motions of the host galaxy and of the lobes in the ISM, the galaxy moved away from where it was inflating the lobes~\citep[][]{ekers1983}. If at the time of the last inflation the galaxy was located where the bridge is, then the flatter spectrum regions at the edges of the lobes could be the remnants of the hotspots of a past FRII activity. Nevertheless, the relative velocity between the galaxy and the lobes would be very high ($\sim 2450$~\kms), because the last inflation stopped $12$ Myr ago and during this time the galaxy should have moved approximately $30$ kpc.

Emission from proton-proton ({\em p--p}) collisions of cosmic rays with the thermal plasma within the filaments of the lobes of \forn, along with inverse Compton scattering, has been invoked to explain the properties of the radio, X-ray and $\gamma$-ray emission of \forn~\citep{seta2013,mckinley2015,ackermann2016}. In the radio band, emission arising from hadronic collisions should result in a single power law spectrum at radio frequencies~\citep[see Eq.~9 in][]{dolag2000}. In \forn, the radio lobes show a clear curvature in the total spectrum (Fig.~\ref{fig:modEW} and Fig.~\ref{fig:modAll}), as well as in most regions within the lobes (at the resolution of $3\arcmin$, see Sect.~\ref{sec:spiBr} and Fig.~\ref{fig:pixJP}). A single power law cannot describe sufficiently well the spectral shape of the lobes of \forn, which suggests that hadronic collisions are not the dominant process characterising their radio emission. The only exception may be region $W_s$ (see Fig.~\ref{fig:pixJP}) which shows the highest break-frequency and flatter spectrum of the West lobe. Interestingly, this region coincides with the peak of the $\gamma$-ray emission~\citep{ackermann2016}. There, it is possible that interactions between the relativistic plasma and high energy particles of the IGM are occurring, as suggested by~\cite{mckinley2015}.

\subsection{The flickering activity of \forn}
\label{sec:flick}

In Section~\ref{sec:spModAll}, we studied the radio spectrum of the lobes, and central emission of \forn. We measured the injection index and break-frequency from which we estimated the total radiative age of the lobes and central jets and examined whether these components are still being injected with relativistic particles. Given our results we are now in a position to provide a timescale for the nuclear activity of \forn. 

\begin{figure}
	\begin{center}
		\includegraphics[width=\columnwidth]{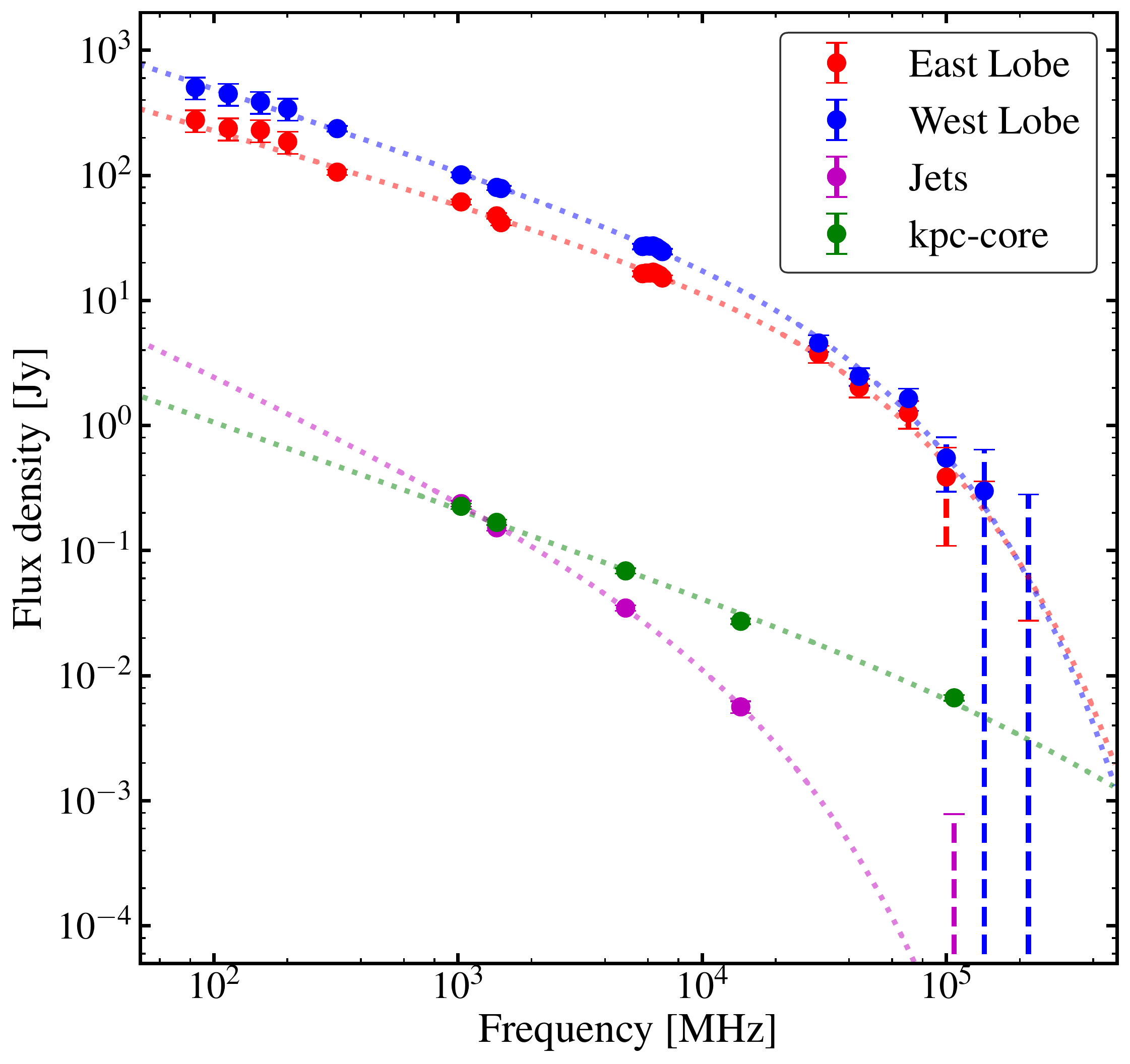}
		\caption{Radio spectrum of the different components of \forn. The East lobe is shown in red, the West lobe in blue. The jets and kpc-core are in magenta and green, respectively. The dashed lines show the model of injection that best fits the flux distributions. The spectral shape of the lobes is very different from the one of the inner components.}
		\label{fig:modAll} 
	\end{center}
\end{figure}

In Fig.~\ref{fig:modAll}, we show the flux density distribution of the East and West lobe and of the kpc-core and jets. The dashed lines show the best fit model to the spectra. For the jets and lobes the best fit model is the CI$_{\rm OFF}$ model, with similar injection index ($\alpha_{\rm inj}\sim 0.6$). While the kpc-core is best described by the CI model. From the spectral modelling, we infer that the total radiative age of the lobes is $\sim 24$ Myr and that they have not been injected with relativistic particles for the last $12$ Myr. The jets show a total radiative age of $\sim 3$ Myr and have been off for most of their lifetime ($t_{\rm OFF}/t_{\rm s}=0.89$). These radiative ages seem to indicate that the AGN turned on again for a very short phase ($\lesssim 1$ Myr), when the lobes had been off for already $\sim 9$ Myr. The spectral index and break frequency maps (Figs.~\ref{fig:breakMapZoom}) confirm that the radiative age of the jets increases radially, with the younger regions closer to the kpc-core, and suggest that the nuclear activity that generated the jets was a short, low-power episode where the plasma ejected by the AGN remained confined in the centre of the galaxy ($r\lesssim6$ kpc). Given the short timescales of the different nuclear activities, likely, \forn\ is rapidly flickering from an active phase to a non-active one.

Another argument in favour of the multiple nuclear activities of \forn\ is given by the location of the jets and lobes in the jet-core luminosity function of radio AGN~\citep{parma1987}. The ratio between the jet power and kpc-core power well fits in the jet-core luminosity function for objects of this radio power. While the ratio between the total radio power ( $2.0\times 10^{25}$~\whz, dominated by the emission of the lobes) and the jet radio power is low for a source of this luminosity, and this would make \forn\ an outlier in the jet radio luminosity function. 

Currently, the central nucleus may be in a new active phase, as suggested by the flat spectral shape of the kpc-core, compatible with the CI model (Fig.~\ref{fig:modCore}). In the centre of \forn, Parkes-Tidbinibilla interferometric (PTI) observations~\citep{jones1994,slee1994} did not detect any emission at 2.2 and 8.4 GHz over scales of $90$ and $27$ mas ($\sim10$ pc and $3$ pc, respectively) down to a $1\sigma$-noise level of $3$ and $6$ mJy. This corresponds to a brightness temperature $\sim 10^5$ K suggesting that, at the time of the observations, the radio emission of the core of \forn\ was very weak ($\lesssim 2.0\times 10^{20}$~\whz). Possibly, the core of \forn\ is variable over short timescales, and new sub-arcsecond resolution observations may provide further insights on its activity.

The radio emission of \forn\ allows us to distinguish three phases of nuclear activity, the one that last injected the lobes, the one that formed the jets and the one of the kpc-core. In the previous Section, we suggest that a possible explanation for the axial-symmetry of the lobes and their flat spectral index and high break-frequency can be that we are currently observing only the last episode of injection, but that multiple episodes of activity may have occurred also in the past. This further corroborates the interpretation that the nuclear activity of \forn\ is flickering. The recurrent activity of \forn\ may fit well in the theoretical scenario of AGN evolution whereby the central engine is active for short periods of time ($10^{4-5}$ years), and that these phases recursively occur over the total lifetime of the AGN \citep[$10^8$ years;~\eg\ ][]{schawinski2015,king2015,morganti2017}.

The environment in which \forn\ is embedded likely plays a crucial role in regulating its flickering nuclear activity. \ngcsix\ is a merger-remnant and the most massive early-type galaxy of a group falling into a cluster. The same MeerKAT observation used in this paper was analysed in spectral line by \citet{serra2019}, leading to the conclusion that the merger occurred between a dominant, gas-poor lenticular and a $\sim 10$ times smaller Milky Way-like galaxy. The cold ISM of this smaller progenitor was partly expelled out to a large radius along tidal tails ($M_{\rm \HI}\sim 7 \times 10^8$~\msun) and partly flowed towards the centre ($M_{\rm \HI}\sim 4 \times 10^7$~\msun). There, it is now mostly found in molecular form~\citep[$6\times 10^8$~\msun][]{horellou2001,morokuma2019}, and has triggered a burst of star formation~\citep[][]{kuntschner2000,silva2008} at about the same time as the formation of the globular clusters~\citep[$1$ -- $3$ Gyr ago;][]{goudfrooij2001,sesto2016,sesto2018}. This merger may have triggered for the first time the AGN, but is not responsible for the recent ($\sim 24$ Myr) last episode of injection of the lobes that we observe now.

After the major merger, \ngcsix\ went through several accretion events and minor mergers of smaller companions~\citep{iodice2017}, for which, at this stage, we do not know the typical gas content. These numerous interactions may have regulated the switching on and off of the multiple episodes of activity that formed the lobes as we see them now. Merger and interaction events are often invoked to explain the triggering of powerful AGN~\citep[\eg][]{heckman1986,hopkins2005,ramosalmeida2012,sabater2013}. Chaotic accretion of cold clouds is also an efficient mechanism to trigger and regulate the recurrent activity of AGN~\citep[\eg][]{gaspari2013,king2015,gaspari2017,gaspari2018}. This mechanism, where cold clouds condensate from turbulence in the hot halo of galaxies, is observed in early-type galaxies, typically living in dense environments such as clusters and groups~\citep[\eg][]{tremblay2016,maccagni2018,juranova2019,storchi-bergmann2019}. Given the distribution of the cold gas in the group of \forn, and the evidence for the occurrence of multiple interaction events, it is possible that one of these triggering mechanisms, or a combination of the two, regulates the flickering activity of \forn.

\section{Summary and conclusions}
\label{sec:conc}

We presented new observations of \forn\ taken at $\sim 1$~GHz with MeerKAT and at $\sim6$~GHz with the SRT. We used these data, along with archival observations of this source between $84$ MHz and $217$ GHz, to characterise the flux density distribution of the lobes and of the central emission of \forn. The spectral modelling of these components revealed that the lobes are not currently being replenished with energetic particles, and that this activity ceased about 12 Myr ago. In the centre, the central emission emission is not the remnant of the active phases that formed the lobes, but is tracing a more recent activity of the AGN. The radiative ageing along the jets suggests that the injection phase was short ($\lesssim 1$~Myr), and that the jets have been off for most of their time ($t_{\rm OFF}\sim 2$~Myr).

 The MeerKAT observation allowed us to obtain crucial information about the lobes of \forn\ and their spectral properties. The filaments appear embedded in a diffuse cocoon forming a double-shell structure. The outer edges of this shell with respect to the AGN have flat spectrum and high break frequency (see Figs.~\ref{fig:spIxMeer} and ~\ref{fig:breakMap}), while the inner regions and the bridge connecting the lobes shows steep spectrum and lower break frequency. The comparison between the radiative age of the lobes and the dynamical age (Fig.~\ref{fig:tsyncOvertad}) suggests that multiple nuclear activities inflated the lobes of high energy particles, and that currently we measure the radiative age of the last of these recurrent episodes, which may also explain the double shell morphology of the lobes.
 
The spectra of the kpc-core and jets (Fig.~\ref{fig:modCore}) suggests that the recent nuclear activity that formed them was shorter and likely less powerful than the previous one that inflated the lobes. As shown in Fig.~\ref{fig:modAll}, currently in \forn\ it is possible to observe three distinct phases of activity. The remnant phase that last injected the lobes, the more recent phase that generated the jets and the current activity of the kpc-core. Given the short timescales of the nuclear activities we identified, \forn\ is likely rapidly flickering from an active nuclear phase to a non-active one. 
 
 Further information about the recurrent activity of \forn\ and, in particular about the last activity that generated the central emission, may be found in the study of the kinematics and distribution of the cold gas in the centre of the galaxy, along the jets. There, the molecular gas shows kinematics deviating from regular rotation, hinting that part of the gas may be involved in the feeding of the AGN~\citep{morokuma2019}. Neutral hydrogen clouds have also been detected at the same locations as the molecular gas~\citep[][]{serra2019}. The analysis of its distribution and kinematics in relation to the nuclear activity of \forn\ will be presented in a following paper.

\begin{acknowledgements}
This paper is dedicated to the memory of Sergio Colafrancesco. The authors thank the anonymous referee for the useful comments and suggestions. The authors wish to thank Fernando Camilo for useful comments on an early draft of this paper. We are grateful to the full MeerKAT team at SARAO for their work on building and commissioning MeerKAT. This project has received funding from the European Research Council (ERC) under the European Union’s Horizon 2020 research and innovation programme (grant agreement no. 679627). PK is partially supported by the BMBF project 05A17PC2 for D-MeerKAT. The MeerKAT telescope is operated by the South African Radio Astronomy Observatory, which is a facility of the National Research Foundation, an agency of the Department of Science and Innovation. The development of the SARDARA backend has been funded by the Autonomous Region of Sardinia (RAS) using resources from the Regional Law 7/2007 “Promotion of the scientific research and technological innovation in Sardinia” in the context of the research project CRP-49231 (year 2011, PI Possenti): “High resolution sampling of the Universe in the radio band: an unprecedented instrument to understand the fundamental laws of the nature”. This paper makes use of the following ALMA data: ADS/JAO.ALMA\#2017.1.00129.S. ALMA is a partnership of ESO (representing its member states). NSF (USA) and NINS (Japan), together with NRC (Canada), MOST and ASAIA (Taiwan), and KASI (Republic of Korea), in cooperation with the Republic of Chile. The Joint ALMA Observatory is operated by ESO, AUI/NRAO and NAOJ. The VLA images at 4.8 and 14.4 GHz have been produced as part of the NRAO VLA Archive Survey, (c) AUI/NRAO. The National Radio Astronomy Observatory (NRAO) is a facility of the National Science Foundation, operated under cooperative agreement by Associated Universities, Inc.
\end{acknowledgements}

%
%


\bibliographystyle{aa} 
\bibliography{fornaxABib.bib} 

\begin{thebibliography}{146}
\expandafter\ifx\csname natexlab\endcsname\relax\def\natexlab#1{#1}\fi

\bibitem[{{Ackermann} {et~al.}(2016){Ackermann}, {Ajello}, {Baldini}, {Ballet},
  {Barbiellini}, {Bastieri}, {Bellazzini}, {Bissaldi}, {Blandford}, {Bloom},
  {Bonino}, {Brandt}, {Bregeon}, {Bruel}, {Buehler}, {Buson}, {Caliandro},
  {Cameron}, {Caragiulo}, {Caraveo}, {Cavazzuti}, {Cecchi}, {Charles},
  {Chekhtman}, {Cheung}, {Chiaro}, {Ciprini}, {Cohen}, {Cohen-Tanugi},
  {Costanza}, {Cutini}, {D'Ammand o}, {Davis}, {de Angelis}, {de Palma},
  {Desiante}, {Digel}, {Di Lalla}, {Di Mauro}, {Di Venere}, {Favuzzi}, {Fegan},
  {Ferrara}, {Focke}, {Fukazawa}, {Funk}, {Fusco}, {Gargano}, {Gasparrini},
  {Georganopoulos}, {Giglietto}, {Giordano}, {Giroletti}, {Godfrey}, {Green},
  {Grenier}, {Guiriec}, {Hays}, {Hewitt}, {Hill}, {Jogler}, {J{\'o}hannesson},
  {Kensei}, {Kuss}, {Larsson}, {Latronico}, {Li}, {Li}, {Longo}, {Loparco},
  {Lubrano}, {Magill}, {Maldera}, {Manfreda}, {Mayer}, {Mazziotta},
  {McConville}, {McEnery}, {Michelson}, {Mitthumsiri}, {Mizuno}, {Monzani},
  {Morselli}, {Moskalenko}, {Murgia}, {Negro}, {Nuss}, {Ohno}, {Ohsugi},
  {Orienti}, {Orlando}, {Ormes}, {Paneque}, {Perkins}, {Pesce-Rollins},
  {Piron}, {Pivato}, {Porter}, {Rain{\`o}}, {Rando}, {Razzano}, {Reimer},
  {Reimer}, {Schmid}, {Sgr{\`o}}, {Simone}, {Siskind}, {Spada}, {Spandre},
  {Spinelli}, {Stawarz}, {Takahashi}, {Thayer}, {Thompson}, {Torres}, {Tosti},
  {Troja}, {Vianello}, {Wood}, {Wood}, {Zimmer}, \& {Fermi LAT
  Collaboration}}]{ackermann2016}
{Ackermann}, M., {Ajello}, M., {Baldini}, L., {et~al.} 2016, \apj, 826, 1

\bibitem[{{Anderson} {et~al.}(2018){Anderson}, {Gaensler}, {Heald},
  {O'Sullivan}, {Kaczmarek}, \& {Feain}}]{anderson2018}
{Anderson}, C.~S., {Gaensler}, B.~M., {Heald}, G.~H., {et~al.} 2018, \apj, 855,
  41

\bibitem[{{Asad} {et~al.}(2019){Asad}, {Girard}, {de Villers}, {Lehmensiek},
  {Ansah-Narh}, {Iheanetu}, {Smirnov}, {Santos}, {Jonas}, {de Villiers},
  {Thorat}, {Hugo}, {Makhathini}, {Camilo}, {Jozsa}, \& {Sirothia}}]{asad2019}
{Asad}, K.~M.~B., {Girard}, J.~N., {de Villers}, M., {et~al.} 2019, arXiv
  e-prints, arXiv:1904.07155

\bibitem[{{Baars} {et~al.}(1977){Baars}, {Genzel}, {Pauliny-Toth}, \&
  {Witzel}}]{baars1977}
{Baars}, J.~W.~M., {Genzel}, R., {Pauliny-Toth}, I.~I.~K., \& {Witzel}, A.
  1977, \aap, 500, 135

\bibitem[{{Battistelli} {et~al.}(2019){Battistelli}, {Fatigoni}, {Murgia},
  {Buzzelli}, {Carretti}, {Castangia}, {Concu}, {Cruciani}, {de Bernardis},
  {Genova-Santos}, {Govoni}, {Guidi}, {Lamagna}, {Luzzi}, {Masi}, {Melis},
  {Paladini}, {Piacentini}, {Poppi}, {Radiconi}, {Rebolo}, {Rubino-Martin},
  {Tarchi}, \& {Vacca}}]{battistelli2019}
{Battistelli}, E.~S., {Fatigoni}, S., {Murgia}, M., {et~al.} 2019, \apjl, 877,
  L31

\bibitem[{{Bennett} {et~al.}(2013){Bennett}, {Larson}, {Weiland}, {Jarosik},
  {Hinshaw}, {Odegard}, {Smith}, {Hill}, {Gold}, {Halpern}, {Komatsu}, {Nolta},
  {Page}, {Spergel}, {Wollack}, {Dunkley}, {Kogut}, {Limon}, {Meyer}, {Tucker},
  \& {Wright}}]{bennett2013}
{Bennett}, C.~L., {Larson}, D., {Weiland}, J.~L., {et~al.} 2013, \apjs, 208, 20

\bibitem[{{Bernardi} {et~al.}(2013){Bernardi}, {Greenhill}, {Mitchell}, {Ord},
  {Hazelton}, {Gaensler}, {de Oliveira-Costa}, {Morales}, {Udaya Shankar},
  {Subrahmanyan}, {Wayth}, {Lenc}, {Williams}, {Arcus}, {Arora}, {Barnes},
  {Bowman}, {Briggs}, {Bunton}, {Cappallo}, {Corey}, {Deshpand e}, {deSouza},
  {Emrich}, {Goeke}, {Herne}, {Hewitt}, {Johnston-Hollitt}, {Kaplan}, {Kasper},
  {Kincaid}, {Koenig}, {Kratzenberg}, {Lonsdale}, {Lynch}, {McWhirter},
  {Morgan}, {Oberoi}, {Pathikulangara}, {Prabu}, {Remillard}, {Rogers},
  {Roshi}, {Salah}, {Sault}, {Srivani}, {Stevens}, {Tingay}, {Waterson},
  {Webster}, {Whitney}, {Williams}, \& {Wyithe}}]{bernardi2013}
{Bernardi}, G., {Greenhill}, L.~J., {Mitchell}, D.~A., {et~al.} 2013, \apj,
  771, 105

\bibitem[{{Bolatto} {et~al.}(2013){Bolatto}, {Warren}, {Leroy}, {Walter},
  {Veilleux}, {Ostriker}, {Ott}, {Zwaan}, {Fisher}, {Weiss}, {Rosolowsky}, \&
  {Hodge}}]{bolatto2013}
{Bolatto}, A.~D., {Warren}, S.~R., {Leroy}, A.~K., {et~al.} 2013, \nat, 499,
  450

\bibitem[{{Booth} \& {Schaye}(2009)}]{booth2009}
{Booth}, C.~M. \& {Schaye}, J. 2009, \mnras, 398, 53

\bibitem[{{Bower} {et~al.}(2006){Bower}, {Benson}, {Malbon}, {Helly}, {Frenk},
  {Baugh}, {Cole}, \& {Lacey}}]{bower2006}
{Bower}, R.~G., {Benson}, A.~J., {Malbon}, R., {et~al.} 2006, \mnras, 370, 645

\bibitem[{{Bridle} {et~al.}(1991){Bridle}, {Baum}, {Fomalont}, {Parma},
  {Fanti}, \& {Ekers}}]{bridle1991}
{Bridle}, A.~H., {Baum}, S.~A., {Fomalont}, E.~B., {et~al.} 1991, \aap, 245,
  371

\bibitem[{{Bridle} \& {Perley}(1984)}]{bridle1984}
{Bridle}, A.~H. \& {Perley}, R.~A. 1984, \araa, 22, 319

\bibitem[{{Brienza} {et~al.}(2016){Brienza}, {Godfrey}, {Morganti}, {Vilchez},
  {Maddox}, {Murgia}, {Orru}, {Shulevski}, {Best}, {Br{\"u}ggen}, {Harwood},
  {Jamrozy}, {Jarvis}, {Mahony}, {McKean}, \& {R{\"o}ttgering}}]{brienza2016}
{Brienza}, M., {Godfrey}, L., {Morganti}, R., {et~al.} 2016, \aap, 585, A29

\bibitem[{{Brienza} {et~al.}(2018){Brienza}, {Morganti}, {Murgia}, {Vilchez},
  {Adebahr}, {Carretti}, {Concu}, {Govoni}, {Harwood}, {Intema}, {Loi},
  {Melis}, {Paladino}, {Poppi}, {Shulevski}, {Vacca}, \&
  {Valente}}]{brienza2018}
{Brienza}, M., {Morganti}, R., {Murgia}, M., {et~al.} 2018, \aap, 618, A45

\bibitem[{{Cameron}(1971)}]{cameron1971}
{Cameron}, M.~J. 1971, \mnras, 152, 439

\bibitem[{{Camilo} {et~al.}(2018){Camilo}, {Scholz}, {Serylak}, {Buchner},
  {Merryfield}, {Kaspi}, {Archibald}, {Bailes}, {Jameson}, {van Straten},
  {Sarkissian}, {Reynolds}, {Johnston}, {Hobbs}, {Abbott}, {Adam}, {Adams},
  {Alberts}, {Andreas}, {Asad}, {Baker}, {Baloyi}, {Bauermeister}, {Baxana},
  {Bennett}, {Bernardi}, {Booisen}, {Booth}, {Botha}, {Boyana}, {Brederode},
  {Burger}, {Cheetham}, {Conradie}, {Conradie}, {Davidson}, {De Bruin}, {de
  Swardt}, {de Villiers}, {de Villiers}, {de Villiers}, {de Villiers}, {De
  Waal}, {Dikgale}, {du Toit}, {du Toit}, {Esterhuyse}, {Fanaroff}, {Fataar},
  {Foley}, {Foster}, {Fourie}, {Gamatham}, {Gatsi}, {Geschke}, {Goedhart},
  {Grobler}, {Gumede}, {Hlakola}, {Hokwana}, {Hoorn}, {Horn}, {Horrell},
  {Hugo}, {Isaacson}, {Jacobs}, {Jansen van Rensburg}, {Jonas}, {Jordaan},
  {Joubert}, {Joubert}, {J{\'o}zsa}, {Julie}, {Julius}, {Kapp}, {Karastergiou},
  {Karels}, {Kariseb}, {Karuppusamy}, {Kasper}, {Knox-Davies}, {Koch},
  {Kotz{\'e}}, {Krebs}, {Kriek}, {Kriel}, {Kusel}, {Lamoor}, {Lehmensiek},
  {Liebenberg}, {Liebenberg}, {Lord}, {Lunsky}, {Mabombo}, {Macdonald},
  {Macfarlane}, {Madisa}, {Mafhungo}, {Magnus}, {Magozore}, {Mahgoub}, {Main},
  {Makhathini}, {Malan}, {Malgas}, {Manley}, {Manzini}, {Marais}, {Marais},
  {Marais}, {Maree}, {Martens}, {Matshawule}, {Matthysen}, {Mauch}, {McNally},
  {Merry}, {Millenaar}, {Mjikelo}, {Mkhabela}, {Mnyand u}, {Moeng}, {Mokone},
  {Monama}, {Montshiwa}, {Moss}, {Mphego}, {New}, {Ngcebetsha}, {Ngoasheng},
  {Niehaus}, {Ntuli}, {Nzama}, {Obies}, {Obrocka}, {Ockards}, {Olyn}, {Oozeer},
  {Otto}, {Padayachee}, {Passmoor}, {Patel}, {Paula}, {Peens-Hough},
  {Pholoholo}, {Prozesky}, {Rakoma}, {Ramaila}, {Rammala}, {Ramudzuli},
  {Rasivhaga}, {Ratcliffe}, {Reader}, {Renil}, {Richter}, {Robyntjies},
  {Rosekrans}, {Rust}, {Salie}, {Sambu}, {Schollar}, {Schwardt}, {Seranyane},
  {Sethosa}, {Sharpe}, {Siebrits}, {Sirothia}, {Slabber}, {Smirnov}, {Smith},
  {Sofeya}, {Songqumase}, {Spann}, {Stappers}, {Steyn}, {Steyn}, {Strong},
  {Struthers}, {Stuart}, {Sunnylall}, {Swart}, {Taljaard}, {Tasse}, {Taylor},
  {Theron}, {Thondikulam}, {Thorat}, {Tiplady}, {Toruvanda}, {van Aardt}, {van
  Balla}, {van den Heever}, {van der Byl}, {van der Merwe}, {van der Merwe},
  {van Niekerk}, {van Rooyen}, {van Staden}, {van Tonder}, {van Wyk}, {Wait},
  {Walker}, {Wallace}, {Welz}, {Williams}, {Xaia}, {Young}, \&
  {Zitha}}]{camilo2018}
{Camilo}, F., {Scholz}, P., {Serylak}, M., {et~al.} 2018, \apj, 856, 180

\bibitem[{{Cantiello} {et~al.}(2013){Cantiello}, {Grado}, {Blakeslee},
  {Raimondo}, {Di Rico}, {Limatola}, {Brocato}, {Della Valle}, \&
  {Gilmozzi}}]{cantiello2013}
{Cantiello}, M., {Grado}, A., {Blakeslee}, J.~P., {et~al.} 2013, \aap, 552,
  A106

\bibitem[{{Carilli} {et~al.}(1991){Carilli}, {Perley}, {Dreher}, \&
  {Leahy}}]{carilli1991}
{Carilli}, C.~L., {Perley}, R.~A., {Dreher}, J.~W., \& {Leahy}, J.~P. 1991,
  \apj, 383, 554

\bibitem[{{Carlqvist}(2010)}]{carlqvist2010}
{Carlqvist}, P. 2010, \apss, 327, 267

\bibitem[{{Churazov} {et~al.}(2001){Churazov}, {Br{\"u}ggen}, {Kaiser},
  {B{\"o}hringer}, \& {Forman}}]{churazov2001}
{Churazov}, E., {Br{\"u}ggen}, M., {Kaiser}, C.~R., {B{\"o}hringer}, H., \&
  {Forman}, W. 2001, \apj, 554, 261

\bibitem[{{Ciotti} {et~al.}(2010){Ciotti}, {Ostriker}, \& {Proga}}]{ciotti2010}
{Ciotti}, L., {Ostriker}, J.~P., \& {Proga}, D. 2010, \apj, 717, 708

\bibitem[{{Ciotti} \& {Ziaee Lorzad}(2018)}]{ciotti2018}
{Ciotti}, L. \& {Ziaee Lorzad}, A. 2018, \mnras, 473, 5476

\bibitem[{{Comerford} {et~al.}(2017){Comerford}, {Barrows},
  {M{\"u}ller-S{\'a}nchez}, {Nevin}, {Greene}, {Pooley}, {Stern}, \&
  {Harrison}}]{comerford2017}
{Comerford}, J.~M., {Barrows}, R.~S., {M{\"u}ller-S{\'a}nchez}, F., {et~al.}
  2017, \apj, 849, 102

\bibitem[{{Croton} {et~al.}(2006){Croton}, {Springel}, {White}, {De Lucia},
  {Frenk}, {Gao}, {Jenkins}, {Kauffmann}, {Navarro}, \& {Yoshida}}]{croton2006}
{Croton}, D.~J., {Springel}, V., {White}, S. D.~M., {et~al.} 2006, \mnras, 365,
  11

\bibitem[{{de Gasperin} {et~al.}(2014){de Gasperin}, {Intema}, {Williams},
  {Br{\"u}ggen}, {Murgia}, {Beck}, \& {Bonafede}}]{deGasperin2014}
{de Gasperin}, F., {Intema}, H.~T., {Williams}, W., {et~al.} 2014, \mnras, 440,
  1542

\bibitem[{{de Gasperin} {et~al.}(2012){de Gasperin}, {Orr{\'u}}, {Murgia},
  {Merloni}, {Falcke}, {Beck}, {Beswick}, {B{\^\i}rzan}, {Bonafede},
  {Br{\"u}ggen}, {Brunetti}, {Chy{\.z}y}, {Conway}, {Croston}, {En{\ss}lin},
  {Ferrari}, {Heald}, {Heidenreich}, {Jackson}, {Macario}, {McKean}, {Miley},
  {Morganti}, {Offringa}, {Pizzo}, {Rafferty}, {R{\"o}ttgering}, {Shulevski},
  {Steinmetz}, {Tasse}, {van der Tol}, {van Driel}, {van Weeren}, {van
  Zwieten}, {Alexov}, {Anderson}, {Asgekar}, {Avruch}, {Bell}, {Bell},
  {Bentum}, {Bernardi}, {Best}, {Breitling}, {Broderick}, {Butcher}, {Ciardi},
  {Dettmar}, {Eisloeffel}, {Frieswijk}, {Gankema}, {Garrett}, {Gerbers},
  {Griessmeier}, {Gunst}, {Hassall}, {Hessels}, {Hoeft}, {Horneffer},
  {Karastergiou}, {K{\"o}hler}, {Koopman}, {Kuniyoshi}, {Kuper}, {Maat},
  {Mann}, {Mevius}, {Mulcahy}, {Munk}, {Nijboer}, {Noordam}, {Paas}, {Pandey},
  {Pand ey}, {Polatidis}, {Reich}, {Schoenmakers}, {Sluman}, {Smirnov},
  {Sobey}, {Stappers}, {Swinbank}, {Tagger}, {Tang}, {van Bemmel}, {van
  Cappellen}, {van Duin}, {van Haarlem}, {van Leeuwen}, {Vermeulen}, {Vocks},
  {White}, {Wise}, {Wucknitz}, \& {Zarka}}]{deGasperin2012}
{de Gasperin}, F., {Orr{\'u}}, E., {Murgia}, M., {et~al.} 2012, \aap, 547, A56

\bibitem[{{Dolag} \& {En{\ss}lin}(2000)}]{dolag2000}
{Dolag}, K. \& {En{\ss}lin}, T.~A. 2000, \aap, 362, 151

\bibitem[{{Donohoe} \& {Smith}(2016)}]{donohoe2016}
{Donohoe}, J. \& {Smith}, M.~D. 2016, \mnras, 458, 558

\bibitem[{{Drinkwater} {et~al.}(2001){Drinkwater}, {Gregg}, \&
  {Colless}}]{drinkwater2001}
{Drinkwater}, M.~J., {Gregg}, M.~D., \& {Colless}, M. 2001, \apjl, 548, L139

\bibitem[{{Duah Asabere} {et~al.}(2016){Duah Asabere}, {Horellou}, {Jarrett},
  \& {Winkler}}]{duha2016}
{Duah Asabere}, B., {Horellou}, C., {Jarrett}, T.~H., \& {Winkler}, H. 2016,
  \aap, 592, A20

\bibitem[{{Egron} {et~al.}(2017){Egron}, {Pellizzoni}, {Iacolina}, {Loru},
  {Marongiu}, {Righini}, {Cardillo}, {Giuliani}, {Mulas}, {Murtas}, {Simeone},
  {Concu}, {Melis}, {Trois}, {Pilia}, {Navarrini}, {Vacca}, {Ricci}, {Serra},
  {Bachetti}, {Buttu}, {Perrodin}, {Buffa}, {Deiana}, {Gaudiomonte}, {Fara},
  {Ladu}, {Loi}, {Marongiu}, {Migoni}, {Pisanu}, {Poppi}, {Saba}, {Urru},
  {Valente}, \& {Vargiu}}]{egron2017}
{Egron}, E., {Pellizzoni}, A., {Iacolina}, M.~N., {et~al.} 2017, \mnras, 470,
  1329

\bibitem[{{Ekers} {et~al.}(1983){Ekers}, {Goss}, {Wellington}, {Bosma},
  {Smith}, \& {Schweizer}}]{ekers1983}
{Ekers}, R.~D., {Goss}, W.~M., {Wellington}, K.~J., {et~al.} 1983, \aap, 127,
  361

\bibitem[{{Ellis} \& {Hamilton}(1966)}]{ellis1966}
{Ellis}, G.~R.~A. \& {Hamilton}, P.~A. 1966, \apj, 143, 227

\bibitem[{{Fabian}(2012)}]{fabian2012}
{Fabian}, A.~C. 2012, \araa, 50, 455

\bibitem[{{Fanaroff} \& {Riley}(1974)}]{fanaroff1974}
{Fanaroff}, B.~L. \& {Riley}, J.~M. 1974, \mnras, 167, 31P

\bibitem[{{Finlay} \& {Jones}(1973)}]{finlay1973}
{Finlay}, E.~A. \& {Jones}, B.~B. 1973, Aust. J. Phys., 26, 389

\bibitem[{{Fluetsch} {et~al.}(2019){Fluetsch}, {Maiolino}, {Carniani},
  {Marconi}, {Cicone}, {Bourne}, {Costa}, {Fabian}, {Ishibashi}, \&
  {Venturi}}]{fluetsch2019}
{Fluetsch}, A., {Maiolino}, R., {Carniani}, S., {et~al.} 2019, \mnras, 483,
  4586

\bibitem[{{Fomalont} {et~al.}(1989){Fomalont}, {Ebneter}, {van Breugel}, \&
  {Ekers}}]{fomalont1989}
{Fomalont}, E.~B., {Ebneter}, K.~A., {van Breugel}, W. J.~M., \& {Ekers}, R.~D.
  1989, \apjl, 346, L17

\bibitem[{{Galametz} {et~al.}(2014){Galametz}, {Albrecht}, {Kennicutt},
  {Aniano}, {Bertoldi}, {Calzetti}, {Croxall}, {Dale}, {Draine}, {Engelbracht},
  {Gordon}, {Hinz}, {Hunt}, {Kirkpatrick}, {Murphy}, {Roussel}, {Skibba},
  {Walter}, {Weiss}, \& {Wilson}}]{galametz2014}
{Galametz}, M., {Albrecht}, M., {Kennicutt}, R., {et~al.} 2014, \mnras, 439,
  2542

\bibitem[{{Galametz} {et~al.}(2012){Galametz}, {Kennicutt}, {Albrecht},
  {Aniano}, {Armus}, {Bertoldi}, {Calzetti}, {Crocker}, {Croxall}, {Dale},
  {Donovan Meyer}, {Draine}, {Engelbracht}, {Hinz}, {Roussel}, {Skibba},
  {Tabatabaei}, {Walter}, {Weiss}, {Wilson}, \& {Wolfire}}]{galametz2012}
{Galametz}, M., {Kennicutt}, R.~C., {Albrecht}, M., {et~al.} 2012, \mnras, 425,
  763

\bibitem[{{Gardner} \& {Whiteoak}(1971)}]{gardner1971}
{Gardner}, F.~F. \& {Whiteoak}, J.~B. 1971, Aust. J. Phys., 24, 899

\bibitem[{{Gaspari} {et~al.}(2019){Gaspari}, {Eckert}, {Ettori}, {Tozzi},
  {Bassini}, {Rasia}, {Brighenti}, {Sun}, {Borgani}, {Johnson}, {Tremblay},
  {Stone}, {Temi}, {Yang}, {Tombesi}, \& {Cappi}}]{gaspari2019}
{Gaspari}, M., {Eckert}, D., {Ettori}, S., {et~al.} 2019, \apj, 884, 169

\bibitem[{{Gaspari} {et~al.}(2018){Gaspari}, {McDonald}, {Hamer}, {Brighenti},
  {Temi}, {Gendron-Marsolais}, {Hlavacek-Larrondo}, {Edge}, {Werner}, {Tozzi},
  {Sun}, {Stone}, {Tremblay}, {Hogan}, {Eckert}, {Ettori}, {Yu}, {Biffi}, \&
  {Planelles}}]{gaspari2018}
{Gaspari}, M., {McDonald}, M., {Hamer}, S.~L., {et~al.} 2018, \apj, 854, 167

\bibitem[{{Gaspari} {et~al.}(2013){Gaspari}, {Ruszkowski}, \&
  {Oh}}]{gaspari2013}
{Gaspari}, M., {Ruszkowski}, M., \& {Oh}, S.~P. 2013, \mnras, 432, 3401

\bibitem[{{Gaspari} {et~al.}(2017){Gaspari}, {Temi}, \&
  {Brighenti}}]{gaspari2017}
{Gaspari}, M., {Temi}, P., \& {Brighenti}, F. 2017, \mnras, 466, 677

\bibitem[{{Geldzahler} \& {Fomalont}(1978)}]{geldzhaler1978}
{Geldzahler}, B.~J. \& {Fomalont}, E.~B. 1978, \aj, 83, 1047

\bibitem[{{Geldzahler} \& {Fomalont}(1984)}]{geldzahler1984}
{Geldzahler}, B.~J. \& {Fomalont}, E.~B. 1984, \aj, 89, 1650

\bibitem[{{Gizani} \& {Leahy}(2003)}]{gizani2003}
{Gizani}, N. A.~B. \& {Leahy}, J.~P. 2003, \mnras, 342, 399

\bibitem[{{Goudfrooij} {et~al.}(2001){Goudfrooij}, {Mack}, {Kissler-Patig},
  {Meylan}, \& {Minniti}}]{goudfrooij2001}
{Goudfrooij}, P., {Mack}, J., {Kissler-Patig}, M., {Meylan}, G., \& {Minniti},
  D. 2001, \mnras, 322, 643

\bibitem[{{Grillmair} {et~al.}(1999){Grillmair}, {Forbes}, {Brodie}, \&
  {Elson}}]{grillmair1999}
{Grillmair}, C.~J., {Forbes}, D.~A., {Brodie}, J.~P., \& {Elson}, R. A.~W.
  1999, \aj, 117, 167

\bibitem[{{Hardcastle} \& {Worrall}(2000)}]{hardcastle2000}
{Hardcastle}, M.~J. \& {Worrall}, D.~M. 2000, \mnras, 319, 562

\bibitem[{{Harwood} {et~al.}(2013){Harwood}, {Hardcastle}, {Croston}, \&
  {Goodger}}]{harwood2013}
{Harwood}, J.~J., {Hardcastle}, M.~J., {Croston}, J.~H., \& {Goodger}, J.~L.
  2013, \mnras, 435, 3353

\bibitem[{{Heckman} {et~al.}(1986){Heckman}, {Smith}, {Baum}, {van Breugel},
  {Miley}, {Illingworth}, {Bothun}, \& {Balick}}]{heckman1986}
{Heckman}, T.~M., {Smith}, E.~P., {Baum}, S.~A., {et~al.} 1986, \apj, 311, 526

\bibitem[{{Hogan} {et~al.}(2015){Hogan}, {Edge}, {Geach}, {Grainge},
  {Hlavacek-Larrondo}, {Hovatta}, {Karim}, {McNamara}, {Rumsey}, {Russell},
  {Salom{\'e}}, {Aller}, {Aller}, {Benford}, {Fabian}, {Readhead}, {Sadler}, \&
  {Saunders}}]{hogan2015}
{Hogan}, M.~T., {Edge}, A.~C., {Geach}, J.~E., {et~al.} 2015, \mnras, 453, 1223

\bibitem[{{Hopkins} {et~al.}(2005){Hopkins}, {Hernquist}, {Cox}, {Di Matteo},
  {Martini}, {Robertson}, \& {Springel}}]{hopkins2005}
{Hopkins}, P.~F., {Hernquist}, L., {Cox}, T.~J., {et~al.} 2005, \apj, 630, 705

\bibitem[{{Horellou} {et~al.}(2001){Horellou}, {Black}, {van Gorkom}, {Combes},
  {van der Hulst}, \& {Charmandaris}}]{horellou2001}
{Horellou}, C., {Black}, J.~H., {van Gorkom}, J.~H., {et~al.} 2001, \aap, 376,
  837

\bibitem[{{Hurley-Walker} {et~al.}(2017){Hurley-Walker}, {Callingham},
  {Hancock}, {Franzen}, {Hindson}, {Kapi{\'n}ska}, {Morgan}, {Offringa},
  {Wayth}, {Wu}, {Zheng}, {Murphy}, {Bell}, {Dwarakanath}, {For}, {Gaensler},
  {Johnston-Hollitt}, {Lenc}, {Procopio}, {Staveley-Smith}, {Ekers}, {Bowman},
  {Briggs}, {Cappallo}, {Deshpande}, {Greenhill}, {Hazelton}, {Kaplan},
  {Lonsdale}, {McWhirter}, {Mitchell}, {Morales}, {Morgan}, {Oberoi}, {Ord},
  {Prabu}, {Shankar}, {Srivani}, {Subrahmanyan}, {Tingay}, {Webster},
  {Williams}, \& {Williams}}]{hurleywalker2017}
{Hurley-Walker}, N., {Callingham}, J.~R., {Hancock}, P.~J., {et~al.} 2017,
  \mnras, 464, 1146

\bibitem[{{Iodice} {et~al.}(2017){Iodice}, {Spavone}, {Capaccioli}, {Peletier},
  {Richtler}, {Hilker}, {Mieske}, {Limatola}, {Grado}, {Napolitano},
  {Cantiello}, {D'Abrusco}, {Paolillo}, {Venhola}, {Lisker}, {Van de Ven},
  {Falcon-Barroso}, \& {Schipani}}]{iodice2017}
{Iodice}, E., {Spavone}, M., {Capaccioli}, M., {et~al.} 2017, \apj, 839, 21

\bibitem[{{Isobe} {et~al.}(2006){Isobe}, {Makishima}, {Tashiro}, {Itoh},
  {Iyomoto}, {Takahashi}, \& {Kaneda}}]{isobe2006}
{Isobe}, N., {Makishima}, K., {Tashiro}, M., {et~al.} 2006, \apj, 645, 256

\bibitem[{{Iyomoto} {et~al.}(1998){Iyomoto}, {Makishima}, {Tashiro}, {Inoue},
  {Kaneda}, {Matsumoto}, \& {Mizuno}}]{iyomoto1998}
{Iyomoto}, N., {Makishima}, K., {Tashiro}, M., {et~al.} 1998, \apjl, 503, L31

\bibitem[{{Jaffe} \& {Perola}(1973)}]{jaffe1973}
{Jaffe}, W.~J. \& {Perola}, G.~C. 1973, \aap, 26, 423

\bibitem[{{Jamrozy} {et~al.}(2004){Jamrozy}, {Klein}, {Mack}, {Gregorini}, \&
  {Parma}}]{jamrozy2004}
{Jamrozy}, M., {Klein}, U., {Mack}, K.~H., {Gregorini}, L., \& {Parma}, P.
  2004, \aap, 427, 79

\bibitem[{{Jonas} \& {MeerKAT Team}(2016)}]{jonas2016}
{Jonas}, J. \& {MeerKAT Team}. 2016, in Proceedings of MeerKAT Science: On the
  Pathway to the SKA. 25-27 May, 1

\bibitem[{{Jones} \& {Preston}(2001)}]{jones2001}
{Jones}, D.~L. \& {Preston}, R.~A. 2001, \aj, 122, 2940

\bibitem[{{Jones} \& {McAdam}(1992)}]{jones1992}
{Jones}, P.~A. \& {McAdam}, W.~B. 1992, \apjs, 80, 137

\bibitem[{{Jones} {et~al.}(1994){Jones}, {McAdam}, \& {Reynolds}}]{jones1994}
{Jones}, P.~A., {McAdam}, W.~B., \& {Reynolds}, J.~E. 1994, \mnras, 268, 602

\bibitem[{{J{\'o}zsa} {et~al.}(2009){J{\'o}zsa}, {Garrett}, {Oosterloo},
  {Rampadarath}, {Paragi}, {van Arkel}, {Lintott}, {Keel}, {Schawinski}, \&
  {Edmondson}}]{jozsa2009}
{J{\'o}zsa}, G.~I.~G., {Garrett}, M.~A., {Oosterloo}, T.~A., {et~al.} 2009,
  \aap, 500, L33

\bibitem[{{Jur{\'a}{\v{n}}ov{\'a}} {et~al.}(2019){Jur{\'a}{\v{n}}ov{\'a}},
  {Werner}, {Gaspari}, {Lakhchaura}, {Nulsen}, {Sun}, {Canning}, {Allen},
  {Simionescu}, {Oonk}, {Connor}, \& {Donahue}}]{juranova2019}
{Jur{\'a}{\v{n}}ov{\'a}}, A., {Werner}, N., {Gaspari}, M., {et~al.} 2019,
  \mnras, 484, 2886

\bibitem[{{Kaneda} {et~al.}(1995){Kaneda}, {Tashiro}, {Ikebe}, {Ishisaki},
  {Kubo}, {Makshima}, {Ohashi}, {Saito}, {Tabara}, \& {Takahashi}}]{kaneda1995}
{Kaneda}, H., {Tashiro}, M., {Ikebe}, Y., {et~al.} 1995, \apjl, 453, L13

\bibitem[{{Kardashev}(1962)}]{kardashev1962}
{Kardashev}, N.~S. 1962, \sovast, 6, 317

\bibitem[{{King} \& {Nixon}(2015)}]{king2015}
{King}, A. \& {Nixon}, C. 2015, \mnras, 453, L46

\bibitem[{{Kolokythas} {et~al.}(2015){Kolokythas}, {O'Sullivan}, {Giacintucci},
  {Raychaudhury}, {Ishwara-Chand ra}, {Worrall}, \&
  {Birkinshaw}}]{kolokythas2015}
{Kolokythas}, K., {O'Sullivan}, E., {Giacintucci}, S., {et~al.} 2015, \mnras,
  450, 1732

\bibitem[{{Komissarov} \& {Gubanov}(1994)}]{komissarov1994}
{Komissarov}, S.~S. \& {Gubanov}, A.~G. 1994, \aap, 285, 27

\bibitem[{{Krause} {et~al.}(2019){Krause}, {Shabala}, {Hardcastle}, {Bicknell},
  {B{\"o}hringer}, {Chon}, {Nawaz}, {Sarzi}, \& {Wagner}}]{krause2019}
{Krause}, M. G.~H., {Shabala}, S.~S., {Hardcastle}, M.~J., {et~al.} 2019,
  \mnras, 482, 240

\bibitem[{{Kuntschner}(2000)}]{kuntschner2000}
{Kuntschner}, H. 2000, \mnras, 315, 184

\bibitem[{{Ku{\'z}micz} {et~al.}(2017){Ku{\'z}micz}, {Jamrozy},
  {Kozie{\l}-Wierzbowska}, \& {We{\.z}gowiec}}]{kuzmicz2017}
{Ku{\'z}micz}, A., {Jamrozy}, M., {Kozie{\l}-Wierzbowska}, D., \&
  {We{\.z}gowiec}, M. 2017, \mnras, 471, 3806

\bibitem[{{Laing} {et~al.}(1999){Laing}, {Parma}, {de Ruiter}, \&
  {Fanti}}]{laing1999}
{Laing}, R.~A., {Parma}, P., {de Ruiter}, H.~R., \& {Fanti}, R. 1999, \mnras,
  306, 513

\bibitem[{{Lanz} {et~al.}(2010){Lanz}, {Jones}, {Forman}, {Ashby}, {Kraft}, \&
  {Hickox}}]{lanz2010}
{Lanz}, L., {Jones}, C., {Forman}, W.~R., {et~al.} 2010, \apj, 721, 1702

\bibitem[{{Leahy} \& {Williams}(1984)}]{leahy1984}
{Leahy}, J.~P. \& {Williams}, A.~G. 1984, \mnras, 210, 929

\bibitem[{{Lintott} {et~al.}(2009){Lintott}, {Schawinski}, {Keel}, {van Arkel},
  {Bennert}, {Edmondson}, {Thomas}, {Smith}, {Herbert}, {Jarvis}, {Virani},
  {Andreescu}, {Bamford}, {Land}, {Murray}, {Nichol}, {Raddick}, {Slosar},
  {Szalay}, \& {Vandenberg}}]{lintott2009}
{Lintott}, C.~J., {Schawinski}, K., {Keel}, W., {et~al.} 2009, \mnras, 399, 129

\bibitem[{{Maccagni} {et~al.}(2018){Maccagni}, {Morganti}, {Oosterloo}, {Oonk},
  \& {Emonts}}]{maccagni2018}
{Maccagni}, F.~M., {Morganti}, R., {Oosterloo}, T.~A., {Oonk}, J.~B.~R., \&
  {Emonts}, B.~H.~C. 2018, \aap, 614, A42

\bibitem[{{Mackie} \& {Fabbiano}(1998)}]{mackie1998}
{Mackie}, G. \& {Fabbiano}, G. 1998, \aj, 115, 514

\bibitem[{{Marconi} {et~al.}(2004){Marconi}, {Risaliti}, {Gilli}, {Hunt},
  {Maiolino}, \& {Salvati}}]{marconi2004}
{Marconi}, A., {Risaliti}, G., {Gilli}, R., {et~al.} 2004, \mnras, 351, 169

\bibitem[{{McGee} {et~al.}(1955){McGee}, {Slee}, \& {Stanley}}]{mcgee1955}
{McGee}, R.~X., {Slee}, O.~B., \& {Stanley}, G.~J. 1955, Aust. J. Phys., 8, 347

\bibitem[{{McKinley} {et~al.}(2015){McKinley}, {Yang}, {L{\'o}pez-Caniego},
  {Briggs}, {Hurley-Walker}, {Wayth}, {Offringa}, {Crocker}, {Bernardi},
  {Procopio}, {Gaensler}, {Tingay}, {Johnston-Hollitt}, {McDonald}, {Bell},
  {Bhat}, {Bowman}, {Cappallo}, {Corey}, {Deshpande}, {Emrich}, {Ewall-Wice},
  {Feng}, {Goeke}, {Greenhill}, {Hazelton}, {Hewitt}, {Hindson}, {Jacobs},
  {Kaplan}, {Kasper}, {Kratzenberg}, {Kudryavtseva}, {Lenc}, {Lonsdale},
  {Lynch}, {McWhirter}, {Mitchell}, {Morales}, {Morgan}, {Oberoi}, {Ord},
  {Pindor}, {Prabu}, {Riding}, {Rogers}, {Roshi}, {Udaya Shankar}, {Srivani},
  {Subrahmanyan}, {Waterson}, {Webster}, {Whitney}, {Williams}, \&
  {Williams}}]{mckinley2015}
{McKinley}, B., {Yang}, R., {L{\'o}pez-Caniego}, M., {et~al.} 2015, \mnras,
  446, 3478

\bibitem[{{McNamara} \& {Nulsen}(2012)}]{mcnamara2012}
{McNamara}, B.~R. \& {Nulsen}, P.~E.~J. 2012, New J. Phys., 14, 055023

\bibitem[{{Melis} {et~al.}(2018){Melis}, {Concu}, {Trois}, {Possenti},
  {Bocchinu}, {Bolli}, {Burgay}, {Carretti}, {Castangia}, {Casu}, {Pestellini},
  {Corongiu}, {D'Amico}, {Egron}, {Govoni}, {Iacolina}, {Murgia}, {Pellizzoni},
  {Perrodin}, {Pilia}, {Pisanu}, {Poddighe}, {Poppi}, {Porceddu}, {Tarchi},
  {Vacca}, {Aresu}, {Bachetti}, {Barbaro}, {Casula}, {Ladu}, {Leurini}, {Loi},
  {Loru}, {Marongiu}, {Maxia}, {Mazzarella}, {Migoni}, {Montisci}, {Valente},
  \& {Vargiu}}]{melis2018}
{Melis}, A., {Concu}, R., {Trois}, A., {et~al.} 2018, JAI, 7, 1850004

\bibitem[{{Monceau-Baroux} {et~al.}(2014){Monceau-Baroux}, {Porth}, {Meliani},
  \& {Keppens}}]{monceau-baroux2014}
{Monceau-Baroux}, R., {Porth}, O., {Meliani}, Z., \& {Keppens}, R. 2014, \aap,
  561, A30

\bibitem[{{Morganti}(2017)}]{morganti2017}
{Morganti}, R. 2017, Nat. Astron., 1, 596

\bibitem[{{Morganti} {et~al.}(1993){Morganti}, {Killeen}, \&
  {Tadhunter}}]{morganti1993}
{Morganti}, R., {Killeen}, N.~E.~B., \& {Tadhunter}, C.~N. 1993, \mnras, 263,
  1023

\bibitem[{{Morokuma-Matsui} {et~al.}(2019){Morokuma-Matsui}, {Serra},
  {Maccagni}, {For}, {Wang}, {Bekki}, {Morokuma}, {Egusa}, {Espada}, {Miura},
  {Nakanishi}, {Koribalski}, \& {Takeuchi}}]{morokuma2019}
{Morokuma-Matsui}, K., {Serra}, P., {Maccagni}, F.~M., {et~al.} 2019, \pasj,
  71, 85

\bibitem[{{Mullin} {et~al.}(2008){Mullin}, {Riley}, \&
  {Hardcastle}}]{mullin2008}
{Mullin}, L.~M., {Riley}, J.~M., \& {Hardcastle}, M.~J. 2008, \mnras, 390, 595

\bibitem[{{Murgia}(2003)}]{murgia2003}
{Murgia}, M. 2003, \pasa, 20, 19

\bibitem[{{Murgia} {et~al.}(2010{\natexlab{a}}){Murgia}, {Eckert}, {Govoni},
  {Ferrari}, {Pand ey-Pommier}, {Nevalainen}, \& {Paltani}}]{murgia2010b}
{Murgia}, M., {Eckert}, D., {Govoni}, F., {et~al.} 2010{\natexlab{a}}, \aap,
  514, A76

\bibitem[{{Murgia} {et~al.}(1999){Murgia}, {Fanti}, {Fanti}, {Gregorini},
  {Klein}, {Mack}, \& {Vigotti}}]{murgia1999}
{Murgia}, M., {Fanti}, C., {Fanti}, R., {et~al.} 1999, \aap, 345, 769

\bibitem[{{Murgia} {et~al.}(2016){Murgia}, {Govoni}, {Carretti}, {Melis},
  {Concu}, {Trois}, {Loi}, {Vacca}, {Tarchi}, {Castangia}, {Possenti},
  {Bocchinu}, {Burgay}, {Casu}, {Pellizzoni}, {Pisanu}, {Poddighe}, {Poppi},
  {D'Amico}, {Bachetti}, {Corongiu}, {Egron}, {Iacolina}, {Ladu}, {Marongiu},
  {Migoni}, {Perrodin}, {Pilia}, {Valente}, \& {Vargiu}}]{murgia2016}
{Murgia}, M., {Govoni}, F., {Carretti}, E., {et~al.} 2016, \mnras, 461, 3516

\bibitem[{{Murgia} {et~al.}(2010{\natexlab{b}}){Murgia}, {Govoni}, {Feretti},
  \& {Giovannini}}]{murgia2010a}
{Murgia}, M., {Govoni}, F., {Feretti}, L., \& {Giovannini}, G.
  2010{\natexlab{b}}, \aap, 509, A86

\bibitem[{{Murgia} {et~al.}(2012){Murgia}, {Markevitch}, {Govoni}, {Parma},
  {Fanti}, {de Ruiter}, \& {Mack}}]{murgia2012}
{Murgia}, M., {Markevitch}, M., {Govoni}, F., {et~al.} 2012, \aap, 548, A75

\bibitem[{{Murgia} {et~al.}(2011){Murgia}, {Parma}, {Mack}, {de Ruiter},
  {Fanti}, {Govoni}, {Tarchi}, {Giacintucci}, \& {Markevitch}}]{murgia2011}
{Murgia}, M., {Parma}, P., {Mack}, K.~H., {et~al.} 2011, \aap, 526, A148

\bibitem[{{Myers} \& {Spangler}(1985)}]{myers1985}
{Myers}, S.~T. \& {Spangler}, S.~R. 1985, \apj, 291, 52

\bibitem[{{Nagino} \& {Matsushita}(2009)}]{nagino2009}
{Nagino}, R. \& {Matsushita}, K. 2009, \aap, 501, 157

\bibitem[{{Noordam} \& {Smirnov}(2010)}]{noordam2010}
{Noordam}, J.~E. \& {Smirnov}, O.~M. 2010, \aap, 524, A61

\bibitem[{{Offringa} {et~al.}(2014){Offringa}, {McKinley}, {Hurley-Walker},
  {Briggs}, {Wayth}, {Kaplan}, {Bell}, {Feng}, {Neben}, {Hughes}, {Rhee},
  {Murphy}, {Bhat}, {Bernardi}, {Bowman}, {Cappallo}, {Corey}, {Deshpand e},
  {Emrich}, {Ewall-Wice}, {Gaensler}, {Goeke}, {Greenhill}, {Hazelton},
  {Hindson}, {Johnston-Hollitt}, {Jacobs}, {Kasper}, {Kratzenberg}, {Lenc},
  {Lonsdale}, {Lynch}, {McWhirter}, {Mitchell}, {Morales}, {Morgan},
  {Kudryavtseva}, {Oberoi}, {Ord}, {Pindor}, {Procopio}, {Prabu}, {Riding},
  {Roshi}, {Shankar}, {Srivani}, {Subrahmanyan}, {Tingay}, {Waterson},
  {Webster}, {Whitney}, {Williams}, \& {Williams}}]{offringa2014}
{Offringa}, A.~R., {McKinley}, B., {Hurley-Walker}, N., {et~al.} 2014, \mnras,
  444, 606

\bibitem[{{Offringa} \& {Smirnov}(2017)}]{offringa2017}
{Offringa}, A.~R. \& {Smirnov}, O. 2017, \mnras, 471, 301

\bibitem[{{Orr{\`u}} {et~al.}(2010){Orr{\`u}}, {Murgia}, {Feretti}, {Govoni},
  {Giovannini}, {Lane}, {Kassim}, \& {Paladino}}]{orru2010}
{Orr{\`u}}, E., {Murgia}, M., {Feretti}, L., {et~al.} 2010, \aap, 515, A50

\bibitem[{{Pacholczyk}(1970)}]{pacholczyk1970}
{Pacholczyk}, A.~G. 1970, {Radio astrophysics. Nonthermal processes in galactic
  and extragalactic sources}

\bibitem[{{Parma} {et~al.}(1987){Parma}, {Fanti}, {Fanti}, {Morganti}, \& {de
  Ruiter}}]{parma1987}
{Parma}, P., {Fanti}, C., {Fanti}, R., {Morganti}, R., \& {de Ruiter}, H.~R.
  1987, \aap, 181, 244

\bibitem[{{Parma} {et~al.}(2007){Parma}, {Murgia}, {de Ruiter}, {Fanti},
  {Mack}, \& {Govoni}}]{parma2007}
{Parma}, P., {Murgia}, M., {de Ruiter}, H.~R., {et~al.} 2007, \aap, 470, 875

\bibitem[{{Parma} {et~al.}(1999){Parma}, {Murgia}, {Morganti}, {Capetti}, {de
  Ruiter}, \& {Fanti}}]{parma1999}
{Parma}, P., {Murgia}, M., {Morganti}, R., {et~al.} 1999, \aap, 344, 7

\bibitem[{{Partridge} {et~al.}(2016){Partridge}, {L{\'o}pez-Caniego}, {Perley},
  {Stevens}, {Butler}, {Rocha}, {Walter}, \& {Zacchei}}]{planckScale}
{Partridge}, B., {L{\'o}pez-Caniego}, M., {Perley}, R.~A., {et~al.} 2016, \apj,
  821, 61

\bibitem[{{Perley} \& {Butler}(2013{\natexlab{a}})}]{perley2013a}
{Perley}, R.~A. \& {Butler}, B.~J. 2013{\natexlab{a}}, \apjs, 204, 19

\bibitem[{{Perley} \& {Butler}(2013{\natexlab{b}})}]{perley2013b}
{Perley}, R.~A. \& {Butler}, B.~J. 2013{\natexlab{b}}, \apjs, 206, 16

\bibitem[{{Perley} \& {Butler}(2017)}]{perley2017}
{Perley}, R.~A. \& {Butler}, B.~J. 2017, \apjs, 230, 7

\bibitem[{{Petry} \& {CASA Development Team}(2012)}]{petry2012}
{Petry}, D. \& {CASA Development Team}. 2012, in Astronomical Society of the
  Pacific Conference Series, Vol. 461, Astronomical Data Analysis Software and
  Systems XXI, ed. P.~{Ballester}, D.~{Egret}, \& N.~P.~F. {Lorente}, 849

\bibitem[{{Piddington} \& {Trent}(1956)}]{piddington1956}
{Piddington}, J.~H. \& {Trent}, G.~H. 1956, Aust. J. Phys., 9, 481

\bibitem[{{Planck Collaboration} {et~al.}(2014){Planck Collaboration},
  {Abergel}, {Ade}, {Aghanim}, {Alves}, {Aniano}, {Armitage-Caplan}, {Arnaud},
  {Ashdown}, {Atrio-Barand ela}, {Aumont}, {Baccigalupi}, {Banday}, {Barreiro},
  {Bartlett}, {Battaner}, {Benabed}, {Beno{\^\i}t}, {Benoit-L{\'e}vy},
  {Bernard}, {Bersanelli}, {Bielewicz}, {Bobin}, {Bock}, {Bonaldi}, {Bond},
  {Borrill}, {Bouchet}, {Boulanger}, {Bridges}, {Bucher}, {Burigana}, {Butler},
  {Cardoso}, {Catalano}, {Chamballu}, {Chary}, {Chiang}, {Chiang},
  {Christensen}, {Church}, {Clemens}, {Clements}, {Colombi}, {Colombo},
  {Combet}, {Couchot}, {Coulais}, {Crill}, {Curto}, {Cuttaia}, {Danese},
  {Davies}, {Davis}, {de Bernardis}, {de Rosa}, {de Zotti}, {Delabrouille},
  {Delouis}, {D{\'e}sert}, {Dickinson}, {Diego}, {Dole}, {Donzelli},
  {Dor{\'e}}, {Douspis}, {Draine}, {Dupac}, {Efstathiou}, {En{\ss}lin},
  {Eriksen}, {Falgarone}, {Finelli}, {Forni}, {Frailis}, {Fraisse},
  {Franceschi}, {Galeotta}, {Ganga}, {Ghosh}, {Giard}, {Giardino},
  {Giraud-H{\'e}raud}, {Gonz{\'a}lez-Nuevo}, {G{\'o}rski}, {Gratton},
  {Gregorio}, {Grenier}, {Gruppuso}, {Guillet}, {Hansen}, {Hanson}, {Harrison},
  {Helou}, {Henrot-Versill{\'e}}, {Hern{\'a}ndez-Monteagudo}, {Herranz},
  {Hildebrand t}, {Hivon}, {Hobson}, {Holmes}, {Hornstrup}, {Hovest},
  {Huffenberger}, {Jaffe}, {Jaffe}, {Jewell}, {Joncas}, {Jones}, {Juvela},
  {Keih{\"a}nen}, {Keskitalo}, {Kisner}, {Knoche}, {Knox}, {Kunz},
  {Kurki-Suonio}, {Lagache}, {L{\"a}hteenm{\"a}ki}, {Lamarre}, {Lasenby},
  {Laureijs}, {Lawrence}, {Leonardi}, {Le{\'o}n-Tavares}, {Lesgourgues},
  {Levrier}, {Liguori}, {Lilje}, {Linden-V{\o}rnle}, {L{\'o}pez-Caniego},
  {Lubin}, {Mac{\'\i}as-P{\'e}rez}, {Maffei}, {Maino}, {Mand olesi}, {Maris},
  {Marshall}, {Martin}, {Mart{\'\i}nez-Gonz{\'a}lez}, {Masi}, {Massardi},
  {Matarrese}, {Matthai}, {Mazzotta}, {McGehee}, {Melchiorri}, {Mendes},
  {Mennella}, {Migliaccio}, {Mitra}, {Miville-Desch{\^e}nes}, {Moneti},
  {Montier}, {Morgante}, {Mortlock}, {Munshi}, {Murphy}, {Naselsky}, {Nati},
  {Natoli}, {Netterfield}, {N{\o}rgaard-Nielsen}, {Noviello}, {Novikov},
  {Novikov}, {Osborne}, {Oxborrow}, {Paci}, {Pagano}, {Pajot}, {Paladini},
  {Paoletti}, {Pasian}, {Patanchon}, {Perdereau}, {Perotto}, {Perrotta},
  {Piacentini}, {Piat}, {Pierpaoli}, {Pietrobon}, {Plaszczynski},
  {Pointecouteau}, {Polenta}, {Ponthieu}, {Popa}, {Poutanen}, {Pratt},
  {Pr{\'e}zeau}, {Prunet}, {Puget}, {Rachen}, {Reach}, {Rebolo}, {Reinecke},
  {Remazeilles}, {Renault}, {Ricciardi}, {Riller}, {Ristorcelli}, {Rocha},
  {Rosset}, {Roudier}, {Rowan-Robinson}, {Rubi{\~n}o-Mart{\'\i}n}, {Rusholme},
  {Sandri}, {Santos}, {Savini}, {Scott}, {Seiffert}, {Shellard}, {Spencer},
  {Starck}, {Stolyarov}, {Stompor}, {Sudiwala}, {Sunyaev}, {Sureau}, {Sutton},
  {Suur-Uski}, {Sygnet}, {Tauber}, {Tavagnacco}, {Terenzi}, {Toffolatti},
  {Tomasi}, {Tristram}, {Tucci}, {Tuovinen}, {T{\"u}rler}, {Umana},
  {Valenziano}, {Valiviita}, {Van Tent}, {Verstraete}, {Vielva}, {Villa},
  {Vittorio}, {Wade}, {Wandelt}, {Welikala}, {Ysard}, {Yvon}, {Zacchei}, \&
  {Zonca}}]{planck2013XI}
{Planck Collaboration}, {Abergel}, A., {Ade}, P.~A.~R., {et~al.} 2014, \aap,
  571, A11

\bibitem[{{Planck Collaboration} {et~al.}(2016{\natexlab{a}}){Planck
  Collaboration}, {Adam}, {Ade}, {Aghanim}, {Akrami}, {Alves}, {Arg{\"u}eso},
  {Arnaud}, {Arroja}, {Ashdown}, {Aumont}, {Baccigalupi}, {Ballardini}, {Band
  ay}, {Barreiro}, {Bartlett}, {Bartolo}, {Basak}, {Battaglia}, {Battaner},
  {Battye}, {Benabed}, {Beno{\^\i}t}, {Benoit-L{\'e}vy}, {Bernard},
  {Bersanelli}, {Bertincourt}, {Bielewicz}, {Bikmaev}, {Bock}, {B{\"o}hringer},
  {Bonaldi}, {Bonavera}, {Bond}, {Borrill}, {Bouchet}, {Boulanger}, {Bucher},
  {Burenin}, {Burigana}, {Butler}, {Calabrese}, {Cardoso}, {Carvalho},
  {Casaponsa}, {Castex}, {Catalano}, {Challinor}, {Chamballu}, {Chary},
  {Chiang}, {Chluba}, {Chon}, {Christensen}, {Church}, {Clemens}, {Clements},
  {Colombi}, {Colombo}, {Combet}, {Comis}, {Contreras}, {Couchot}, {Coulais},
  {Crill}, {Cruz}, {Curto}, {Cuttaia}, {Danese}, {Davies}, {Davis}, {de
  Bernardis}, {de Rosa}, {de Zotti}, {Delabrouille}, {Delouis}, {D{\'e}sert},
  {Di Valentino}, {Dickinson}, {Diego}, {Dolag}, {Dole}, {Donzelli},
  {Dor{\'e}}, {Douspis}, {Ducout}, {Dunkley}, {Dupac}, {Efstathiou},
  {Eisenhardt}, {Elsner}, {En{\ss}lin}, {Eriksen}, {Falgarone}, {Fantaye},
  {Farhang}, {Feeney}, {Fergusson}, {Fernandez-Cobos}, {Feroz}, {Finelli},
  {Florido}, {Forni}, {Frailis}, {Fraisse}, {Franceschet}, {Franceschi},
  {Frejsel}, {Frolov}, {Galeotta}, {Galli}, {Ganga}, {Gauthier},
  {G{\'e}nova-Santos}, {Gerbino}, {Ghosh}, {Giard}, {Giraud-H{\'e}raud},
  {Giusarma}, {Gjerl{\o}w}, {Gonz{\'a}lez-Nuevo}, {G{\'o}rski}, {Grainge},
  {Gratton}, {Gregorio}, {Gruppuso}, {Gudmundsson}, {Hamann}, {Handley},
  {Hansen}, {Hanson}, {Harrison}, {Heavens}, {Helou}, {Henrot-Versill{\'e}},
  {Hern{\'a}ndez-Monteagudo}, {Herranz}, {Hildebrandt}, {Hivon}, {Hobson},
  {Holmes}, {Hornstrup}, {Hovest}, {Huang}, {Huffenberger}, {Hurier},
  {Ili{\'c}}, {Jaffe}, {Jaffe}, {Jin}, {Jones}, {Juvela}, {Karakci},
  {Keih{\"a}nen}, {Keskitalo}, {Khamitov}, {Kiiveri}, {Kim}, {Kisner},
  {Kneissl}, {Knoche}, {Knox}, {Krachmalnicoff}, {Kunz}, {Kurki-Suonio},
  {Lacasa}, {Lagache}, {L{\"a}hteenm{\"a}ki}, {Lamarre}, {Langer}, {Lasenby},
  {Lattanzi}, {Lawrence}, {Le Jeune}, {Leahy}, {Lellouch}, {Leonardi},
  {Le{\'o}n-Tavares}, {Lesgourgues}, {Levrier}, {Lewis}, {Liguori}, {Lilje},
  {Lilley}, {Linden-V{\o}rnle}, {Lindholm}, {Liu}, {L{\'o}pez-Caniego},
  {Lubin}, {Ma}, {Mac{\'\i}as-P{\'e}rez}, {Maggio}, {Maino}, {Mak},
  {Mandolesi}, {Mangilli}, {Marchini}, {Marcos-Caballero}, {Marinucci},
  {Maris}, {Marshall}, {Martin}, {Martinelli}, {Mart{\'\i}nez-Gonz{\'a}lez},
  {Masi}, {Matarrese}, {Mazzotta}, {McEwen}, {McGehee}, {Mei}, {Meinhold},
  {Melchiorri}, {Melin}, {Mendes}, {Mennella}, {Migliaccio}, {Mikkelsen},
  {Millea}, {Mitra}, {Miville-Desch{\^e}nes}, {Molinari}, {Moneti}, {Montier},
  {Moreno}, {Morgante}, {Mortlock}, {Moss}, {Mottet}, {M{\"u}nchmeyer},
  {Munshi}, {Murphy}, {Narimani}, {Naselsky}, {Nastasi}, {Nati}, {Natoli},
  {Negrello}, {Netterfield}, {N{\o}rgaard-Nielsen}, {Noviello}, {Novikov},
  {Novikov}, {Olamaie}, {Oppermann}, {Orlando}, {Oxborrow}, {Paci}, {Pagano},
  {Pajot}, {Paladini}, {Pandolfi}, {Paoletti}, {Partridge}, {Pasian},
  {Patanchon}, {Pearson}, {Peel}, {Peiris}, {Pelkonen}, {Perdereau}, {Perotto},
  {Perrott}, {Perrotta}, {Pettorino}, {Piacentini}, {Piat}, {Pierpaoli},
  {Pietrobon}, {Plaszczynski}, {Pogosyan}, {Pointecouteau}, {Polenta}, {Popa},
  {Pratt}, {Pr{\'e}zeau}, {Prunet}, {Puget}, {Rachen}, {Racine}, {Reach},
  {Rebolo}, {Reinecke}, {Remazeilles}, {Renault}, {Renzi}, {Ristorcelli},
  {Rocha}, {Roman}, {Romelli}, {Rosset}, {Rossetti}, {Rotti}, {Roudier},
  {Rouill{\'e} d'Orfeuil}, {Rowan-Robinson}, {Rubi{\~n}o-Mart{\'\i}n},
  {Ruiz-Granados}, {Rumsey}, {Rusholme}, {Said}, {Salvatelli}, {Salvati},
  {Sandri}, {Sanghera}, {Santos}, {Saunders}, {Sauv{\'e}}, {Savelainen},
  {Savini}, {Schaefer}, {Schammel}, {Scott}, {Seiffert}, {Serra}, {Shellard},
  {Shimwell}, {Shiraishi}, {Smith}, {Souradeep}, {Spencer}, {Spinelli},
  {Stanford}, {Stern}, {Stolyarov}, {Stompor}, {Strong}, {Sudiwala}, {Sunyaev},
  {Sutter}, {Sutton}, {Suur-Uski}, {Sygnet}, {Tauber}, {Tavagnacco}, {Terenzi},
  {Texier}, {Toffolatti}, {Tomasi}, {Tornikoski}, {Tramonte}, {Tristram},
  {Troja}, {Trombetti}, {Tucci}, {Tuovinen}, {T{\"u}rler}, {Umana},
  {Valenziano}, {Valiviita}, {Van Tent}, {Vassallo}, {Vibert}, {Vidal}, {Viel},
  {Vielva}, {Villa}, {Wade}, {Walter}, {Wand elt}, {Watson}, {Wehus},
  {Welikala}, {Weller}, {White}, {White}, {Wilkinson}, {Yvon}, {Zacchei},
  {Zibin}, \& {Zonca}}]{planck2016I}
{Planck Collaboration}, {Adam}, R., {Ade}, P.~A.~R., {et~al.}
  2016{\natexlab{a}}, \aap, 594, A1

\bibitem[{{Planck Collaboration} {et~al.}(2016{\natexlab{b}}){Planck
  Collaboration}, {Adam}, {Ade}, {Aghanim}, {Alves}, {Arnaud}, {Ashdown},
  {Aumont}, {Baccigalupi}, {Banday}, {Barreiro}, {Bartlett}, {Bartolo},
  {Battaner}, {Benabed}, {Beno{\^\i}t}, {Benoit-L{\'e}vy}, {Bernard},
  {Bersanelli}, {Bielewicz}, {Bock}, {Bonaldi}, {Bonavera}, {Bond}, {Borrill},
  {Bouchet}, {Boulanger}, {Bucher}, {Burigana}, {Butler}, {Calabrese},
  {Cardoso}, {Catalano}, {Challinor}, {Chamballu}, {Chary}, {Chiang},
  {Christensen}, {Clements}, {Colombi}, {Colombo}, {Combet}, {Couchot},
  {Coulais}, {Crill}, {Curto}, {Cuttaia}, {Danese}, {Davies}, {Davis}, {de
  Bernardis}, {de Rosa}, {de Zotti}, {Delabrouille}, {D{\'e}sert}, {Dickinson},
  {Diego}, {Dole}, {Donzelli}, {Dor{\'e}}, {Douspis}, {Ducout}, {Dupac},
  {Efstathiou}, {Elsner}, {En{\ss}lin}, {Eriksen}, {Falgarone}, {Fergusson},
  {Finelli}, {Forni}, {Frailis}, {Fraisse}, {Franceschi}, {Frejsel},
  {Galeotta}, {Galli}, {Ganga}, {Ghosh}, {Giard}, {Giraud-H{\'e}raud},
  {Gjerl{\o}w}, {Gonz{\'a}lez-Nuevo}, {G{\'o}rski}, {Gratton}, {Gregorio},
  {Gruppuso}, {Gudmundsson}, {Hansen}, {Hanson}, {Harrison}, {Helou},
  {Henrot-Versill{\'e}}, {Hern{\'a}ndez-Monteagudo}, {Herranz}, {Hildebrandt},
  {Hivon}, {Hobson}, {Holmes}, {Hornstrup}, {Hovest}, {Huffenberger}, {Hurier},
  {Jaffe}, {Jaffe}, {Jones}, {Juvela}, {Keih{\"a}nen}, {Keskitalo}, {Kisner},
  {Kneissl}, {Knoche}, {Kunz}, {Kurki-Suonio}, {Lagache},
  {L{\"a}hteenm{\"a}ki}, {Lamarre}, {Lasenby}, {Lattanzi}, {Lawrence}, {Le
  Jeune}, {Leahy}, {Leonardi}, {Lesgourgues}, {Levrier}, {Liguori}, {Lilje},
  {Linden-V{\o}rnle}, {L{\'o}pez-Caniego}, {Lubin}, {Mac{\'\i}as-P{\'e}rez},
  {Maggio}, {Maino}, {Mandolesi}, {Mangilli}, {Maris}, {Marshall}, {Martin},
  {Mart{\'\i}nez-Gonz{\'a}lez}, {Masi}, {Matarrese}, {McGehee}, {Meinhold},
  {Melchiorri}, {Mendes}, {Mennella}, {Migliaccio}, {Mitra},
  {Miville-Desch{\^e}nes}, {Moneti}, {Montier}, {Morgante}, {Mortlock}, {Moss},
  {Munshi}, {Murphy}, {Naselsky}, {Nati}, {Natoli}, {Netterfield},
  {N{\o}rgaard-Nielsen}, {Noviello}, {Novikov}, {Novikov}, {Orlando},
  {Oxborrow}, {Paci}, {Pagano}, {Pajot}, {Paladini}, {Paoletti}, {Partridge},
  {Pasian}, {Patanchon}, {Pearson}, {Perdereau}, {Perotto}, {Perrotta},
  {Pettorino}, {Piacentini}, {Piat}, {Pierpaoli}, {Pietrobon}, {Plaszczynski},
  {Pointecouteau}, {Polenta}, {Pratt}, {Pr{\'e}zeau}, {Prunet}, {Puget},
  {Rachen}, {Reach}, {Rebolo}, {Reinecke}, {Remazeilles}, {Renault}, {Renzi},
  {Ristorcelli}, {Rocha}, {Rosset}, {Rossetti}, {Roudier},
  {Rubi{\~n}o-Mart{\'\i}n}, {Rusholme}, {Sandri}, {Santos}, {Savelainen},
  {Savini}, {Scott}, {Seiffert}, {Shellard}, {Spencer}, {Stolyarov}, {Stompor},
  {Strong}, {Sudiwala}, {Sunyaev}, {Sutton}, {Suur-Uski}, {Sygnet}, {Tauber},
  {Terenzi}, {Toffolatti}, {Tomasi}, {Tristram}, {Tucci}, {Tuovinen}, {Umana},
  {Valenziano}, {Valiviita}, {Van Tent}, {Vielva}, {Villa}, {Wade}, {Wandelt},
  {Wehus}, {Wilkinson}, {Yvon}, {Zacchei}, \& {Zonca}}]{planck2016X}
{Planck Collaboration}, {Adam}, R., {Ade}, P.~A.~R., {et~al.}
  2016{\natexlab{b}}, \aap, 594, A10

\bibitem[{{Planck Collaboration} {et~al.}(2016{\natexlab{c}}){Planck
  Collaboration}, {Aghanim}, {Ashdown}, {Aumont}, {Baccigalupi}, {Ballardini},
  {Band ay}, {Barreiro}, {Bartolo}, {Basak}, {Benabed}, {Bernard},
  {Bersanelli}, {Bielewicz}, {Bonavera}, {Bond}, {Borrill}, {Bouchet},
  {Boulanger}, {Burigana}, {Calabrese}, {Cardoso}, {Carron}, {Chiang},
  {Colombo}, {Comis}, {Couchot}, {Coulais}, {Crill}, {Curto}, {Cuttaia}, {de
  Bernardis}, {de Zotti}, {Delabrouille}, {Di Valentino}, {Dickinson}, {Diego},
  {Dor{\'e}}, {Douspis}, {Ducout}, {Dupac}, {Dusini}, {Elsner}, {En{\ss}lin},
  {Eriksen}, {Falgarone}, {Fantaye}, {Finelli}, {Forastieri}, {Frailis},
  {Fraisse}, {Franceschi}, {Frolov}, {Galeotta}, {Galli}, {Ganga},
  {G{\'e}nova-Santos}, {Gerbino}, {Ghosh}, {Giraud-H{\'e}raud},
  {Gonz{\'a}lez-Nuevo}, {G{\'o}rski}, {Gruppuso}, {Gudmundsson}, {Hansen},
  {Helou}, {Henrot-Versill{\'e}}, {Herranz}, {Hivon}, {Huang}, {Jaffe},
  {Jones}, {Keih{\"a}nen}, {Keskitalo}, {Kiiveri}, {Kisner}, {Krachmalnicoff},
  {Kunz}, {Kurki-Suonio}, {Lamarre}, {Langer}, {Lasenby}, {Lattanzi},
  {Lawrence}, {Le Jeune}, {Levrier}, {Lilje}, {Lilley}, {Lindholm},
  {L{\'o}pez-Caniego}, {Ma}, {Mac{\'\i}as-P{\'e}rez}, {Maggio}, {Maino}, {Mand
  olesi}, {Mangilli}, {Maris}, {Martin}, {Mart{\'\i}nez-Gonz{\'a}lez},
  {Matarrese}, {Mauri}, {McEwen}, {Melchiorri}, {Mennella}, {Migliaccio},
  {Miville-Desch{\^e}nes}, {Molinari}, {Moneti}, {Montier}, {Morgante}, {Moss},
  {Natoli}, {Oxborrow}, {Pagano}, {Paoletti}, {Patanchon}, {Perdereau},
  {Perotto}, {Pettorino}, {Piacentini}, {Plaszczynski}, {Polastri}, {Polenta},
  {Puget}, {Rachen}, {Racine}, {Reinecke}, {Remazeilles}, {Renzi}, {Rocha},
  {Rosset}, {Rossetti}, {Roudier}, {Rubi{\~n}o-Mart{\'\i}n}, {Ruiz-Granados},
  {Salvati}, {Sandri}, {Savelainen}, {Scott}, {Sirignano}, {Sirri}, {Soler},
  {Spencer}, {Suur-Uski}, {Tauber}, {Tavagnacco}, {Tenti}, {Toffolatti},
  {Tomasi}, {Tristram}, {Trombetti}, {Valiviita}, {Van Tent}, {Vielva},
  {Villa}, {Vittorio}, {Wandelt}, {Wehus}, {Zacchei}, \&
  {Zonca}}]{planck2016XLVIII}
{Planck Collaboration}, {Aghanim}, N., {Ashdown}, M., {et~al.}
  2016{\natexlab{c}}, \aap, 596, A109

\bibitem[{{Planck Collaboration} {et~al.}(2018){Planck Collaboration},
  {Akrami}, {Ashdown}, {Aumont}, {Baccigalupi}, {Ballardini}, {Band ay},
  {Barreiro}, {Bartolo}, {Basak}, {Benabed}, {Bersanelli}, {Bielewicz}, {Bond},
  {Borrill}, {Bouchet}, {Boulanger}, {Bucher}, {Burigana}, {Calabrese},
  {Cardoso}, {Carron}, {Casaponsa}, {Challinor}, {Colombo}, {Combet}, {Crill},
  {Cuttaia}, {de Bernardis}, {de Rosa}, {de Zotti}, {Delabrouille}, {Delouis},
  {Di Valentino}, {Dickinson}, {Diego}, {Donzelli}, {Dor{\'e}}, {Ducout},
  {Dupac}, {Efstathiou}, {Elsner}, {En{\ss}lin}, {Eriksen}, {Falgarone},
  {Fernandez-Cobos}, {Finelli}, {Forastieri}, {Frailis}, {Fraisse},
  {Franceschi}, {Frolov}, {Galeotta}, {Galli}, {Ganga}, {G{\'e}nova-Santos},
  {Gerbino}, {Ghosh}, {Gonz{\'a}lez-Nuevo}, {G{\'o}rski}, {Gratton},
  {Gruppuso}, {Gudmundsson}, {Hand ley}, {Hansen}, {Helou}, {Herranz}, {Huang},
  {Jaffe}, {Karakci}, {Keih{\"a}nen}, {Keskitalo}, {Kiiveri}, {Kim}, {Kisner},
  {Krachmalnicoff}, {Kunz}, {Kurki-Suonio}, {Lagache}, {Lamarre}, {Lasenby},
  {Lattanzi}, {Lawrence}, {Le Jeune}, {Levrier}, {Liguori}, {Lilje},
  {Lindholm}, {L{\'o}pez-Caniego}, {Lubin}, {Ma}, {Mac{\'\i}as-P{\'e}rez},
  {Maggio}, {Maino}, {Mandolesi}, {Mangilli}, {Marcos-Caballero}, {Martin},
  {Mart{\'\i}nez-Gonz{\'a}lez}, {Matarrese}, {Mauri}, {McEwen}, {Meinhold},
  {Melchiorri}, {Mennella}, {Migliaccio}, {Miville-Desch{\^e}nes}, {Molinari},
  {Moneti}, {Montier}, {Morgante}, {Natoli}, {Oppizzi}, {Pagano}, {Paoletti},
  {Partridge}, {Peel}, {Pettorino}, {Piacentini}, {Polenta}, {Puget}, {Rachen},
  {Reinecke}, {Remazeilles}, {Renzi}, {Rocha}, {Roudier},
  {Rubi{\~n}o-Mart{\'\i}n}, {Ruiz-Granados}, {Salvati}, {Sandri}, {Savelainen},
  {Scott}, {Seljebotn}, {Sirignano}, {Spencer}, {Suur-Uski}, {Tauber},
  {Tavagnacco}, {Tenti}, {Thommesen}, {Toffolatti}, {Tomasi}, {Trombetti},
  {Valiviita}, {Van Tent}, {Vielva}, {Villa}, {Vittorio}, {Wandelt}, {Wehus},
  {Zacchei}, \& {Zonca}}]{planck2018IV}
{Planck Collaboration}, {Akrami}, Y., {Ashdown}, M., {et~al.} 2018, arXiv
  e-prints, arXiv:1807.06208

\bibitem[{{Ramos Almeida} {et~al.}(2012){Ramos Almeida}, {Bessiere},
  {Tadhunter}, {P{\'e}rez-Gonz{\'a}lez}, {Barro}, {Inskip}, {Morganti}, {Holt},
  \& {Dicken}}]{ramosalmeida2012}
{Ramos Almeida}, C., {Bessiere}, P.~S., {Tadhunter}, C.~N., {et~al.} 2012,
  \mnras, 419, 687

\bibitem[{{Roussel} {et~al.}(2007){Roussel}, {Helou}, {Hollenbach}, {Draine},
  {Smith}, {Armus}, {Schinnerer}, {Walter}, {Engelbracht}, {Thornley},
  {Kennicutt}, {Calzetti}, {Dale}, {Murphy}, \& {Bot}}]{roussel2007}
{Roussel}, H., {Helou}, G., {Hollenbach}, D.~J., {et~al.} 2007, \apj, 669, 959

\bibitem[{{Sabater} {et~al.}(2013){Sabater}, {Best}, \&
  {Argudo-Fern{\'a}ndez}}]{sabater2013}
{Sabater}, J., {Best}, P.~N., \& {Argudo-Fern{\'a}ndez}, M. 2013, \mnras, 430,
  638

\bibitem[{{Saikia} \& {Jamrozy}(2009)}]{saikia2009}
{Saikia}, D.~J. \& {Jamrozy}, M. 2009, Bull. Astron. Soc. India, 37, 63

\bibitem[{{Schawinski} {et~al.}(2015){Schawinski}, {Koss}, {Berney}, \&
  {Sartori}}]{schawinski2015}
{Schawinski}, K., {Koss}, M., {Berney}, S., \& {Sartori}, L.~F. 2015, \mnras,
  451, 2517

\bibitem[{{Schweizer}(1980)}]{schweizer1980}
{Schweizer}, F. 1980, \apj, 237, 303

\bibitem[{{Serra} {et~al.}(2016){Serra}, {de Blok}, {Bryan}, {Colafrancesco},
  {Dettmar}, {Frank}, {Govoni}, {Jozsa}, {Kraan-Korteweg}, {Maccagni},
  {Loubser}, {Murgia}, {Oosterloo}, {Peletier}, {Pizzo}, {Richter},
  {Ramatsoku}, {Smith}, {Trager}, {van Gorkom}, \& {Verheijen}}]{serra2016}
{Serra}, P., {de Blok}, W.~J.~G., {Bryan}, G.~L., {et~al.} 2016, in Proceedings
  of MeerKAT Science: On the Pathway to the SKA. 25-27 May, 8

\bibitem[{{Serra} {et~al.}(2019){Serra}, {Maccagni}, {Kleiner}, {de Blok}, {van
  Gorkom}, {Hugo}, {Iodice}, {J{\'o}zsa}, {Kamphuis}, {Kraan-Korteweg}, {Loni},
  {Makhathini}, {Moln{\'a}r}, {Oosterloo}, {Peletier}, {Ramaila}, {Ramatsoku},
  {Smirnov}, {Smith}, {Spavone}, {Thorat}, {Trager}, \& {Venhola}}]{serra2019}
{Serra}, P., {Maccagni}, F.~M., {Kleiner}, D., {et~al.} 2019, \aap, 628, A122

\bibitem[{{Serra} {et~al.}(2015){Serra}, {Westmeier}, {Giese}, {Jurek},
  {Fl{\"o}er}, {Popping}, {Winkel}, {van der Hulst}, {Meyer}, {Koribalski},
  {Staveley-Smith}, \& {Courtois}}]{serra2015}
{Serra}, P., {Westmeier}, T., {Giese}, N., {et~al.} 2015, \mnras, 448, 1922

\bibitem[{{Sesto} {et~al.}(2016){Sesto}, {Faifer}, \& {Forte}}]{sesto2016}
{Sesto}, L.~A., {Faifer}, F.~R., \& {Forte}, J.~C. 2016, \mnras, 461, 4260

\bibitem[{{Sesto} {et~al.}(2018){Sesto}, {Faifer}, {Smith Castelli}, {Forte},
  \& {Escudero}}]{sesto2018}
{Sesto}, L.~A., {Faifer}, F.~R., {Smith Castelli}, A.~V., {Forte}, J.~C., \&
  {Escudero}, C.~G. 2018, \mnras, 479, 478

\bibitem[{{Seta} {et~al.}(2013){Seta}, {Tashiro}, \& {Inoue}}]{seta2013}
{Seta}, H., {Tashiro}, M.~S., \& {Inoue}, S. 2013, \pasj, 65, 106

\bibitem[{{Shabala} {et~al.}(2017){Shabala}, {Deller}, {Kaviraj}, {Middelberg},
  {Turner}, {Ting}, {Allison}, \& {Davis}}]{shabala2016}
{Shabala}, S.~S., {Deller}, A., {Kaviraj}, S., {et~al.} 2017, \mnras, 464, 4706

\bibitem[{{Shain}(1958)}]{shain1958}
{Shain}, C.~A. 1958, Aust. J. Phys., 11, 517

\bibitem[{{Shain} \& {Higgins}(1954)}]{shain1954}
{Shain}, C.~A. \& {Higgins}, C.~S. 1954, Aust. J. of Phys., 7, 130

\bibitem[{{Shimmins}(1971)}]{shimmins1971}
{Shimmins}, A.~J. 1971, Aust. J. Phys. Astrophys. Suppl., 21, 1

\bibitem[{{Shulevski} {et~al.}(2012){Shulevski}, {Morganti}, {Oosterloo}, \&
  {Struve}}]{shulevski2012}
{Shulevski}, A., {Morganti}, R., {Oosterloo}, T., \& {Struve}, C. 2012, \aap,
  545, A91

\bibitem[{{Silva} {et~al.}(2008){Silva}, {Kuntschner}, \&
  {Lyubenova}}]{silva2008}
{Silva}, D.~R., {Kuntschner}, H., \& {Lyubenova}, M. 2008, \apj, 674, 194

\bibitem[{{Slee} {et~al.}(2001){Slee}, {Roy}, {Murgia}, {Andernach}, \&
  {Ehle}}]{slee2001}
{Slee}, O.~B., {Roy}, A.~L., {Murgia}, M., {Andernach}, H., \& {Ehle}, M. 2001,
  \aj, 122, 1172

\bibitem[{{Slee} {et~al.}(1994){Slee}, {Sadler}, {Reynolds}, \&
  {Ekers}}]{slee1994}
{Slee}, O.~B., {Sadler}, E.~M., {Reynolds}, J.~E., \& {Ekers}, R.~D. 1994,
  \mnras, 269, 928

\bibitem[{{Stanghellini} {et~al.}(2005){Stanghellini}, {O'Dea}, {Dallacasa},
  {Cassaro}, {Baum}, {Fanti}, \& {Fanti}}]{stanghellini2005}
{Stanghellini}, C., {O'Dea}, C.~P., {Dallacasa}, D., {et~al.} 2005, \aap, 443,
  891

\bibitem[{{Storchi-Bergmann} \&
  {Schnorr-M{\"u}ller}(2019)}]{storchi-bergmann2019}
{Storchi-Bergmann}, T. \& {Schnorr-M{\"u}ller}, A. 2019, Nat. Astron., 3, 48

\bibitem[{{Tadhunter} {et~al.}(1993){Tadhunter}, {Morganti}, {di Serego
  Alighieri}, {Fosbury}, \& {Danziger}}]{tadhunter1993}
{Tadhunter}, C.~N., {Morganti}, R., {di Serego Alighieri}, S., {Fosbury},
  R.~A.~E., \& {Danziger}, I.~J. 1993, \mnras, 263, 999

\bibitem[{{Tashiro} {et~al.}(2009){Tashiro}, {Isobe}, {Seta}, {Matsuta}, \&
  {Yaji}}]{tashiro2009}
{Tashiro}, M.~S., {Isobe}, N., {Seta}, H., {Matsuta}, K., \& {Yaji}, Y. 2009,
  \pasj, 61, S327

\bibitem[{{Tremblay} {et~al.}(2016){Tremblay}, {Oonk}, {Combes}, {Salom{\'e}},
  {O'Dea}, {Baum}, {Voit}, {Donahue}, {McNamara}, {Davis}, {McDonald}, {Edge},
  {Clarke}, {Galv{\'a}n-Madrid}, {Bremer}, {Edwards}, {Fabian}, {Hamer}, {Li},
  {Maury}, {Russell}, {Quillen}, {Urry}, {Sanders}, \& {Wise}}]{tremblay2016}
{Tremblay}, G.~R., {Oonk}, J.~B.~R., {Combes}, F., {et~al.} 2016, \nat, 534,
  218

\bibitem[{{Woltjer}(1959)}]{woltjer1959}
{Woltjer}, L. 1959, \apj, 130, 38

\end{thebibliography}



\begin{appendix} 

\section{Continuum images of \forn\ from archival data}
\label{appendix:contIms}
\subsection{MWA: 84 - 230 MHz}
\label{sec:dRedMWA}

\begin{figure*}[tbh]
	\begin{center}
		\includegraphics[trim = 0 0 0 0, width=\textwidth]{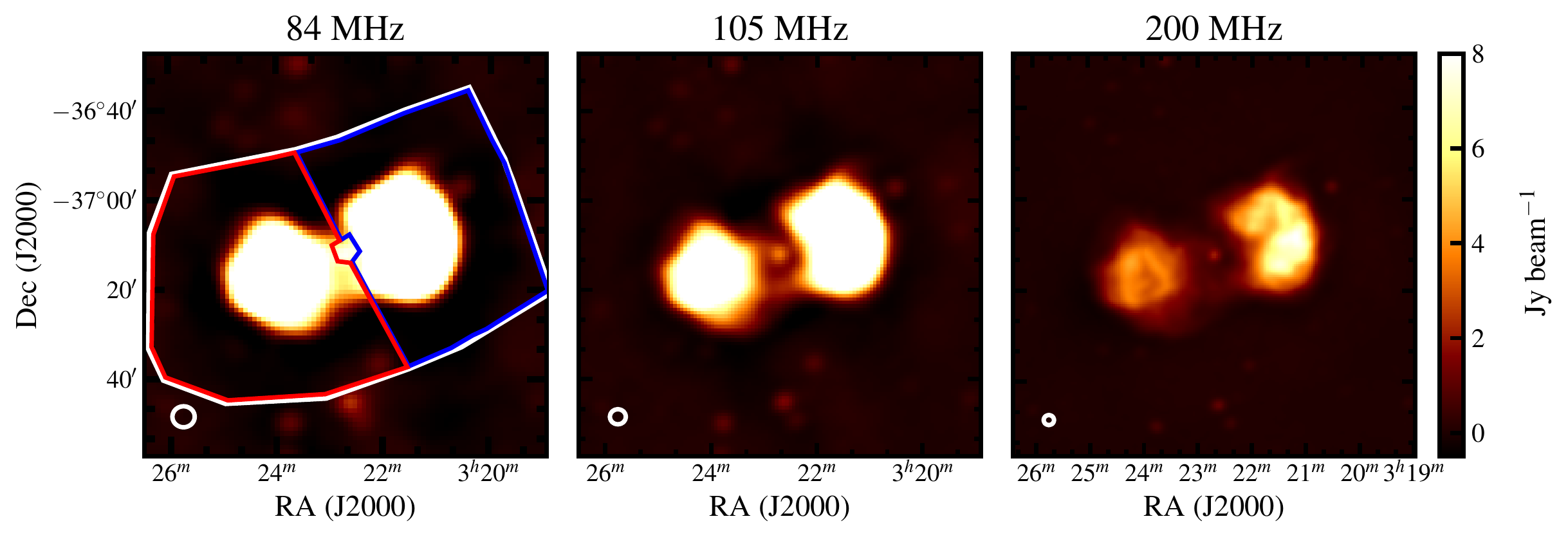}
		\caption{\forn\ seen by the MWA between 70 MHz and 230 MHz. The colour-scale is the same in all panels. In the {\em left panel}, the white contours mark the region where we measure the total flux of the source, the red and blue contours show the regions of the East and West lobes. The PSF of the images is shown in white in the bottom left corner.}
		\label{fig:fornMWA} 
	\end{center}
\end{figure*}

To characterise the radio spectrum of \forn\ at low frequencies ($84$ -- $200$ MHz) we select MWA archival observations from the GLEAM survey. The images are shown in Fig.~\ref{fig:fornMWA}. As reported by \citet{mckinley2015}, self-calibration is known to affect the flux-density scale of MWA images of \forn. For this reason in their work on \forn, the authors calculated a flux-scaling factor of $1.325$. We apply the same scaling factor to our flux measurements obtaining compatible results (see Table~\ref{tab:sedTot} and Fig.~\ref{fig:sedLit}).

\subsection{VLA: $p$-band, 1500, 4800 MHz and 14.4 GHz}
\label{sec:dRedVLA}

\begin{figure*}[tbh]
	\begin{center}
		\includegraphics[trim = 0 0 0 0, width=\textwidth]{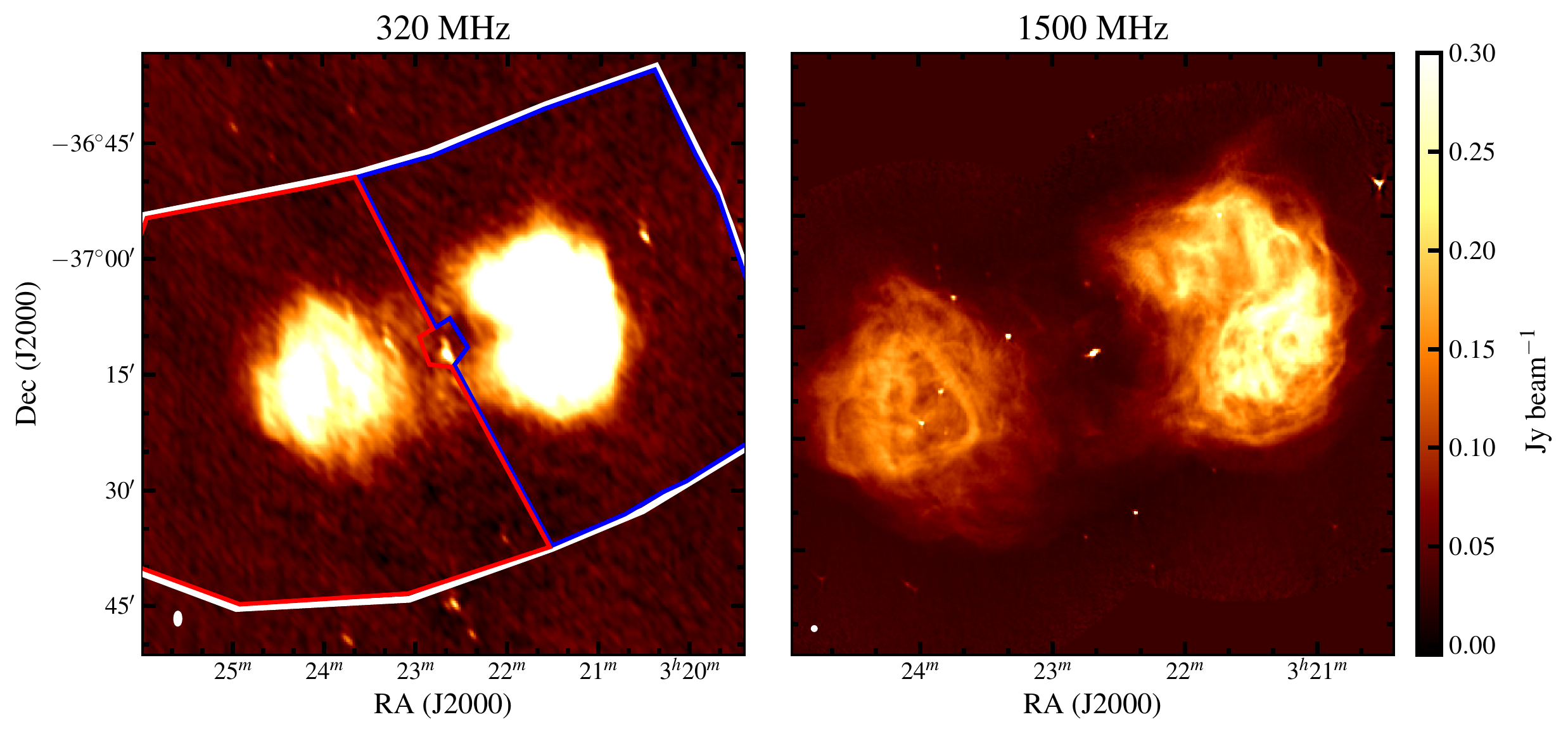}
		\caption{\forn\ seen by the VLA at 320 MHz and 1500 MHz. The colour-scale is the same in both panels. As in the previous Figures, the regions in the left panel show where we measure the flux density of the radio lobes.  The PSF of the images is shown in white in the bottom left corner.}
		\label{fig:fornVLA} 
	\end{center}
\end{figure*}

We select VLA observations of \forn\ to analyse both the flux density of the lobes (at $320$ MHz and $1500$ MHz) and the central emission of \forn\ (at $4.8$ GHz and $14.4$ GHz). The bandwidth, PSF and noise of the images are summarised in Table~\ref{tab:forObs}. The $p$-band observations of \forn\ were taken on May 12th 1989 with configurations B and C. The total integration time was $2.5$ hours. For the purposes of this work, we performed a new data reduction. Calibration and imaging were performed with the Astronomical Image Processing System ({\ttfamily AIPS}). De-convolution was done with multi-scale cleaning, and direction-dependent calibration has been used in the self-calibration phase. The resulting continuum image is shown in the left panel of Fig.~\ref{fig:fornVLA}. 

The image at $1500$ MHz shown in the right panel of the figure as presented for the first time in \citet{fomalont1989}. It is the mosaic of multiple pointings centred on the lobes, with all VLA configurations.

As illustrated in Sect.~\ref{sec:dRed}, the highest resolution observations are chosen to study the central radio emission. Both images at $4.8$ GHz and $14.4$ GHz (see Fig.~\ref{fig:fornCore}) have been taken from the NRAO VLA archive survey images~\footnote{\url{http://www.aoc.nrao.edu/~vlbacald/ArchIndex.shtml}} without applying any further modification. The image at 4.8 GHz was produced on February 21, 2007 from observations taken on October 6, 2002. The image at 14.4 GHz was produced on March 6, 2007 from observations taken on January 24, 2002.

\begin{figure*}[tbh]
	\begin{center}
		\includegraphics[trim = 0 0 0 0, width=\textwidth]{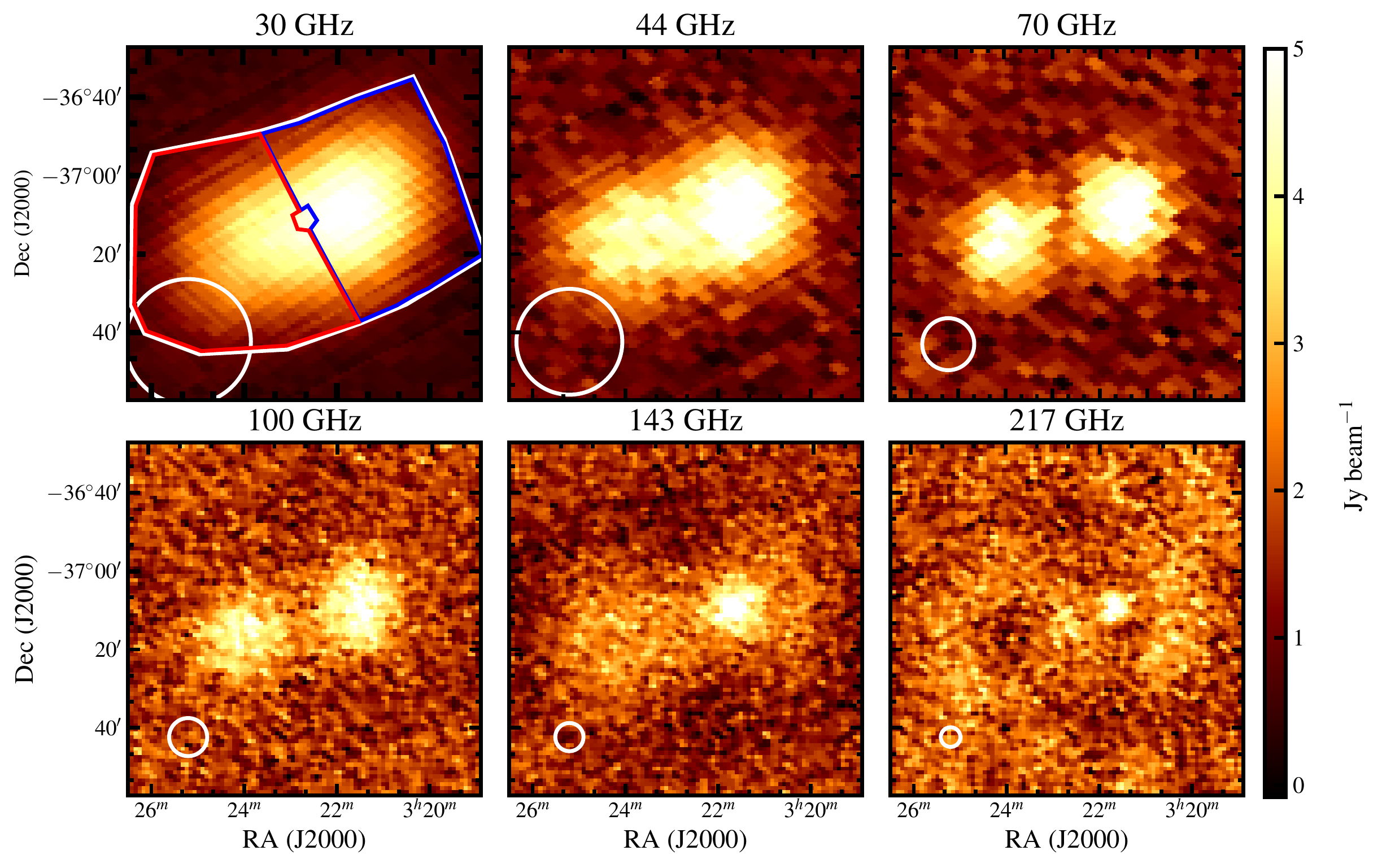}
		\caption{\forn\ seen by the Planck {\em LFI} at 30, 44 and 70 GHz, and by Planck {\em HFI} at $100$, $144$ and $217$ GHz. At $217$ GHz the lobes of \forn\ are undetected. The colour-scale is the same in all panels. The PSF of the images is shown in white in the bottom left corner. The blue and red regions in the top left panel indicate where we measure the flux density of the East and West lobes.}
		\label{fig:fornPlanck} 
	\end{center}
\end{figure*}

\subsection{Planck: 30 GHz - 217 GHz}
\label{sec:dRedPlanck}

We obtained images of \forn\ at $30$, $44$, $70$, $100$, $143$ and $217$ GHz analysing the final release of the \pl\ foreground maps~\citep{planck2018IV}. From these maps, we cut out an image four degree wide centred on \forn.

At frequencies $\gtrsim 100$ GHz thermal dust emission dominates over the non-thermal emission~\citep{planck2013XI,planck2016X}. The {\em Planck} collaboration released full-mission maps of the thermal dust component at $353$ GHz, $545$ GHz and $857$ GHz~\citep{planck2018IV}. The separation of the thermal dust emission from the other components can be done using different methods~\citep[][]{planck2016X,planck2016XLVIII}. Here, we use the products of the Generalized Needlet Internal Linear Combination (GNILC) component separation presented in \citet{planck2016XLVIII}. The SED of thermal dust is well described by a modified blackbody: 

\begin{equation}
\label{eq:dust}
I_\nu = \tau_{\nu_0} B_\nu(T_{\rm obs}) \bigg(\frac{\nu}{\nu_0}\bigg)^{\beta_{\rm obs}} 
\end{equation}

\noindent where $\nu_0=345$ GHz is the reference frequency at which the optical depth $\tau_{\nu_0}$ is estimated, $\beta_{\rm obs}$ is the spectral index of the dust and $T_{\rm obs}$ its temperature. $B_\nu(T_{\rm obs})$ indicates the flux density of the blackbody. Using the maps of the optical depth, temperature and spectral index of the dust emission~\citep{planck2018IV}, produced by the GNILC component separation including also IRAS observations at $100\, \mu$m, we produced maps of the dust thermal emission at $100$ GHz, $143$ GHz and $217$ GHz. These maps were subtracted pixel by pixel from the archival foreground emission maps. 

Analogously to the dust, also free-free emission contributes as diffuse foreground emission and needs to be subtracted from the \pl\ maps to determine the correct flux density of \forn. Using the maps of emission measure and electron temperature~\citep{planck2016X,planck2018IV}, produced by the GNILC component separation, we generated maps of the free-free emission at all HFI frequencies, and we subtracted them from the foreground dust-subtracted maps.

Besides the thermal dust and the free-free components, the Cosmic Infrared Background (CIB) also contributes to the flux of the \pl\ foreground maps. Since this component is diffuse on scales larger than \forn\, we corrected the images of \forn\ subtracting the average value of the CIB reported in \citet{planck2016I}.

\pl\ images are released in {\tt HEALPix} format, in units of $T_{\rm CMB}$ (\ie\ CMB anisotropies, $\Delta T/T$). We regridded the images to {\tt WCS} projection in equatorial coordinates, and we converted them to units of MJy/sr as follows:

\begin{equation}
S_{\nu_{\rm obs}}\, [{\rm MJy/sr}] = S_{\nu_{\rm obs}}\, [T_{\rm CMB}]\,\times 10^{4}\cdot\frac{C}{T_{\rm CMB}}\frac{\sigma}{e^x}\cdot\frac{x\,e^x}{e^x-1}
\end{equation}   

\noindent where $\sigma = \nu_{\rm obs}/29.979$, $x= 1.4388\sigma/T_{\rm CMB}$, $T_{\rm CMB}= 2.725$ K and $C = 2.99\times10^{10}$ cm s$^{-1}$. Then:

\begin{equation}
\begin{split}
S_{\nu_{\rm obs}}\, [{\rm Jy/beam}] =& 2.350443\times10^{-5}\cdot S_{\nu_{\rm obs}}\, [{\rm MJy/sr}] \times\\
& 2\pi\frac{\Theta_{\rm maj,\nu}['']}{2.35482}\frac{\Theta_{\rm min,\nu}['']}{2.35482}
\end{split}
\end{equation}

\noindent Figure~\ref{fig:fornPlanck} shows the centre of the field of view of the \pl\ images of \forn\ between 30 and 217 GHz in units of \Jyb, subtracted of the dust, free-free and CIB backgrounds. In the LFI band the lobes of \forn\ are unresolved, or resolved with at most one resolution element ($70$ GHz), in the HFI band the lobes are resolved, but fall below the sensitivity of the observations at $\gtrsim 143$ GHz.

As illustrated in Sect.~\ref{sec:dRed}, the noise in the \pl\ images surrounding \forn\ is not white. Hence, to make a reliable estimate on the error on the flux of \forn, we measure the noise as the standard deviation from the average flux density measured in the field of view in $100$ regions of the same size of the lobes of \forn\ (Fig.~\ref{fig:fornPlanck}). 

\section{The total flux density distribution of \forn}
\label{appendix:sed}

\begin{figure*}
	\begin{center}
		\includegraphics[trim = 0 0 0 0, width=0.44\textwidth]{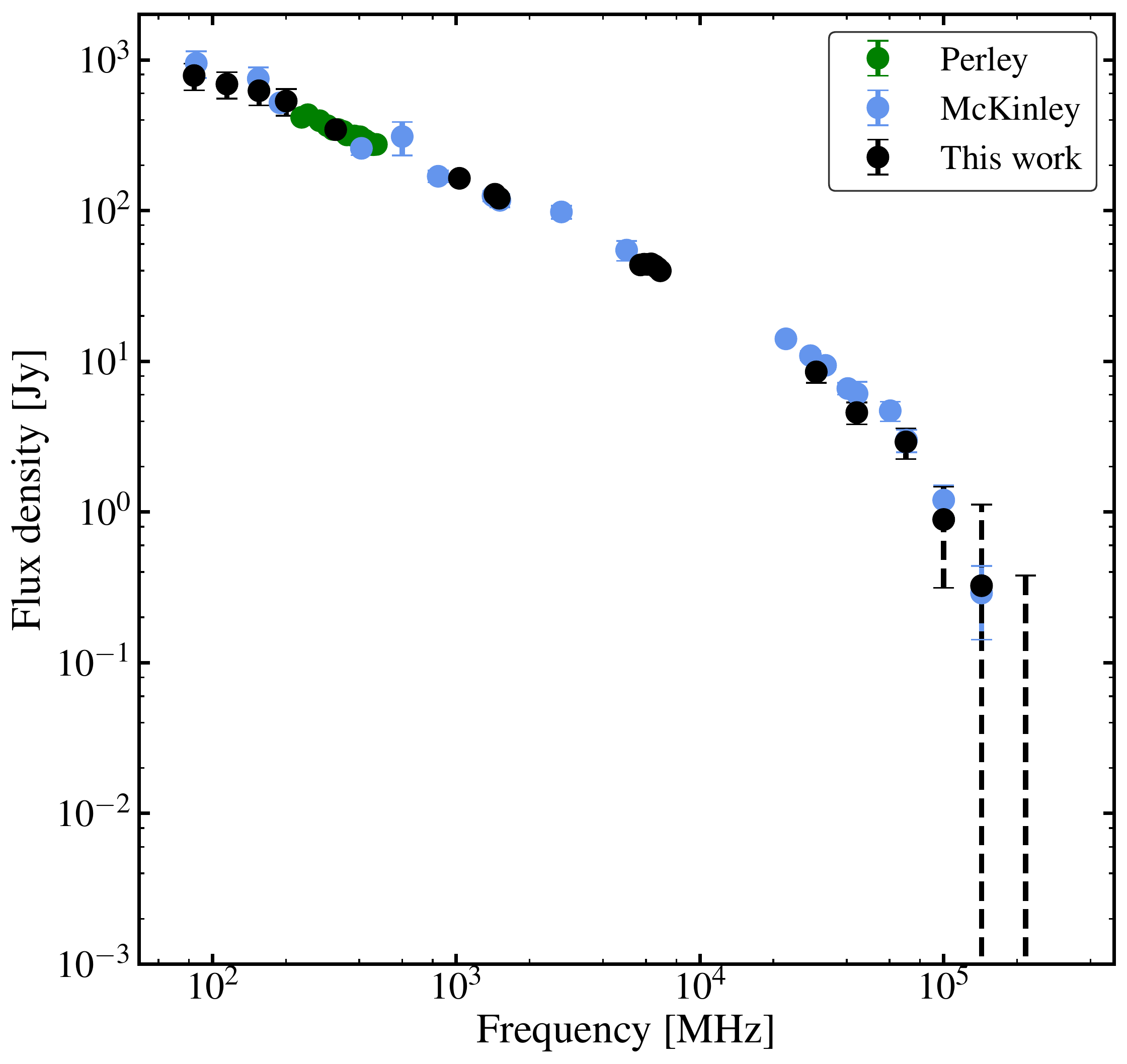}
		\includegraphics[trim = 0 0 0 0, width=0.432\textwidth]{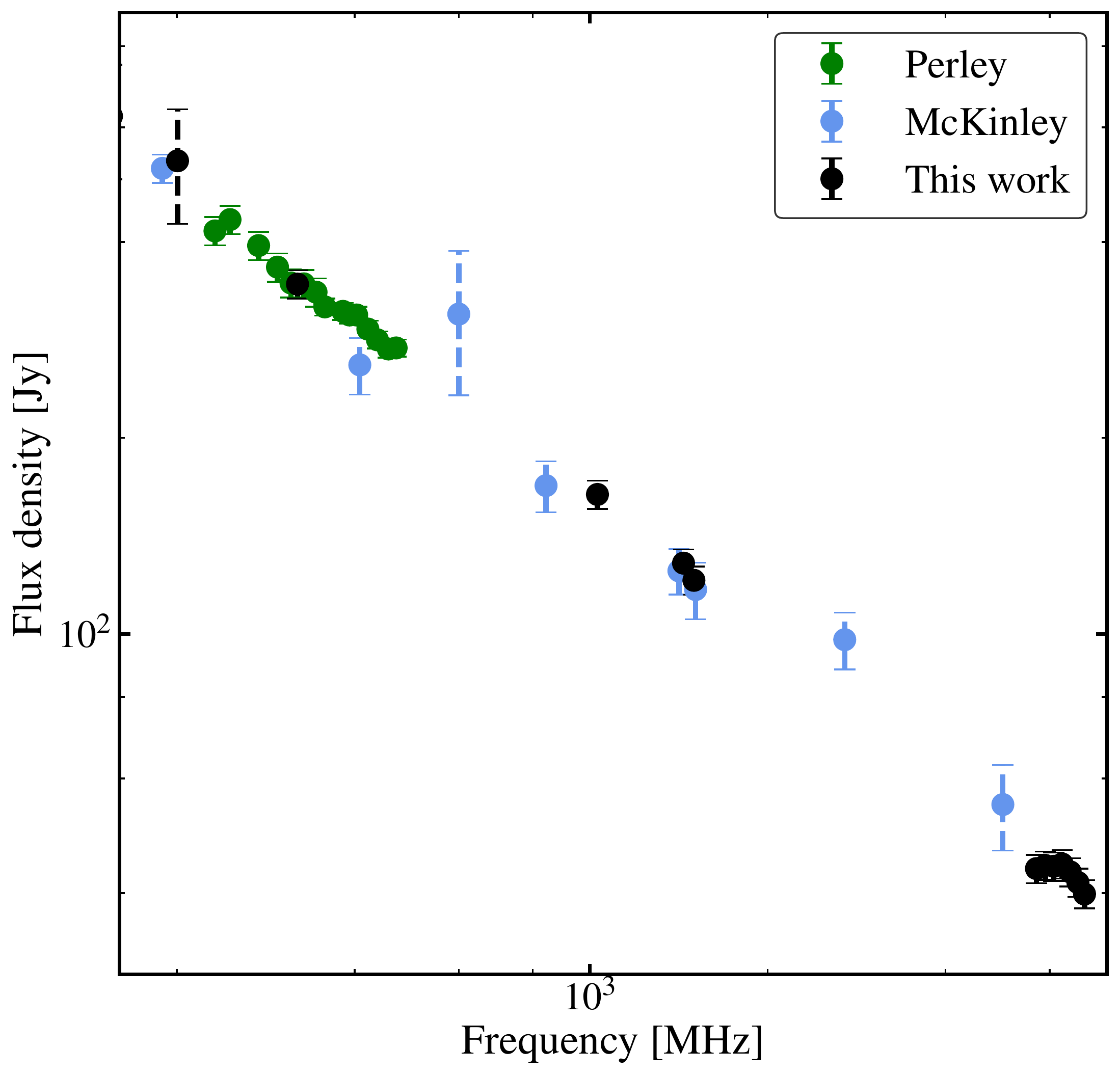}
		\caption{{\em Left panel}: Total flux density distribution of \forn\ between $84$~MHz and $217$~GHz measured in this work (black), between $4.7$~MHz and $143$~GHz by \citet{mckinley2015} (blue) and between $232$~MHz and $470$~MHz by \citet{perley2017} (green). {\em Right panel}: Zoom of the left panel between $200$~MHz and $7$~GHz, the bandwidth used to make the spectral index maps discussed in Sect.~\ref{sec:spModLobes}.}
		\label{fig:sedLit} 
	\end{center}
\end{figure*}

In Table~\ref{tab:sedTot} we show the total flux density distribution of \forn\ measured in this work along with the previously published measurements of \citet{mckinley2015} and \citet{perley2017}. \citet{mckinley2015} show the spectrum of \forn\ between $4.7$ MHz and $1.3\times10^{18}$ MHz, making use of archival observations as well as new MWA, WMAP, \pl\ and {\em Fermi}-LAT observations.\citet{perley2017} focus in extending the flux calibration scale of \forn\ between 200 and 500 MHz. As shown in Fig.~\ref{fig:sedLit}, the measurements of the total flux density of \forn\ presented in this work are compatible within the errors with the measurements previously published. The left panel of the figure shows that the flux measured at 1.50 GHz by the VLA is slightly lower than what expected, compared to the MeerKAT measurement at 1.44 GHz. It is possible that the limited coverage of the VLA over short baselines misses part of the flux of the lobes ($\lesssim 5\%$). The measurements in the low SRT frequencies ($\nu\lesssim6300$ MHz) also seem not to be compatible with the monotonous trend of the spectrum from synchrotron radiation. These errors may be caused by the opacity correction we applied (Eq.\ref{eq:op}), that does not take into account the frequency-dependence of the opacity. This may cause an under-estimate of the flux in the low-frequency end of the band. These errors are negligible for the purposes of this study. We point out that this correction becomes important at SRT only for low-elevation observations, such as the one of \forn. 

\begin{table*}[tbh]
	\caption{Total flux density of \forn}
	\centering
	\label{tab:sedTot}
	\begin{tabularx}{0.9\textwidth}{X c c c c}  
		\hline\hline		
		Frequency	& Flux density (this work)	& Flux density (McK)					& Flux density (PeR)	& Telescope/Reference\\
		$[$GHz$]$		&  [Jy]						&	[Jy]								& [Jy]				& [-]\\
		\hline
		$4.70\times10^{-3}$			&							&	13500$\pm20\%$ (*)					& 						&	\citet{ellis1966}		\\
		$1.83\times10^{-2}$		&							&	3500$\pm20\%$ 						&						&	\citet{shain1954}	\\
		$1.97\times10^{-2}$		&							&	4300$\pm20\%$ 						&						&	\citet{shain1958}	\\
		$2.99\times10^{-2}$		&							&	2120$\pm10\%$ 						&						&	\citet{finlay1973}	\\
		$8.40\times10^{-2}$			&	$788\pm158$				&	950$\pm20\%$ 						&						&	MWA	\\
		$0.118$			&	$692\pm138$				&										&						&	MWA	\\
        $0.154$			&	$642\pm125$				&	750$\pm19\%$						&						&	MWA	\\
        $0.189$			&							&	519$\pm5\%$ (*)						&						&	\citet{bernardi2013}	\\
        $0.200$			&	$533\pm107$				&										&						&	MWA	\\
        $0.232$			&							&										&	416$\pm5\%$			&	VLA	\\
		$0.246$			&							&										&	433$\pm5\%$			&	VLA	\\
		$0.275$			&							&										&	395$\pm5\%$			&	VLA	\\
		$0.296$			&							&										&	366$\pm5\%$			&	VLA	\\
		$0.312$			&							&										&	346$\pm5\%$			&	VLA	\\
		$0.320$			& $343\pm17$				&										&						&	VLA	\\
		$0.328$			&							&										&	345$\pm5\%$			&	VLA	\\
		$0.344$			&							&										&	335$\pm5\%$			&	VLA	\\
		$0.356$			&							&										&	318$\pm3\%$			&	VLA	\\
		$0.382$			&							&										&	313$\pm3\%$			&	VLA	\\
		$0.392$			&							&										&	309$\pm3\%$			&	VLA	\\
		$0.400$			&							&	140$\pm10\%$ (*)					&						&	\citet{mcgee1955}	\\
		$0.403$			&							&										&	309$\pm3\%$			&	VLA	\\
		$0.408$			&							&	259$\pm10\%$ 						&						& 	\citet{cameron1971}	\\
		$0.421$			&							&										&	294$\pm3\%$			&	VLA	\\
		$0.437$			&							&										&	283$\pm3\%$			&	VLA	\\
		$0.456$			&							&										&	274$\pm3\%$			&	VLA	\\
		$0.470$			&							&										&	275$\pm3\%$			&	VLA	\\
		$0.600$			&							&	310$\pm25\%$						&						&	\citet{piddington1956}	\\
		$0.843$			&							&	169$\pm9\%$							&						&	\citet{jones1992}		\\
		$1.03$		&	$163\pm8$			&										&						&	\meer \\
		$1.42$		&							&	125$\pm8\%$							&						&	VLA		\\
		$1.44$		&	$128\pm	6$			&										&						&	\meer		\\		
		$1.50$		&	$121\pm6$			&	117$\pm10\%$						&						&	VLA	\\
		$2.70$		&							&	98$\pm10\%$							&						&	\citet{shimmins1971}	\\
		$5.00$		&							&	54$\pm5\%$							&						&	\citet{gardner1971}	\\
		$5.70$		&	$44\pm2$			&										&						&	SRT	\\
		$5.90$		&	$44\pm2$			&										&						&	SRT	\\
		$6.10$		&	$44\pm2$			&										&						&	SRT	\\
		$6.30$		&	$44\pm2$			&										&						&	SRT	\\
		$6.50$		&	$43\pm2$			&										&						&	SRT	\\
		$6.70$		&	$41\pm2$			&										&						&	SRT	\\
		$6.87$		&	$39\pm1$			&										&						&	SRT	\\
		$22.5$		&							&	14$\pm5\%$							&						&	WMAP$^{\dag}$	\\
		$30.0$		&	$8.5\pm1.3$			&	11$\pm7\%$							&						&	\pl	\\
		$32.8$		&							&	9.4$\pm6\%$							&						&	WMAP	\\
		$40.4$ 		&							&	6.6$\pm9\%$							&						&	\pl	\\
		$44.0$		&	$4.6\pm0.7$				&	6.1$\pm20\%$						&						&	WMAP	\\
		$60.4$		&							&	4.7$\pm15\%$						&						&	WMAP	\\
		$70.0$		&	$2.9\pm0.5$			&	3.0$\pm17\%$						&						&	\pl	\\
		$100$		&	$0.9\pm0.4$			&	1.2$\pm25\%$						&						&	\pl	\\
		$143$		&	$0.3\pm0.5$			&	0.29$\pm51\%$						&						&	\pl	\\
		$217$		&	$(0.7)$			&										&						&	\pl	\\		
		\hline                           
	\end{tabularx}
	\tablefoot{Column (1) shows the frequencies at which we measure the flux density of \forn. Column (2) shows the flux densities measured in this work. Column (3) shows the flux densities shown in \citet{mckinley2015}. Column (4) shows the flux densities shown in \citet{perley2017}. Column (5) indicates the telescope used for the measurements, or the reference to the published fluxes. (*) indicates flux densities that have not been included in the study of the SED. \\ $^\dag$ WMAP: Wilkinson Microwave Anisotropy Probe \citep{bennett2013}.}
\end{table*}

\end{appendix}

\end{document}